\documentclass[aps,floatfix,nofootinbib,notitlepage,superscriptaddress,twocolumn]{revtex4-1}

\usepackage{booktabs}
\usepackage{lipsum}
\usepackage{graphicx}
\usepackage{amsmath,amssymb}
\usepackage{braket}
\usepackage{mathtools}
\usepackage{hyperref}
\usepackage{color}
\usepackage{float}
\usepackage{hyperref}
\usepackage[normalem]{ulem}
\usepackage{colonequals}
\usepackage{siunitx}
\usepackage{enumitem}
\usepackage{bbm}
\usepackage[dvipsnames]{xcolor}
\usepackage{tabularx}

\newcommand{\ov}{\overline}
\newcommand{\Dan}[1]{{\color{red} (Dan) #1}}

\newcommand{\Z}{\mathbb{Z}}
\newcommand{\C}{\mathbb{C}}
\newcommand{\F}{\mathcal{F}}
\newcommand{\n}[1]{\left| #1 \right|}
\newcommand{\st}[1]{\left\{ #1 \right\}}

\renewcommand{\v}[1]{\boldsymbol{#1}}
\DeclareMathOperator{\Tr}{Tr}
\DeclareMathOperator{\tr}{tr}

\newcommand{\R}{\mathbb{R}}
\newcommand{\im}{\operatorname{Im}}
\newcommand{\re}{\operatorname{Re}}

\newcommand{\tbg}{TBG}

\newcommand{\dr}{d^2 \v{r}}

\newcommand{\bz}{\mathrm{BZ}}

\newcommand{\Asample}{A_s}

\renewcommand{\c}{\hat{c}}
\renewcommand{\d}{\hat{c}^\dagger}

\renewcommand{\k}{\v{k}}
\newcommand{\q}{\v{q}}
\newcommand{\br}{\v{r}}
\newcommand{\bk}{{\v k}}
\newcommand{\bq}{{\v q}}

\newcommand{\bA}{{\v A}}

\newcommand{\bnabla}{{\v \nabla}}

\renewcommand{\H}{\hat{H}^{(3)}}

\DeclarePairedDelimiter\abs{\lvert}{\rvert}%
\DeclarePairedDelimiter\norm{\lVert}{\rVert}%
\makeatletter
\let\oldabs\abs
\def\abs{\@ifstar{\oldabs}{\oldabs*}}
\let\oldnorm\norm
\def\norm{\@ifstar{\oldnorm}{\oldnorm*}}
\makeatother

\begin{document}

\title{Field-tuned and zero-field fractional Chern insulators in magic angle graphene}

\author{Daniel Parker}
\affiliation{Department of Physics, Harvard University, Cambridge, MA 02138, USA}

\author{Patrick Ledwith}
\affiliation{Department of Physics, Harvard University, Cambridge, MA 02138, USA}

\author{Eslam Khalaf}
\affiliation{Department of Physics, Harvard University, Cambridge, MA 02138, USA}

\author{Tomohiro Soejima}
\affiliation{Department of Physics, University of California, Berkeley, CA 94720, USA}

\author{Johannes Hauschild}
\affiliation{Department of Physics, University of California, Berkeley, CA 94720, USA}

\author{Yonglong Xie}
\affiliation{Department of Physics, Harvard University, Cambridge, MA 02138, USA}
\affiliation{Department of Physics, Massachusetts Institute of Technology, Cambridge, MA 02139, USA}

\author{Andrew Pierce}
\affiliation{Department of Physics, Harvard University, Cambridge, MA 02138, USA}

\author{Michael P. Zaletel}
\affiliation{Department of Physics, University of California, Berkeley, CA 94720, USA}
\affiliation{Materials Sciences Division, Lawrence Berkeley National Laboratory, Berkeley, California 94720, USA}

\author{Amir Yacoby}
\affiliation{Department of Physics, Harvard University, Cambridge, MA 02138, USA}
\affiliation{John A. Paulson School of Applied Sciences and Engineering, Harvard University, Cambridge Massachusetts 02138 USA}

\author{Ashvin Vishwanath}
\affiliation{Department of Physics, Harvard University, Cambridge, MA 02138, USA}

\begin{abstract}
In contrast to the fractional quantum Hall  (FQH) effect, where electron density fixes the applied magnetic field, 
fractional Chern insulators (FCIs) can realize FQH states in comparatively weak or even zero magnetic fields.  Previous theoretical work highlighted magic angle graphene as a promising FCI platform, satisfying the twin requirements of flat bands and lowest-Landau-level-like quantum geometry.  Indeed, recent experiments have  demonstrated FCIs in magic angle graphene  with weak magnetic fields. Here we conduct a detailed theoretical study of the most prominent FCI state observed, and  clarify the role of the magnetic field in stabilizing this state. We introduce two new technical tools: first, we generalize the notion of ideal quantum geometry to Hofstadter minibands and, second, we extend the Hartree-Fock theory of magic-angle graphene to finite field, to account for the interaction generated bandwidth. We show that magnetic field both dramatically reduces the effective bandwidth and improves the quantum geometry for hosting FCIs. Using density matrix renormalization group (DMRG) simulations of a microscopic model of magic angle graphene,  we establish the regime of bandwidth and quantum geometry indicators where FCIs are stabilized. Further characterizing the finite-field bands by the same quantities we show how a zero-field charge density wave state gives way to an FCI state at a magnetic flux consistent with experiment. We also speculate on the other FCIs seen in the same experiments, including  anomalous incompressible states and  even-denominator fractions which may host non-Abelian states. Finally, when bandwidth is the limiting factor, we propose a range of experimental parameters where FCIs should appear at zero magnetic field.
\end{abstract}

\maketitle

\section{Introduction}

The fractional quantum Hall (FQHE) effect is a dramatic manifestation of strong correlations \cite{FQHbook}. On a fundamental level, the FQHE was the first experimental realization of anyonic excitations:
quasiparticles with fractional statistics and quantum numbers including fractional charge. On a practical level, the FQHE has been proposed as a platform for fault-tolerant
topological quantum computers. However, this possibility is hindered by the relatively large external magnetic field required along with small electron density, low temperatures, and clean samples.

Fractional Chern insulators realize the celebrated fractional quantum hall effect in a crystalline setting \cite{FCINeupertChamon, FCIDonnaSheng, FCIBernevigRegnault}. Just as Chern insulators promote the integer quantum Hall effect to lattice systems, fractional Chern insulators (FCIs) obtained on partially filling a Chern band realize an intrinsic topological phase --- including anyonic quasiparticles --- without continuous translation symmetry and potentially at weak or even zero applied magnetic field. The potential for enhanced energy scales and for making hybrid devices with superconductors  has made the realization of a weak field FCI a central goal in the field of topological materials.

More generally, FCIs can be defined as FQH states enriched by lattice translation symmetry. If $ \nu= t\phi+s$ labels the trajectory of an incompressible phase with a filling of $\nu$ electrons per unit cell, and  $\phi$ flux-quanta per unit cell, then FCIs correspond to the case where both $t$ and $s$ are fractional, while lattice  translation symmetry is preserved. Such FCI states were observed \cite{spanton2018observation} in Hofstadter bands of a bilayer graphene (BLG)
heterostructure aligned with hexagonal boron nitride (hBN) at relatively large ($>25$ T) magnetic fields. In that setup, the band topology originates from the magnetic field, thus precluding FCIs in the zero-field
limit.

A route to realizing a weak field FCI phase builds on the possibility of reproducing the physics of the lowest Landau level (LLL).
The first step in this direction was the classic Haldane model \cite{HaldaneModel}, a lattice model constructed to give bands that are topologically equivalent to Landau levels, but without a magnetic field.
More than two decades later, this problem was revisited with a flurry of numerical works \cite{FCINeupertChamon, FCIDonnaSheng, FCIBernevigRegnault, FCIBergholtzLauchli, FCIZoology} on Haldane-like lattice models with interactions. At fractional filling, these models often realize an FCI phase, providing a proof of principle that FCIs could be realized. These numerical works, as well as subsequent analytical studies \cite{royBandGeometryFractional2014,jacksonGeometricStabilityTopological2015,parameswaran2013fractional,bergholtz2013topological,claassen2015position,Lee2017,Simon2020,meraKahlerGeometryChern2021,ozawaRelationsTopologyQuantum2021,Zhang2021,MeraDirac,MeraOzawaEngineering2021,varjas2021topological}, 
identified conditions beyond band topology which favor FCIs: (i) an isolated band with a small bandwidth, separated by a bandgap larger than the interaction strength, so that the projected interaction dominates the physics; 
(ii) a uniform distribution of the Berry curvature $\mathcal{F}$; and (iii) a relation called the ``trace condition" between the Berry curvature and the quantum (Fubini-Study) metric $g$ of the wavefunctions.
These conditions measure how close the system is to the lowest Landau level, which is a special point in the space of Chern bands (Fig. \ref{fig:overview}a) and is where the fractional quantum Hall effect is known to be realized. Note that conditions (i) and (ii) are satisfied by all Landau levels - the condition (iii) is crucial because it specifically selects the physics of the \emph{lowest} Landau level (LLL). An early result by \cite{royBandGeometryFractional2014} (see also \cite{varjas2021topological}) showed that the conditions of vanishing bandwidth, uniform Berry curvature, and the trace condition $\tr g = \n{\mathcal F}$ are sufficient to reproduce the physics of the LLL exactly. However, despite attempts \cite{okamoto}, the question of constructing experimentally realistic models of FCIs has remained an open one.

\begin{figure}
    \centering
    \includegraphics[width=0.8\linewidth]{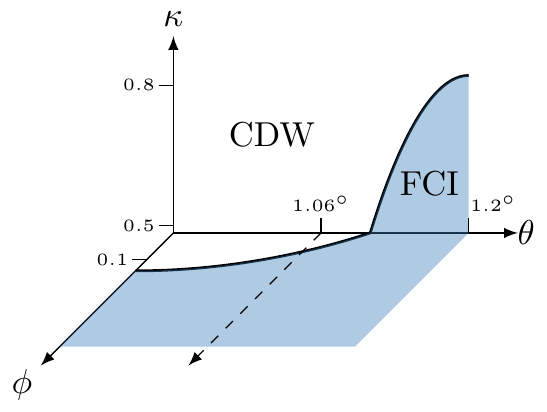}
    \caption{Schematic phase diagram of FCIs in \tbg{} at filling $\nu=3+\tfrac{2}{3} + \tfrac{1}{3}\phi$. Here $\theta$ is the twist angle, $\kappa$ is the chiral ratio, and $\phi$ is the number of magnetic flux quanta per unit cell. The sample measured in \cite{xie2021fractional} falls approximately along the dashed line.}
    \label{fig:my_label}
\end{figure}

Shortly after the discovery of correlated states in magic angle twisted bilayer graphene (TBG) \cite{PabloMott, PabloSC}, several theoretical works recognized that its non-interacting band structure possesses a subtle topological character \cite{PoTBG, songAllMagicAngles2019, HejaziTopological,  AhnStiefelWhitney}. For a single valley and spin species, the \tbg{} band structure consists of two narrow bands connected by a pair of Dirac cones with the \emph{same} chirality. An immediate consequence is that the most natural way to open a gap --- a sublattice potential, generated by alignment with an hBN substrate --- yields bands with Chern numbers $\pm 1$ \cite{BultinckAnomalousHall, ZhangAnomalousHall} (Regular graphene, by contrast, requires a Haldane mass to generate non-trival Chern bands; a sublattice potential leads to an atomic insulator). This led to the prediction \cite{BultinckAnomalousHall, ZhangAnomalousHall} that a spin and valley polarized state in TBG with a sublattice potential will the exhibit the quantum anomalous Hall effect. This prediction was borne out in experiments, first through the observation of orbital magnetization and anomalous Hall effect in Ref.~\cite{SharpeQAH}, then followed by the observation of $e^2/h$ quantization plateau for the anomalous Hall effect in the vicinity of $\nu = 3$ \cite{YoungQAH} in TBG samples aligned with hBN.

These theoretical and experimental findings suggest that TBG naturally exhibits two of the three requirements for realizing FCI phases: the right band topology and the right energetics, with an isolated Chern band whose bandwidth is much smaller than the interaction scale. The remaining requirements of good quantum geometry also turn out, quite remarkably, to be realized in the \tbg{} wavefunctions.
The special character of the TBG wavefunctions was first identified in Ref.~\cite{TarnopolskyChiralModel} which found that these wavefunctions have a 
special holomorphic structure, 
similar to the wavefunctions of the LLL, in a certain limit. This limit, dubbed the chiral limit, is obtained by selectively turning off the Moir\'e tunneling terms between the two TBG layers which connect the same (AA or BB) sublattice. Later on, Ref.~\cite{ledwithFractionalChernInsulator2020} investigated the quantum geometry of the flatband wavefunctions in this limit and found that they satisfy almost ideal conditions for realizing FCIs: the trace  condition holds exactly, while the Berry curvature is relatively uniform. Relatedly, in this limit, one may use the spatial dependence of the wavefunctions to write down ``Laughlin wavefunctions" that are zero energy ground states of short range interaction potentials \cite{ledwithFractionalChernInsulator2020,WangExactLandauLevel}. Upon deviating from the chiral limit, one can investigate how band geometry changes as we change the ratio of same-sublattice to opposite-sublattice tunneling, henceforth denoted by $\kappa$. As shown in Ref.~\cite{ledwithFractionalChernInsulator2020}, these band geometry indicators increase away from the chiral limit $\kappa = 0$, but remain favorable for FCIs up to around $\kappa \approx 0.6 - 0.7$, then increasing rapidly beyond this point. These analytic considerations were corroborated by parallel \cite{repellinChernBandsTwisted2019, abouelkomsan2020particle} and subsequent numerical works \cite{wilhelm2021interplay} which identified an FCI phase at small values of $\kappa$ that destabilizes around $\kappa \approx 0.7$.

While these results suggest that the main limiting factor to realizing FCIs in TBG is the quantum geometry controlled by the chiral ratio $\kappa$, another complication arises from the interaction-generated dispersion. Although the bare bandwidth of the TBG bands obtained from the Bistritzer-Macdonald model is only a few meV, significantly smaller than the interaction scale, several sources of dispersion including renormalization effects from the remote bands \cite{XieMacdonald,repellin2020ferromagnetism, ShangNematic, bultinck2020ground} and strain \cite{BiFuStrain, parker2020strain} can alter the dispersion. More importantly, at non-zero integer fillings, the  dispersion is modified due to a Hartree term arising from the filled bands,  which gives rise to a significant bandwidth. For instance, at $\nu = 3$, the electron dispersion posses a strong dip at the $\Gamma$ point with a total bandwidth of $\sim 30$ meV \cite{XieMacdonald, ShangNematic, GuineaHF, TBGV, PierceCDW, VafekBernevigCascade}, which is comparable to or even exceeds the interaction scale. One goal of this work is to investigate how this dispersion of \tbg{} affects the possibility of FCI phases.

These questions have become especially topical in light of the recent experiment which reports FCI phases in hBN-aligned TBG at several fractional fillings, particularly between $\nu = 3$ and $4$ \cite{xie2021fractional}. While such phases have not been seen at zero magnetic field, the field needed to stabilize them is around $5-6$T --- only a fraction ($\sim20$\%) of the field required to thread one flux quantum per Moir\'e unit cell. Equivalently, to realize a $1/3$ Laughlin fractional quantum Hall state at the density of holes at $\nu=3\frac23$, requires a five times larger field than the ${\sim} 5 \si{\tesla}$ required to induce the FCI, justifying the `weak field' label.  This fact, combined with the simultaneous observation of the Chern insulator at the neighboring $\nu = 3$ state, which is most simply explained as having spontaneous spin and valley polarization, strongly suggest that the topology is provided by the native TBG bands rather than the magnetic field induced Hofstadter bands. In this picture, the magnetic field plays the less fundamental role of improving the energetic and band geometric conditions that favor FCI states, rather than being the fundamental source of band topology. This would mean that FCI phases at $3 \le \nu \le 4$ in \tbg{} are {\em adiabatically connected} to a zero field state. In principle, slightly altering the system could allow an FCI in \tbg{} at zero field. Motivated by this tantalizing possibility, this study examines the landscape of FCIs in \tbg{} in detail, and will arrive at concrete experimental parameters where FCIs may be observable without external field.

The remainder of this work is organized as follows. To orient ourselves, we begin with a density matrix renormalization group (DMRG) study in Section \ref{sec:DMRG_FCIs} on a microscopic model of \tbg{} at zero field with filling $\nu=3+\tfrac{2}{3}$. We find a large FCI phase, terminating slightly away from our estimate of  experimentally-realistic parameters. To explain its extent, we define the ``FCI indicators", which quantify how suitable the quantum geometry and interaction-induced bandwidth are for FCIs.

We then study the effect of finite field. As the FCI indicators have traditionally been studied in the setting of a single band, we must upgrade them to work in the setting of interacting Hofstadter spectra with arbitrarily many bands. Sec. \ref{sec:multiband_FCI_indicators} generalizes the quantum geometry to this setting. In Section \ref{sec:finite_field} we construct a model for \tbg{} at finite field --- including Hartree-Fock corrections --- and compute the quantum geometry and interaction-induced bandwidth, allowing us to infer a phase diagram.

We find that even small magnetic fields are sufficient to improve the quantum geometry and collapse the ``Hartree dip." We explain the rapid collapse of the Hartree dip with external field using a Berry phase corrected semiclassical analysis and obtain excellent quantitative agreement with the exact calculation.

Our study culminates in Section \ref{sec:large_angles}, where we apply the clear physical picture gleaned from these analytic and numerical techniques to determine where FCIs may appear at zero field. Specifically, we identify a range of parameters, which seem possible to access in experiment, where FCIs appear numerically at zero field. Finally, Section \ref{sec:expt_comparison} discusses FCIs found in TBG in the experiment of  Ref.~\cite{xie2021fractional}; we divide the FCIs into two classes, those that have a ``conventional" explanation in that they may survive under adiabatic continuity to the LLL and those that lie beyond this picture, either through breaking translation symmetry or through the formation of an exotic FCI. We briefly comment on the states with even denominators. We conclude in Section \ref{sec:conclusions}. Extensive technical Appendices contain all details of our models, calculations, and classification.

\section{DMRG Study}
\label{sec:DMRG_FCIs}

\subsection{Hamiltonian}

\begin{figure}
    \begin{center}
    	\includegraphics[width=\linewidth]{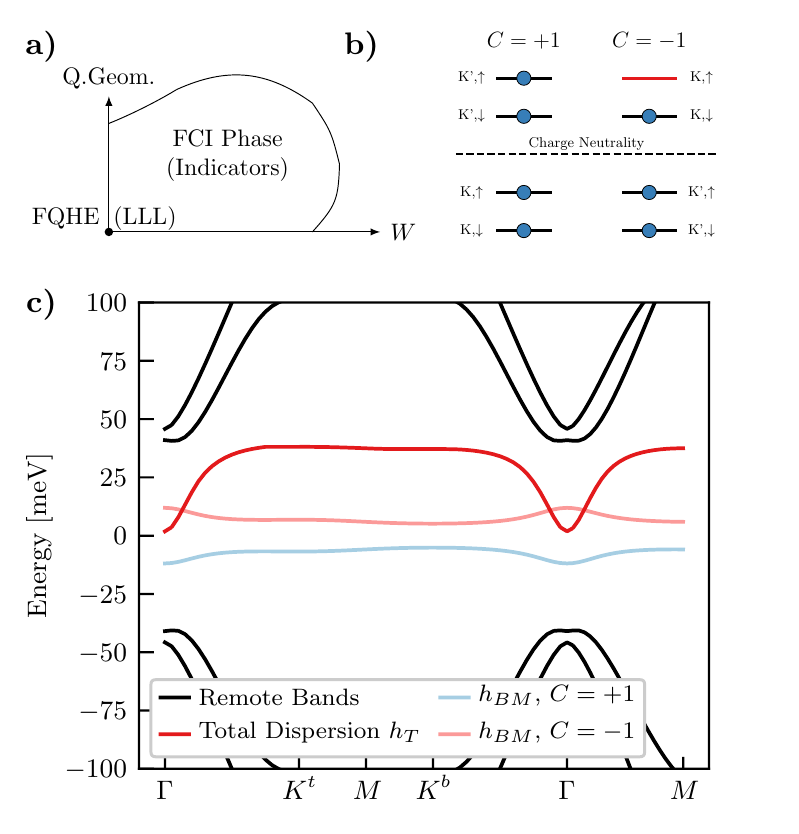}
    \end{center}
	\caption{
		\textbf{a)} Schematic phase diagram of an FCI phase. \textbf{b)} The spin- and valley-polarized $\nu=3$ state relevant to FCIs in magic-angle graphene.
	\textbf{c)} Bandstructure of hBN-aligned TBG from the BM model. Active bands are colored, and the dark red line is the total dispersion $\hat{h}_{T}$, accounting for interactions, which sets the relevant bandwidth for FCIs.}
    \label{fig:overview}
\end{figure}

To orient ourselves in the landscape of FCIs in \tbg{}, we begin with a DMRG study of the FCI at $\nu=3+\tfrac{2}{3}$ in a realistic model of TBG \cite{soejima2020efficient,parker2020strain}. We employ a standard model for \tbg{}: Coulomb interactions in the active (flat) bands. Here we recall its essential points;  full details are given in Appendices \ref{app:continuum_formalism}--\ref{app:Hamiltonian_construction}, and is reviewed in \cite{ledwith2021strong}. Here and below we assume spin- and valley-polarization in the interval $3 \le \nu \le 4$, allowing us to consider a single active band (Fig. \ref{fig:overview}b) --- we discuss this point in detail shortly. The dispersion of this well-isolated Chern band is shown in Fig. \ref{fig:overview}c. The model
\begin{subequations}
\begin{align}
	\H &= \lambda \hat{h}_{\mathrm{T}} + \frac{1}{2A} \sum_{\v{q}} V_{\v{q}} : \delta\widehat{\rho}_{\v{q}} \delta\widehat{\rho}_{-\v{q}}:\\
	\hat{h}_{T} &= \hat{h}_{BM} + \hat{h}_{HF}[P] 
\end{align}\label{eq:projected_Hamiltonian}\end{subequations}
where $V_{\q}$ represents gate-screened Coulomb interactions and $\delta\widehat{\rho}_{\v{q}}$ is the charge density at wavevector $\v{q}$, measured relative to the $\nu=3$ polarized state. The total dispersion $\hat{h}_T$ is composed of the BM band $\hat{h}_{BM}$ with an mass from hBN alignment and a contribution to account for the background charge density of all the filled bands at the Hartree-Fock level: $\hat{h}_{H/F} = \sum_{k} \hat{c}_{\k}^\dagger h_{H/F}(\v{k}) \hat{c}_{\k}$ where
\begin{subequations}\begin{align}
	h_H(\v{k}) &= \frac{1}{A}\sum_{\v{G}} V_{\v{G}} \Lambda_{\v{G}}(\v{k}) \Tr[P^T \Lambda_{\v{G}}^\dagger]\\
	h_F(\v{k}) &= -\frac{1}{A}\sum_{\v{q}} V_{\v{q}} \Lambda_{\v{q}}(\v{k}) P^T(\v{k}+\v{q}) \Lambda_{\v{q}}(\v{k})^\dagger.
\end{align}\label{eq:Hartree_Fock_potentials}\end{subequations}
where $\Lambda_{\q}(\k) = \braket{u_{\k}|u_{\k+\q}}$ are form factors and $A$ is the sample area. Here $P$ is a correlation matrix in the space of active and remote bands of all flavors that encodes the background charge in the ``infinite temperature" subtraction scheme, which ensures that the effective dispersion at the charge neutral point $\nu=0$ matches the BM dispersion (see Appendix \ref{app:Hamiltonian_construction} for details). The contributions from $P$ are approximately three times the Hartree potential for a single band plus the Fock potential of a single band.

Our main assumption in restricting to a single active band is that the parent state at $\nu=3$ is given by completely filling seven of the eight non-interacting Chern bands of the BM Hamiltonian. In the chiral limit this is an exact eigenstate of the Hamiltonian (we must also assume that the bare dispersion $\hat{h}_{BM}$ is dominated by the Chern-diagonal hBN potential which is an excellent approximation near the magic angle). Away from the chiral limit, the projector to this band deforms adiabatically and, in principle, one needs to perform a fully self-consistent calculation to obtain the projector onto the single active band. However, it is known that the projector obtained from the non-interacting bands gives an accurate approximation to the self-consistent result \cite{bultinck2020ground,TBG1V} at zero field. We have verified this is also the case at finite field, which justifies using one-shot Hartree-Fock in Eq. \eqref{eq:projected_Hamiltonian}, which we will employ throughout.

Crucially, the total dispersion $\hat{h}_T$ is the physically relevant single-particle bandstructure for FCIs. As we have used charge density relative to the $\nu=3$ state and normal ordering in Eq. \eqref{eq:projected_Hamiltonian}, $\hat{h}_T$ gives the spectrum of single-electron excitations above the $\nu=3$ state in the restricted single-band Hilbert space.\footnote{We note that one could also consider the picture of doping the $\nu=4$ state with polarized holes. However, such holes would enter at the top of the electron band where the Berry curvature is almost vanishing, akin to doping a trivial band, and are thus not well-suited to form FCIs.} Eq. \eqref{eq:projected_Hamiltonian} fixes a decomposition of $\H$ into ``dispersion" and ``interaction" parts in the correct manner to apply psuedopotential arguments. Therefore we regard $\hat{h}_T$ as ``the" total dispersion. We underscore that the bandwidth of $\hat{h}_T$, shown in Fig. \ref{fig:overview}c, is dramatically larger than the corresponding BM bands.

The key parameters for the model are as follows.
\begin{itemize}
    \item \textit{The twist angle} $\theta$ ---  we work near the magic angle $\theta \approx 1.06^\circ$.
    \item \textit{The chiral ratio} $\kappa$ --- the ratio of intra- to inter-sublattice tunneling. The chiral limit is $\kappa =0$. 
    \item \textit{The dispersion scale} $\lambda$ --- an artificial parameter that reduces the dispersion from its physical value at $\lambda = 1$ to the limit of vanishing bandwidth $\lambda=0$.
\end{itemize}
We pause to comment on estimates of the chiral ratio parameter $\kappa$, which plays an important role. For a pristine, unrelaxed structure $\kappa = 1$. However, both in-plane lattice relaxation, which expands AB regions and contracts AA regions \cite{CarrKaxiras}, as well as out-of-plane relaxation, which increases the interlayer separation in AA regions relative to AB regions \cite{NamRelaxation},  reduce $\kappa$. Initial \textit{ab-initio} estimates based on out-of-plane relaxation alone \cite{Koshino2018outofplane} yielded $\kappa\approx 0.8$. Contributions from in-plane relaxation  \cite{NamRelaxation,Carr2018relax, CarrKaxiras, Carr2020review, Guinea2019Relax,TBorNotTB} further reduces this value, so $\kappa$ estimates vary from $0.5-0.8$ - see for example Ref.~\cite{Guinea2019Relax} in which the values of $\kappa$ vary dramatically for different choices of inter-atomic potentials and tight-binding parameterizations; such choices can be motivated by electronic screening or reduction of out-of-plane relaxation by the hBN substrate. A recent calculation in Ref. \cite{TBorNotTB} uses first principles calculations for these parameterizations \cite{Fang2016,Carr2018PressureDep} and yields $\kappa = 0.53$, although this calculation also depends on the chosen density
functional. We note that STM experiments can visualize the size of AA and AB regions and has been used to estimate a value of $\kappa$ around 0.8 \cite{xie2019spectroscopic}. However, we stress that this estimate only accounts for in-plane relaxation and neglects the effects of out-of-plane relaxation. The latter alone is sufficient to reduce $\kappa$ by around $15-20$\%  \cite{Koshino2018outofplane}. Ref. \cite{Das2021} extracts $\kappa$ through the dispersion of the remote bands and finds $\kappa = 0.7-0.8$. However, interaction renormalization of the band structure parameters, strain, or particle-hole symmetry breaking can alter the estimate. We also note that Ref. \cite{vafek2020renormalization} found $\kappa$ to decreases under renormalization group flow. We therefore treat $\kappa$ as a variable here, with  $0.5 <\kappa<0.8$ being our current best estimate of the physical range. Recently, Ref. \cite{TBorNotTB} proposed a multilayer structures realizing smaller $\kappa$ values. For the remaining parameters, we use tunneling strength $w_1 = \SI{110}{meV}$, gate distance $d = \SI{25}{\nano\meter}$, relative permitivity $\epsilon_r = 10$, and top layer hBN mass $m_t = \SI{30}{meV}$.

\subsection{DMRG}

Psuedopotential arguments \cite{ledwithFractionalChernInsulator2020} at $\kappa =0$ and the limit of vanishing bandwidth $\lambda =0$ find a Laughlin-like ground state at $\nu=3+\tfrac{2}{3}$. Conversely, large dispersions can lead to charge-density waves or Fermi liquids, and the competition between these possibilities is a matter of detailed energetics. To assess this competition, we study the phase diagram of $\H$ at $\nu=3+\tfrac{2}{3}$ with infinite DMRG.

As DMRG is limited to quasi-1d systems but Eq. \eqref{eq:projected_Hamiltonian} is a 2d continuum model with long-range interactions, we must use the specialized methods developed in \cite{parker2020local, soejima2020efficient, parker2020strain}. To overcome the dimensional barrier, we use an infinite cylinder geometry, previously employed to find FCI states with DMRG on simple lattice models \cite{grushin2015characterization,motruk2016density}, and in FQHE studies \cite{zaletel2013topological,zaletel2015infinite}. In particular, we take $L_y=8$ horizontal cuts through the Brillouin zone, and Fourier transform along the $x$-direction to produce Wannier-Qi states \cite{qi2011generic}. This gives real-space along the cylinder and (discrete) $y$-momenta around the cylinder. Unfortunately, expressing $\H$ in this basis leads to a matrix product operator (MPO) with bond dimension $\chi \sim 10^4 - 10^5$ --- far too large to use in practice. We apply the MPO compression technique of \cite{parker2020local, soejima2020efficient, parker2020strain} to reduce the bond dimension to $\chi \approx 500$ with errors of $10^{-3}$ \si{meV} or less.

\begin{figure}
	\begin{center}
    \includegraphics[width=\linewidth]{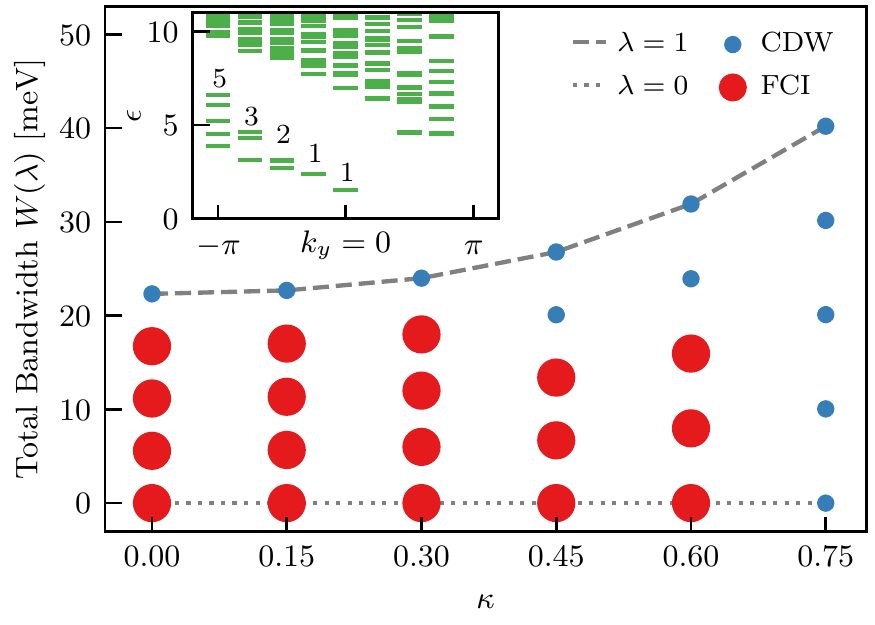}
\end{center}
	\caption{DMRG phase diagram of hBN-aligned magic angle graphene Eq. \eqref{eq:projected_Hamiltonian} at filling $\nu=3+\tfrac{2}{3}$ as a function of $\kappa$ and the bandwidth $W(\lambda)$ for $\lambda \in [0,1]$. Inset: entanglement spectrum at $\kappa=0.3,\lambda=0$ in the electric charge zero sector. The counts match those expected from the chiral edge of a Laughlin state. Parameters: $\theta = 1.06^\circ, m_t = \SI{30}{meV}$, $\epsilon_r = 10$, $d=\SI{25}{nm}$, $\chi = 256$.}
    \label{fig:dmrg}
\end{figure}

Fig. \ref{fig:dmrg} shows the phase diagram we find within the $(\kappa,\lambda)$ plane at $\nu= 3+\tfrac{2}{3}$ assuming spin- and valley-polarization.  With the full dispersion $\lambda = 1$ a Hartree-Fock picture suggests a charge-density wave (CDW) order is preferred \cite{VafekBraiding, PierceCDW}. Most of the system's bandwidth comes from the Hartree dip at the $\Gamma$ point of ${\sim}\SI{30}{meV}$ (Fig. \ref{fig:FCI_indicators}a). At two-thirds filling, it is energetically favorable to break translation symmetry by tripling the unit cell, so that most of the spectral weight near $\Gamma$ goes into a mini-band that is pushed below the Fermi level \cite{PierceCDW}. We define an order parameter for translation symmetry along the cylinder $O_T =  \sum_{k_y} \sum_{x=1}^{3} \braket{ \hat{n}_{x+1 (mod 3) ,k_y}} - \braket{\hat{n}_{x,k_y}}$ in terms of the Fermions $\hat{c}_{x,k_y}$ in unit cell $x$ and momentum $k_y$. Here the unit cell is $3L_y$, so this measures the magnitude of density fluctuations between unit cells, which vanishes if translation symmetry is preserved. This is large at $\lambda=1$, suggesting CDW order is present. For $\kappa > 0.6$, $O_T$ is large for all values of $\lambda$, suggesting CDW order there as well. Some regions of the phase diagram may also be metallic. This result is consistent with the CDW phase seen in experiment at filling $\nu=3+\tfrac{2}{3}$ in small magnetic fields \cite{xie2021fractional}.

For $\lambda \lesssim 3/4$ and $\kappa \le 0.6$, a phase is observed with small translation-breaking $O_T$, small entanglement entropy, and a short correlation length. To identify it, we consider its entanglement spectrum. One of the most distinctive features of an FCI is a chiral edge mode described by a chiral conformal field theory (CFT). If one makes an entanglement cut in a system, the entanglement Hamiltonian will be in the same universality class as the edge Hamiltonian, and the lowest energy states in the entanglement spectrum will match the CFT \cite{li2008entanglement}. For the case of $\nu = 1/3$, one expects the lowest states to have counts $\st{1,1,2,3,5,7,\dots}$ (which come from the CFT partition function) and chiral momentum around the cut \cite{motruk2016density}. The inset of Fig. \ref{fig:dmrg} shows this expected counting, confirming the presence of an FCI phase.\footnote{As the Wannier-Qi basis states used in DMRG change when flux is added \cite{soejima2020efficient}, one cannot easily perform a flux-pumping experiment as a secondary verification of the presence of the FCI. Nevertheless, the edge spectrum alone is a highly non-trivial check.} As there is residual CDW order inside the FCI phase (the default expectation on a finite cylinder, even with psuedopotential interactions), it is difficult to assess the phase transition from FCI to CDW order \cite{seidel2005incompressible}. Appendix \ref{app:numerical_details} gives further numerical details.

\subsection{FCI Indicators}

\begin{figure}
    \begin{center}
		\hspace{-1em}
		\includegraphics[width=\linewidth]{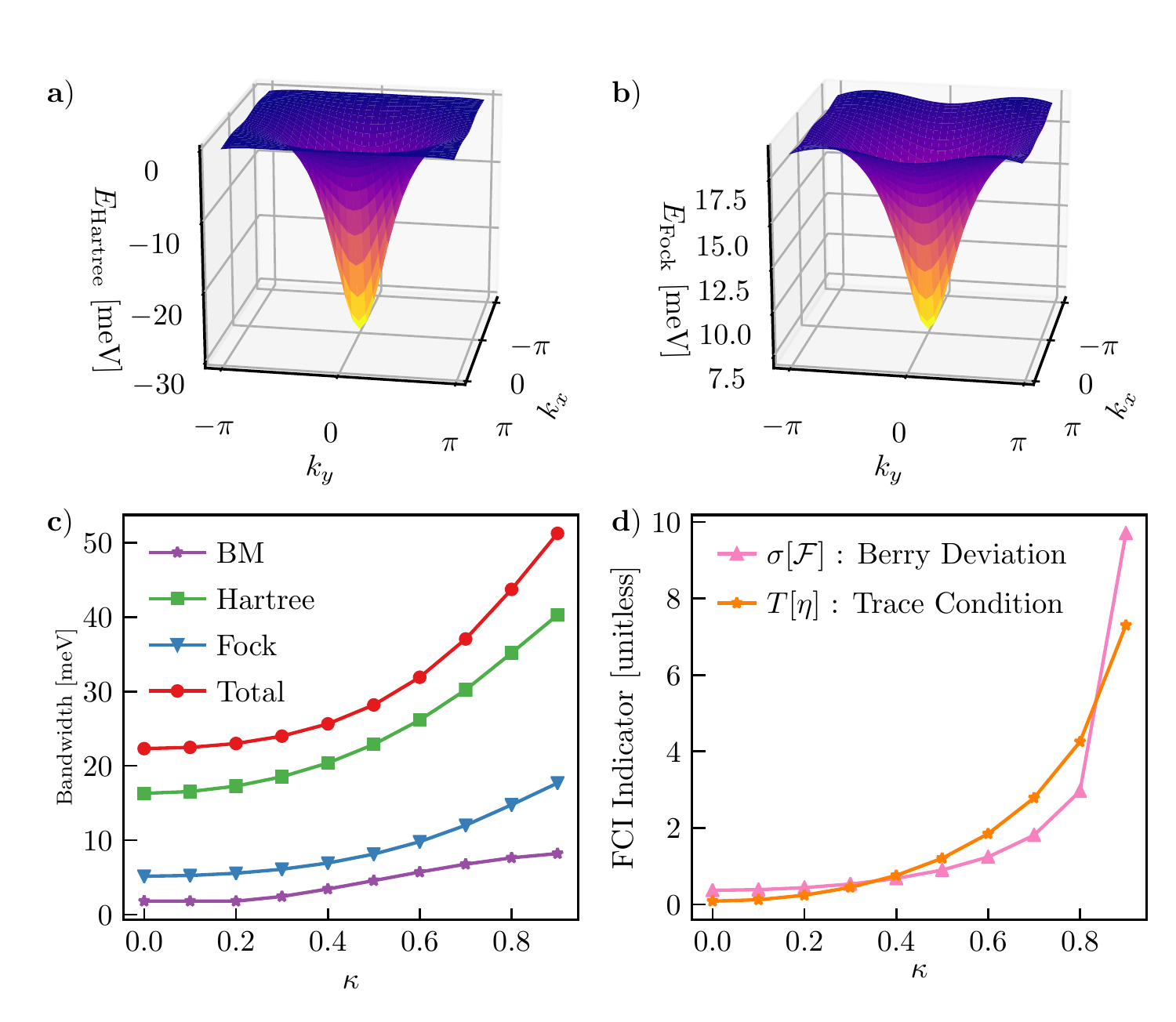}
    \end{center}
	\caption{\textbf{a)} Hartree potential $\hat{h}_H$ in the moir\'e Brillouin zone for $\nu=3$. The dip is centered at $\Gamma$. \textbf{b)} The Fock potential $\hat{h}_F$ for $\nu=3$. Its shape is roughly the same as the Hartree potential, but $\sim 1/3$ as large. \textbf{c)} Bandwidths of each part of the total dispersion as a function of $\kappa$. Note that the BM dispersion has a small peak at $\Gamma$, while Hartree-Fock contributions have corresponding dip. \textbf{d)} Measures of the non-ideality of the quantum geometry, defined in the text. The chiral limit $\kappa=0$ has almost ideal quantum geometry, and the geometry becomes quickly worse for $\kappa > 0.7$. Parameters match Fig. \ref{fig:dmrg}.}
    \label{fig:FCI_indicators}
\end{figure}

The DMRG phase diagram above tells us that \tbg{} does indeed have an FCI phase which is destroyed when either the total bandwidth or the chiral ratio $\kappa$ is too large. A clear physical picture for these phase boundaries is provided by the ``FCI indicators", which we now motivate and define.

At face value, the FCI indicators quantify how close the spectrum and wavefunction of a band is to the lowest Landau level. The logic is that adding interactions to a band close to the LLL should have a ground state adiabatically connected to the Laughlin wavefunction, and thus inside an FCI phase (see Fig. \ref{fig:overview}a). When is a band close to the LLL? The spectrum is simple to characterize: it's close to the LLL if the bandwidth is small. The wavefunction is more complex; we characterize it in a gauge-invariant, model-independent manner in terms of the Berry curvature $\mathcal{F}$ and the quantum metric $g$. After introducing these notions, we define the FCI indicators I1-I3 that precisely quantify the distance between a band and the lowest Landau level \cite{royBandGeometryFractional2014,jacksonGeometricStabilityTopological2015,Simon2020}.

Consider a projector $\mathcal{P}_{\k} = \ket{u_{\k}}\bra{u_{\k}}$ to a single band of a continuum model,\footnote{The FCI indicators are most useful in continuum models \cite{simon2020contrasting}.} and define the quantum geometric tensor
\begin{equation}
    \eta_{\mu\nu}(\k) = A\braket{\partial_{\nu} u_{\k}| (1-\mathcal{P}_{\k}) | \partial_{\mu} u_{\k}},
\end{equation}
where $A$ is the area of the Brillouin zone, and $\mu,\nu \in \{k_x,k_y\}$. Its real and imaginary parts 
\begin{equation}
    g(\v{k}) = \re \eta(\v{k}), \qquad \mathcal{F}(\v{k}) = 2 \im \eta^{xy}(\v{k}), 
\end{equation}
are respectively the Berry curvature and the quantum metric.\footnote{The quantum (or Fubini-Study) metric is a $2\times 2$ real, symmetric matrix that encodes the distance between Bloch states: $\braket{u_{\k}|u_{\k + \delta\k}} = 1 - A^{-1} \sum_{\mu,\nu=x,y} g^{\mu\nu}(\k) \delta k_{\mu} \delta k_{\nu}$.} In the lowest Landau level, the spectrum is completely flat in $\v{k}$, and $\eta$ takes the special form
\begin{equation}
    \eta_{\text{LLL}}(\k) = \frac{1}{2}\begin{pmatrix}
        1 & -i\\
        i & 1
    \end{pmatrix}.
\end{equation}
All bands obey the inequality\footnote{We use a lower case $\tr$ for the $\v{k}$-space indices $\mu,\nu$.}
\begin{equation}
    \tr g \geq \abs{\F}.
    \label{eq:trace_condition}
\end{equation}
However, there are only a few rare examples of bands that saturate the inequality --- and the LLL is one of them.\footnote{In the $n$th Landau level, the Berry curvature is unchanged while the metric is $g_n = (2n+1) g_{\text{LLL}}$, avoiding saturation for $n > 0$.} Such bands are said to obey the \textit{trace condition} $\tr g = \n{\mathcal{F}}$.

We now define the three FCI indicators:
\begin{itemize}
	\item[I1.] the bandwidth $W$ of $\hat{h}_T$,
	\item[I2.] the standard deviation of the Berry curvature 
	\begin{equation}
		\sigma[\mathcal{F}] := \left[\int d^2 \k \, (\tfrac{1}{2\pi}\mathcal{F}(\k) - C)^2 \right]^{1/2}, 
		\label{eq:berry_deviation_single_band}
	\end{equation}
	\item[I3.] the failure of the trace condition
		\begin{equation}
			T[\eta] := \int d^2\k \, \big[ \tr g(\k) - |\mathcal{F}(\k)| \big] \geq 0.
			\label{eq:trace_condition_single_band}
		\end{equation}
\end{itemize}
Each of these is positive semi-definite and vanishes for the LLL. Conversely, Roy has shown \cite{royBandGeometryFractional2014} that if $\sigma[\mathcal{F}] = T[\eta] = 0$, then the density operators $\hat{\rho}_{\q}$ satisfy the GMP algebra \cite{girvinCollectiveExcitationGapFractional1985} which --- together with vanishing bandwidth and Coulomb interactions --- completely reproduces the problem of interacting electrons in the lowest Landau level. We note that if $T[\eta]$ is not identically zero, then I2 and I3 are quasi-independent of the bandwidth. Furthermore, we caution that no strict upper limits on the indicators are known; when their values are ``too large" for an FCI to form depends on the nature of the competing states. In sum, small values of I1-I3 indicate that the system is close to the LLL, and thus favorable to host FCIs.

The FCI indicators, particularly the trace condition I3, are physically important beyond just characterizing distance to the lowest Landau level. For instance, if $T[\eta] = 0$ for a flat band of a continuum model, then analytic arguments show that the Laughlin state is the ground state for sufficiently short range interaction potentials; Chiral \tbg{} is such a model \cite{ledwithFractionalChernInsulator2020,WangExactLandauLevel}. Intriguingly, the trace condition holds if and only if the wavefunctions of the band can be chosen to be holomorphic in $k_\zeta = k_x + i \zeta k_y$ for some $\zeta = \pm 1$. This is the first link in a chain of work \cite{meraKahlerGeometryChern2021,ozawaRelationsTopologyQuantum2021,MeraOzawaEngineering2021,WangExactLandauLevel} connecting the indicators to K\"ahler manifolds and complex geometry.

Our first practical application of the FCI indicators is to explain the extent of the FCI phase in DMRG at the magic angle. Fig. \ref{fig:FCI_indicators}cd shows the FCI indicators as a function of $\kappa$. At the chiral limit $\kappa=0$, the trace condition holds exactly \cite{ledwithFractionalChernInsulator2020} and the Berry deviation is small, so the quantum geometry is highly favorable for FCIs. In fact, the limit of vanishing total bandwidth, $\lambda=0$, is analytically tractable\cite{ledwithFractionalChernInsulator2020} as the flat band wavefunctions are essentially those of an electron in a spatially periodic magnetic field. One may apply pseudopotential arguments to show FCI order is present for sufficiently short-ranged interactions \cite{ledwithFractionalChernInsulator2020}. Fig. \ref{fig:FCI_indicators}d shows that the Berry deviation and trace condition increase slowly until $\kappa\approx 0.7$, and then rapidly at larger $\kappa$. The unfavorable quantum geometry at the magic angle destroys the FCI phase beyond $\kappa \approx 0.7$, even when the bandwidth vanishes. Conversely, the total bandwidth is large even at the chiral limit, due to the background charge density, and increases monotonically with $\kappa$. Comparing with Fig. \ref{fig:dmrg}, where the FCI phase always disappears beyond $\lambda \gtrsim 3/4$, suggests that the bandwidth is too large for FCIs at $\lambda = 1$. The FCI indicators therefore furnish a simplified but convenient physical picture: the primary limiting factor on FCIs at small $\kappa$ is the bandwidth, while the quantum geometry is the main barrier at large $\kappa$.

\section{Multiband FCI Indicators}
\label{sec:multiband_FCI_indicators}

Local compressibility \cite{xie2021fractional} show that applying a magnetic field of \SI{5}{\tesla} can drive a transition to an FCI phase at $\nu=3+\tfrac{2}{3}$. Motivated by this phenomena, the next two sections study the relation between magnetic field and FCIs.

Magnetic fields are notoriously difficult to treat theoretically. Hofstadter's classic work \cite{hofstadterEnergyLevelsWave1976} showed that adding magnetic field to even a simple tight-binding model results in a fractal spectrum with arbitrarily many bands --- now recognized as a general phenomena that occurs in any model. Our strategy to understand the effect of magnetic fields on FCIs is to focus on the FCI indicators, which are significantly easier than exact numerics and provide more insight. However there is an obstacle: previous work has largely considered the FCI indicators I1-I3 for a single band. To overcome this, this section will generalize the FCI indicators to the multiband setting and show how their connection to holomorphic geometry constrains the wavefunctions. The subsequent section will  apply them to \tbg{} at finite field.

Remarkably, the indicators not only generalize naturally to the multiband setting, but their physical interpretation is unchanged. In fact, the generalization is fixed almost entirely by the single principle that they should be unchanged by unnecessary band-folding. We now define the indicators, then proceed to elucidate their properties.

Consider a bandstructure with $N$ bands isolated by a spectral gap across the entire Brillouin zone and let $\mathcal{P}_{\v{k}} = \sum_{a=1}^n \ket{u_{\k}^a}\bra{u_{\k}^a}$ be a projector to those bands of interest. We consider the multiband (non-Abelian) quantum geometric tensor (QGT)
\begin{equation}
	\eta^{ab}_{\mu \nu}(\k) = NA\braket{\partial_\nu u_{\v k}^b| (1 - \mathcal{P}(\v k)|\partial_\mu u_{\v k}^a }
	\label{eq:nonabelian_quantum_geometric_tensor}
\end{equation}
where $A$ is the Brillouin zone area, $\mu,\nu \in \{k_x,k_y\}$ are $\v{k}$-space indices, and $a,b$ run over bands, making $\eta$ matrix-valued. Under gauge transformations $\mathsf{U}: BZ \to U(N)$ acting on the band indices as $\ket{u_{\k}} \mapsto \mathsf{U}_{\v{k}} \ket{u_{\k}}$, the QGT is gauge-covariant: $\eta \mapsto \mathsf{U}_{\v{k}} \eta \mathsf{U}_{\v{k}}^\dagger$. The non-Abelian quantum metric and Berry curvature are defined as the symmetry and antisymmetric parts of $\eta$ as 
\begin{equation}
  g_{\mu \nu}^{ab} = \frac{1}{2}(\eta_{\mu \nu}^{ab} + \eta_{\nu \mu}^{ab}), \quad \F^{ab} = i \sum_{\mu \nu} \varepsilon^{\mu \nu} \eta_{\mu \nu}^{ab},
  \label{eq:non-Abelian_metric_and_curvature}
\end{equation}
in terms of the Levi-Civita symbol $\varepsilon^{\mu \nu}$.

The FCI indicators must be continuous functions of field, and hence can only depend on the energy window of the projector and not the number of bands. Explicitly, the indicators must be invariant under the operation of ``forgetting" some of the translation symmetry, e.g. doubling the unit cell, halving the BZ, and doubling the number of bands. To this end, we define the normalized trace
\begin{equation}
  \Tr[\mathcal{O}] = (N A)^{-1} \sum_{b=1}^N \int d^2 \v{k} \, \mathcal{O}_{bb}(\v{k}).
	\label{eq:normalized_trace}
\end{equation}
The normalization here and in Eq. \eqref{eq:nonabelian_quantum_geometric_tensor} is chosen precisely to satisfy the ``forgetting" property. This also ensures $\Tr[\mathbbm{1}] = 1$ and that all moments $\Tr[\mathcal{F}^n]$ (multiplication in band indices) of the Berry curvature distribution are gauge-invariant and dimensionless. We will use a lowercase ``$\tr$" for the $\v{k}$-space indicies to distinguish from Eq. \eqref{eq:normalized_trace}.

\textit{I1 --- Bandwidth.} The bandwidth is simply the total bandwidth of all $N$ bands. In the presence of interactions, this means the total bandwidth of $\hat{h}_T$, the spectrum of single-particle excitations above the parent state.

\textit{I2 --- Berry Deviation.} To minic the physics of the lowest Landau level, which has a constant effective magnetic field, one must have constant Berry curvature $\mathcal{F}^{ab}(\v{k}) = \mathcal{F}_0 \delta^{ab}$ and $C=\pm1$. As before, we quantify the failure of this condition by the standard deviation of the Berry curvature
\begin{equation}
	\sigma[\mathcal{F}] := \Tr\Big[ (\tfrac{1}{2\pi C} \mathcal{F} -1 )^2 \Big]^{1/2} \ge 0
	\label{eq:berry_deviation}
\end{equation}
where $C = \Tr[\mathcal{F}]/(2\pi)$ is the  total Chern number for the $N$ bands of interest. For us it is equal to $-1$. The physics of the lowest Landau level is reproduced when $\sigma[\mathcal{F}] = 0$, which occurs if, and only if,
\begin{equation}
    \mathcal{F} = 2\pi C \, \textrm{Id}_N
\end{equation}
where $C$ is total Chern number and $\textrm{Id}_N$ is the identity matrix.

\textit{I3 --- Trace Condition.} At an operative level, the multiband trace condition is a direct generalization. Define
\begin{equation}
	T[\eta] := \Tr[\tr g] - \abs{\Tr[\mathcal{F}]} \geq 0.
	\label{eq:multiband_trace_condition_violation}
\end{equation}
\label{sec:finite_field}
\begin{figure*}
    \centering
    \includegraphics[width=0.98\textwidth]{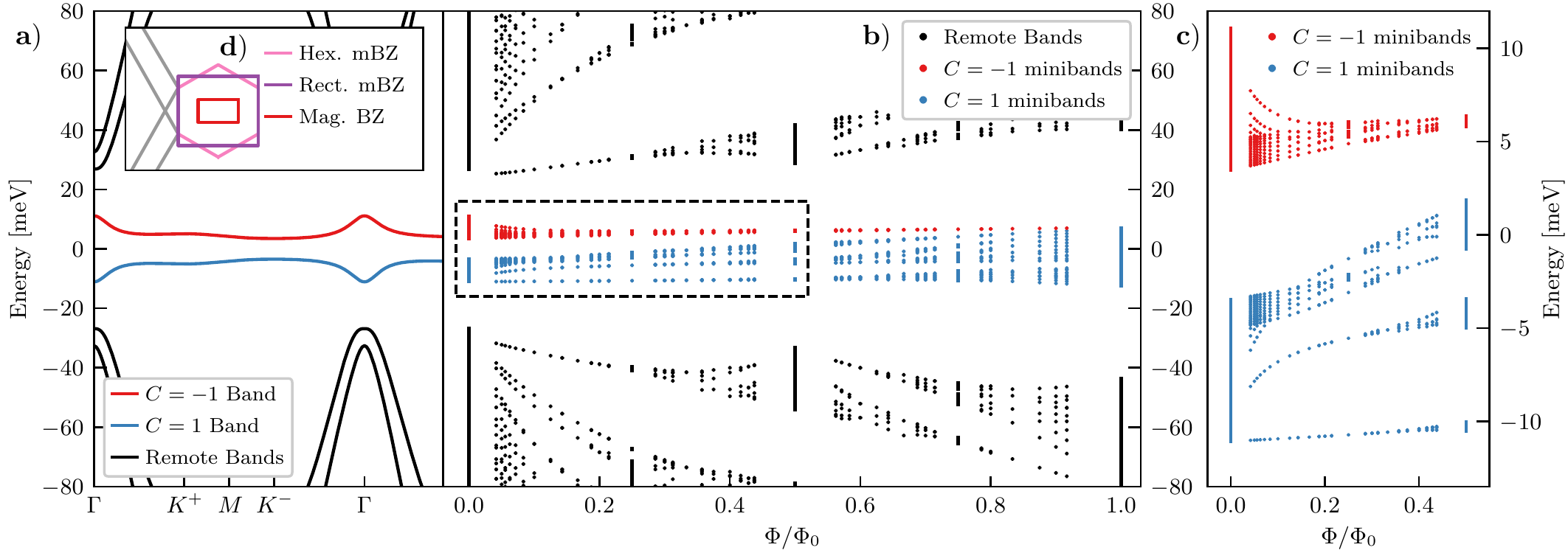}
	\caption{\textbf{a)} Spectrum of $\hat{h}_{BM}$, for comparison.  \textbf{b)} Spectrum of $\hat{h}_{BM}$ at finite field at commensurate flux with $q > 15$ in the $K$ valley. The spectra is sampled over the full magnetic Brillouin zone, and bands with sizable bandwidth are drawn as lines.  \textbf{c)} Detail of the spectrum in the active bands, whose density changes according to the Streda formula. \textbf{d)} Geometry of the moir\'e and magnetic Brillouin zones for $\phi = 1/6$. The hexagonal and rectangular Brillouin zones are equivalent choices for $\phi = 0$. Parameters: $\kappa=0.8$, $\theta=1.06^\circ$, $m_t = \SI{30}{meV}$.}
    \label{fig:moire_butterfly_and_BZ_geometry}
\end{figure*}

The physical interpretation of $\sigma[\mathcal{F}]$ and $T[\eta]$ is unchanged from the single-band case. If $\sigma[\mathcal{F}] = T[\eta] = 0$ then, just as in the single band case, the density matrices $\hat{\rho}_{\q}$ projected to the $N$ bands satisfy the GMP algebra. The proof closely follows the single band case \cite{royBandGeometryFractional2014,varjas2021topological}, and is given in Appendix \ref{app:GMP}; along the way we review the positivity of \eqref{eq:multiband_trace_condition_violation}. One way to interpret this is that, when the trace condition is satisfied and the Berry deviation vanishes, then the projected position operator obeys 
\begin{equation}
    [\mathcal{P} x \mathcal{P}, \mathcal{P} \hat{y} \mathcal{P}] = -i 2\pi C A.
\label{eq:electrons_in_effective_mag_field}
\end{equation}
Intuitively, the electrons feel a uniform effective magnetic field, just as in the lower Landau level. Thus equipped with  the multiband FCI indicators, we now proceed to compute them in \tbg{} at finite field.

\section{Finite Magnetic Field \& FCIs}

This section studies FCIs in \tbg{} at finite field using the multiband FCI indicators. We build on extensions of the BM model to finite field that have been studied by a number of authors \cite{bistritzerMoireBandsTwisted2011,zhangLandauLevelDegeneracy2019,hejaziLandauLevelsTwisted2019}. However, these works have focused mainly on the spectrum and topology, whereas to fulfill our goal of understanding the quantum geometry and the interaction induced dispersion, one is obliged to treat the wavefunctions carefully and systematically. To do this, we employ with a hybrid analytic-numerical basis of continuum wavefunctions, with the dual virtues of compatibility with the magnetic translation algebra, and numerical convenience. We note a recent work \cite{wang2021narrow} that studied the related question of single electron excitations when a finite field is added to the charge neutral point.

Below we introduce the non-interacting BM model at finite field, and discuss the crucial choice of basis. We then derive expressions for the Hartree-Fock contributions to the dispersion, and go on to compute the multiband FCI indicators at finite field. Remarkably, we shall find that even a small finite field leads to a vast improvement in the FCI indicators, stabilizing FCIs in accord with experimental findings \cite{xie2021fractional}. Finally, we combine our results to obtain an extrapolated phase diagram of FCIs in \tbg{} at finite field.

\subsection{The BM Model in a Magnetic Field}

We now summarize the non-interacting model; full details are given in App. \ref{app:continuum_formalism} and \ref{app:TBG_in_B_field}. In a finite magnetic field $B \hat{z}$, the canonical momentum is $\hat{\v{\pi}}= \hat{\v{p}} + e\hat{\v{A}}$ and the kinetic term of the Hamiltonian becomes
\begin{equation}
	\hat{D}_{\gamma} = v_F \sigma^+ \left[ \hat{a}^- - \tfrac{i \gamma}{2} \hbar k_\theta  \right] + h.c.,
	\label{eq:BM_kinetic_finite_field}
\end{equation}
where $\hat{a}^{-} = \hat{\pi}_x - i \hat{\pi}_y$ and $\gamma = \pm 1$ indexes the top/bottom layer; the tunneling and interactions are unchanged from zero-field. Magnetic fields reduce lattice translation to a projective symmetry. A magnetic generalization of Bloch's theorem is only recovered when the commensurability condition
\begin{equation}
    \phi := \frac{BA}{\Phi_0} = \frac{p}{q} \in \mathbb{Q}
    \label{eq:commensurability_condition}
\end{equation}
is satisfied, so that the magnetic flux through the original unit cell of area $A_{uc}$ is a rational number. One may then work with a magnetic Brillouin zone which is $q$ times smaller than the original (Fig. \ref{fig:moire_butterfly_and_BZ_geometry}d), and each band is folded into $q$ Hofstadter minibands.

Just as at zero field, one is obliged to solve the BM model numerically and the choice of a computational basis is therefore crucial. Since $\hat{h}_{BM}(\phi)$ is a continuum model, one must choose a basis of functions on the real-space magnetic unit cell (with twisted magnetic boundary conditions). We choose a basis of periodic sums of Landau levels in the Landau gauge, $v_{\k I}(\v{r}) = \braket{\v{r}|v_{\v{k} I}}$, where $I$ is a multi-index running over: Landau levels, $p$-fold covers of the Brillouin zone, sublattices, layers, and valleys (see App. \ref{app:TBG_in_B_field} for details). We then resolve $\hat{h}_{BM}(\phi)$ in this basis as
\begin{equation}
	\hat{h}_{\textrm{BM}}(\phi)_{\v{k}} U_{\v{k}\alpha} = \epsilon_{\v{k}\alpha} U_{\v{k}\alpha},\quad  \ket{u_{\v{k}\alpha}} = \sum_I 
	\ket{v_{\v{k}I}} U_{\v{k}\alpha}^I,
\end{equation}
where the $U$'s are the coefficients of eigenstate $\alpha$ in the computational basis. By selecting a maximum number of Landau levels to keep (i.e. a UV cutoff), this becomes a finite dimensional eigenvalue problem, allowing us to solve for the hybrid analytic-numerical eigenstates $\ket{u_{\k}}$. Note that, unlike in tight-binding models, the computational basis vectors are themselves $k$-dependent, and play a crucial role in calculations. For instance, to compute the $k$-derivative of the wavefunctions, one must use the product rule $\ket{\partial^\mu u_{\v{k}\alpha}} = \sum_I \ket{\partial^\mu v_{\v{k}I}} U_{\v{k}\alpha}^I  + \ket{v_{\v{k}I}} \partial^\mu U_{\v{k}\alpha}^I$.

We show the spectrum of $\hat{h}_{BM}(\phi)$ in Fig. \ref{fig:moire_butterfly_and_BZ_geometry}b. One can see that the two flat bands become $2q$ minibands in total. The proportion of minibands adiabatically connected to the $C=-1$ band at zero field is a purely topological quantity set by the Streda formula $\frac{d\nu}{d\phi}\big|_{\mu} = t$ or 
\begin{equation}
\nu = t \phi + s
	\label{eq:streda_formula}
\end{equation}
where $\nu$ is the filling per moir\'{e} unit cell and, in the case of Chern insulators, $t$ is the Chern number. (In FCIs, $t$ may be fractional.) Explicitly, the top $C=-1$ band deforms into $q-p$ minibands, whereas the bottom band becomes $q+p$ minibands, with spectral weight transferred from the top to bottom bands as $\phi$ increases. In fact, as $\phi \to 1$, the top band limits to the lowest Landau level wavefunction with vanishing bandwidth. These features make Fig. \ref{fig:moire_butterfly_and_BZ_geometry}b,c  quite dissimilar to typical plots of the Hofstadter butterfly, which start with Chern zero bands at $\phi=0$.

\subsection{Q. Geometry and Hartree Fock at Finite Field}

\begin{figure*}
    \centering
    \includegraphics[width=\textwidth]{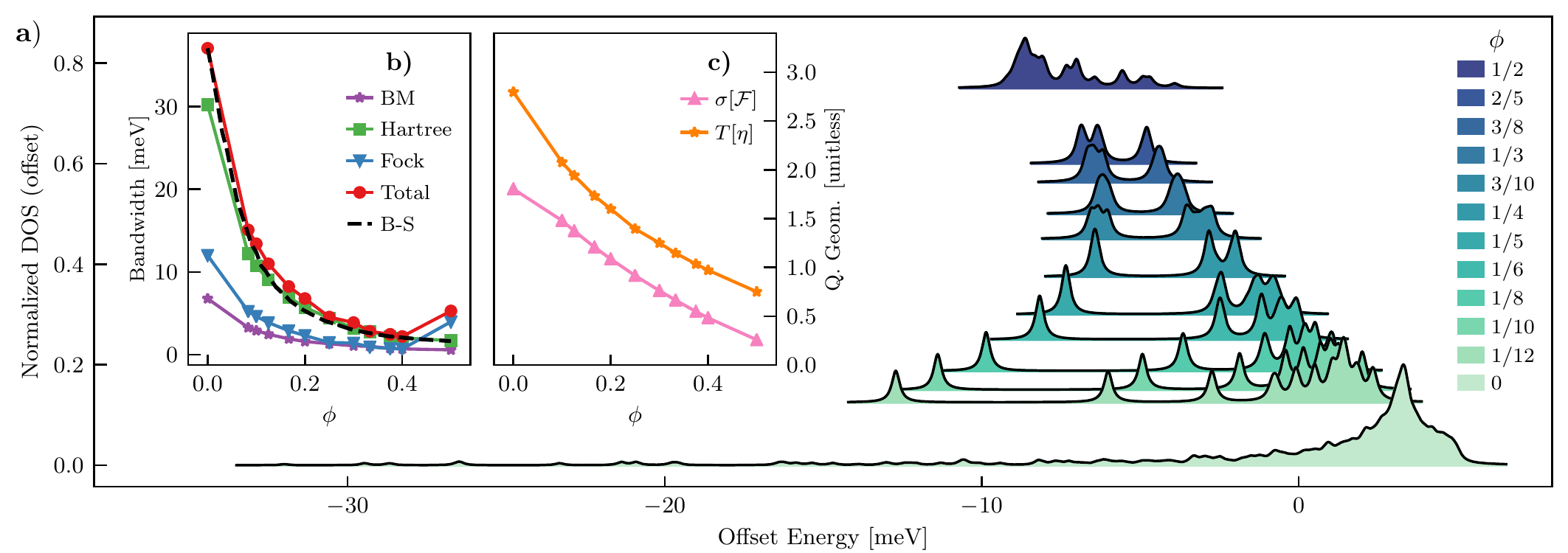}
	\caption{\textbf{a)} Density of states of the total dispersion $\hat{h}_{T}$, including the finite-field Hartree-Fock contribution, at several values of the magnetic field. The Hartree dip of ${\sim}\SI{30}{meV}$ at zero field is greatly suppressed even at small fields, and becomes an isolated Landau level. Data is offset vertically (proportionally to $\phi$) as well as horizontally. \textbf{b)} Bandwidths of various parts of the total dispersion as a function of flux. B-S is the estimated bandwidth from Bohr-Sommerfeld quantization, explained in the text. \textbf{c)} Quantum geometry indicators as a function of flux. Parameters match Fig. \ref{fig:dmrg}, the broadening is $\gamma = 0.25 \si{meV}$, and lines are a guide to the eye.}
    \label{fig:FCI_indicators_finite_field}
\end{figure*}

We now compute the FCI indicators for the total dispersion $\hat{h}_T$ at finite field $\phi$. The total dispersion is the sum of the BM Hamiltonian at finite field, computed above, and the Hartree-Fock contribution from the filled bands. This is given by Eq. \eqref{eq:Hartree_Fock_potentials} just as at zero field --- but using finite field form factors. Form factors are inner products of the real-space wavefunctions and may be computed semi-analytically from the wavefunctions of the finite field BM model as
\begin{equation}
	\Lambda_{\v{q}}(\v{k}) = \braket{u_{\v{k}}|u_{\v{k}+\v{q}}} = U_{\v{k}}^\dagger \lambda_{\v{q}}(\v{k}) U_{\v{k}+\v{q}},
	\label{eq:form_factor_computational_basis}
\end{equation}
where $[\lambda_{\v{q}}(\v{k})]_{IJ} = \braket{v_{\v{k},I}|v_{\v{k}+\v{q},J}}$ is the (analytic) form factor matrix in the computational basis. We note that the form factors of the computational basis vanish for the BM model in the plane-wave basis, and in tight-binding models. All quantities of interest may be computed stably from the numerical eigenstates $U$ in the first magnetic Brillouin zone together with analytic matrix elements (see App \ref{app:TBG_in_B_field}). 

Fig. \ref{fig:FCI_indicators_finite_field} shows the density of states $\mathcal{D}(\epsilon) \propto \tr[\hat{h}_T - \epsilon + i \gamma]$ of the total dispersion (i.e. including Hartree-Fock contributions) as a function of magnetic flux. We have chosen a finite grid of $Nq \times M$ points in the magnetic Brillouin zone and ensure the results are well-converged with the grid spacing, the Landau level (UV) cutoff, and the reciprocal (IR) cutoff $\n{\v{q}} < Q$. The required dimension of $\lambda_{\v{q}}$ can be $10^4$ or more, and grows quickly with $q$, so we restrict our attention to small denominators. At zero flux, the density of states is dominated by a long tail, due to the $\Gamma$ dip of the Hartree potential (Fig. \ref{fig:FCI_indicators}a). However, the bulk of the density of states comes from a small window of energy. As the field increases, the bandwidth decreases dramatically. The bottom of the Hartree dip is squeezed into a single Landau level, which is separated from the rest of the system by a quickly-decreasing gap. Further Landau levels are visible nearer to the center of the distribution, but with much smaller gaps. The total bandwidth decreases monotonically until $\phi \approx 1/2$, whereupon it briefly broadens, before diminishing to zero at $\phi \to 1$ (not shown). The same qualitative pattern appears in the BM bandwidth, if we omit the Hartree Fock contributions, though the broadening at $\phi = 1/2$ is less pronounced.

The quantum geometry may also be computed from the form factors. The multiband Berry curvature is defined as an infintesimal Wilson loop as $e^{i \mathcal{F}(\v{k})}= \lim_{\delta\to 0} \Lambda_{\delta\v{q}_x} \Lambda_{\delta\v{q}_y} \Lambda_{-\delta\v{q}_x} \Lambda_{-\delta\v{q}_y}$ and the metric from a second derivative $g_{\mu\nu}(\v{k}) = 
\tfrac{\partial}{\partial q_{\mu}} 
\tfrac{\partial}{\partial q_{\nu}} 
\Lambda_{\v{q}}(\v{k})\big|_{\v{q}=0}$. Fig. \ref{fig:FCI_indicators_finite_field}b,c shows the multiband Berry deviation and violation of the multiband trace condition as a function of field. These also decrease quickly, and vanish for $\phi \to 1$.

\begin{figure*}
    \centering
    \includegraphics[width=\textwidth]{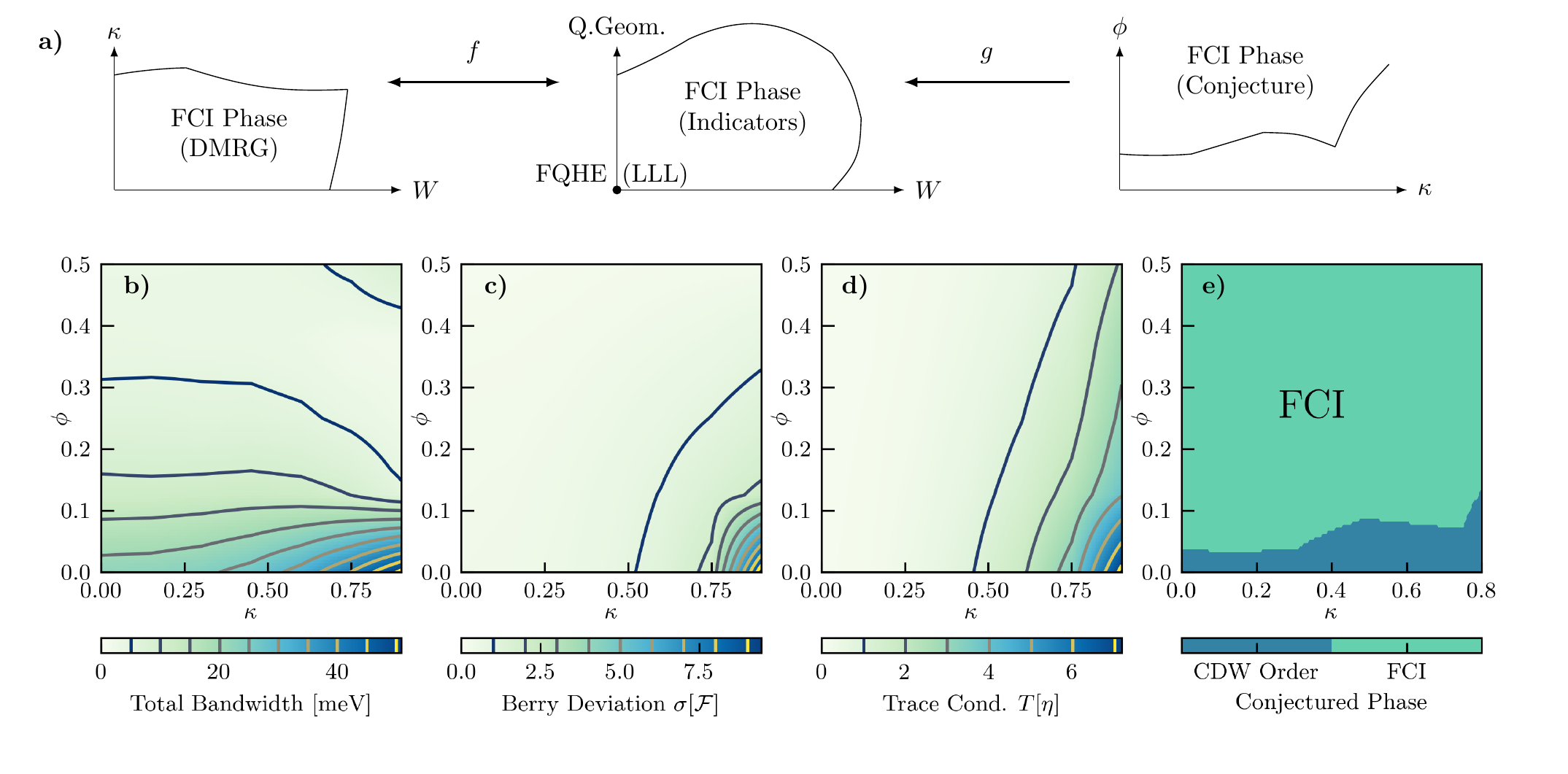}
    \caption{\textbf{a)} FCI phase diagram in various coordinates and the transforms between them. \textbf{b)} Bandwidth of the total dispersion, \textbf{c)} Berry deviation, and \textbf{d)} violation of the trace condition on the $(\kappa,\phi)$ plane. \textbf{e)} Extrapolated extent of the FCI phase at finite field using the method outlined in the text. One can see that a small flux drives a transition to an FCI, at least for sufficiently low $\kappa$. Parameters match Fig. \ref{fig:dmrg}.} 
    \label{fig:thermo_phase_diagram}
\end{figure*}

Surprisingly, the Hartree dip is not only responsible for the large bandwidth at $B=0$, but also its fast decrease with flux. To see this, we perform a semiclassical estimate of the bandwidth through Bohr-Sommerfeld (B-S) quantization. The B-S quantization condition for electrons in a metal at finite field is \cite{onsager1952interpretation,lifshitz1956theory,mikitik1999manifestation}
\begin{equation}
\frac{\ell^2_B}{2\pi}\int_{S(\epsilon)} d^2\v{k} = n+\frac{1}{2} - \frac{1}{2\pi}\int_{S(\epsilon)} \mathcal{F}\, d^2\k,
\label{eq:Bohr-Sommerfeld}
\end{equation}
where $S(\epsilon)$ is the region of the Brillouin enclosed by an orbit at energy $\epsilon$ and $n \in \mathbb{N}$ indexes the Landau level. Given the total dispersion $\hat{h}_T$ at $B=0$, and its corresponding Berry curvature distribution, Eq. \eqref{eq:Bohr-Sommerfeld} is an implicit equation for the energy of the lowest Landau level $\tilde{\epsilon}_{n=0}(\phi)$. The Bohr-Sommerfeld orbit moves up the Hartree dip (Fig. \ref{fig:FCI_indicators}a) quickly with flux, so the estimated energy of the lowest Landau level increases quickly as well, matching the exact result in Fig. \ref{fig:FCI_indicators_finite_field}a. We note the Berry phase term in Eq. \eqref{eq:Bohr-Sommerfeld} increases $\tilde{\epsilon}_0(\phi)$ still further. Estimating the total bandwidth as $W(\phi=0) - \tilde{\epsilon}_{n=0}(\phi)$ (which neglects the small variation of the top of the band with field) gives an unexpectedly accurate estimate for the total bandwidth, shown by a dashed line in Fig. \ref{fig:FCI_indicators_finite_field}b. The quick decrease in bandwidth is yet another consequence of the Hartree dispersion. Any $C=-1$ band will limit to the LLL as $\phi \to 1$, so its bandwidth much eventually decrease, but here the small mass at $\Gamma$ together with distribution of Berry curvature conspire to immediately reduce the bandwidth at small flux. We conclude that applying even a small magnetic field makes both the quantum geometry and bandstructure of \tbg{} dramatically more suitable for hosting FCI states.

\subsection{Finite Field Phase Diagram}

To round out this section, we use the FCI indicators to produce a extrapolated phase diagram of the FCI phase at finite field. Let us make the approximation that the extent of the FCI phase is entirely determined by the values of the FCI indicators $W$ and $T[\eta]$ alone.\footnote{The analagous approximation that the phase only depends on $W$ and $\sigma[F]$ produces a highly similar phase diagram; using $T[\eta]$ give a more conservative estimate for the extent of the FCI phase.}

The phase diagram is constructed as follows. For a given $(\kappa,\phi)$, we may compute the bandwidth $W$ and trace violation $T[\eta]$. At zero field, the trace violation is a monotonic function of $\kappa$, so the finite field value of $T[\eta]$ determines a unique effective zero-field $\kappa$. We may then look up whether this $(\kappa[T],W)$ leads to an FCI through the zero-fieldnm,n the experimental transition observed at $\phi \sim 0.18$ \cite{xie2021fractional}. One feature that our present phase diagram does not capture is the fact that the density of phases with different Hall conductance will differ in a field.

\section{Towards Zero Field FCIs}
\label{sec:large_angles}

\begin{figure}
	\begin{center}
        \includegraphics[width=\linewidth]{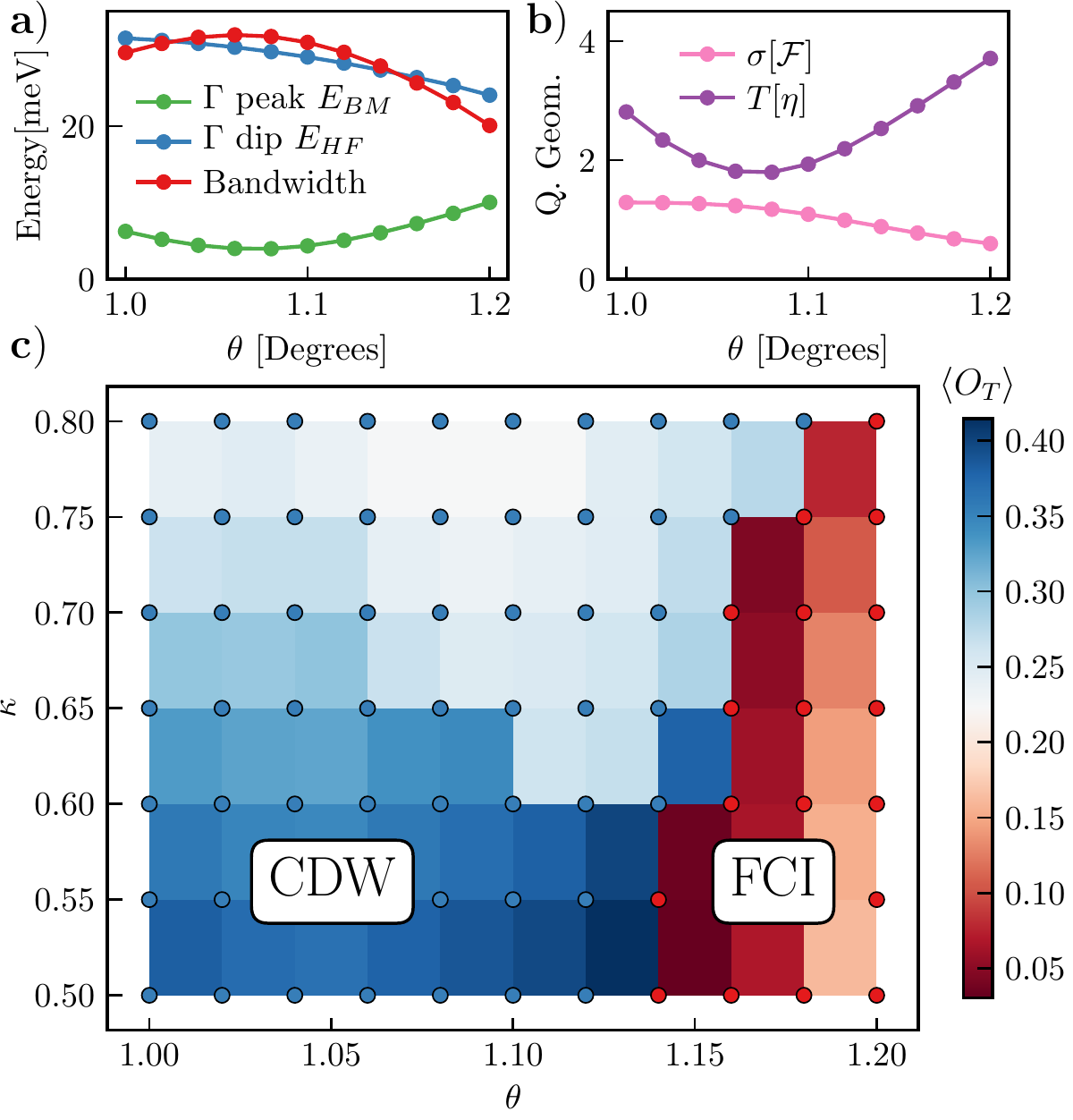}
    \end{center}
\caption{\textbf{a), b)} FCI indicators as a function of the twist angle at $\kappa=0.6$. Peak and dip heights are calculated relative to the mean energy across the BZ.
\textbf{c)} DMRG phase diagram of FCIs in the $(\theta,\kappa)$ plane at $\nu=3+\frac{2}{3}$ assuming spin- and valley-polarization. Red datapoints have entanglement spectra of Laughlin type and are therefore inside the FCI phase. Solid boxes give the magnitude of translation-breaking $\braket{O_T}$, an order parameter for the CDW phase. Parameters: $L_y = 8$, $\chi=256$, $\epsilon_{\textrm{MPO}} = 10^{-5}$. 
} 
	\label{fig:larger_theta}
\end{figure}

The numerically obtained phase diagram in Section \ref{sec:DMRG_FCIs} suggests that the quantum geometry is the main limiting factor for FCIs at the magic angle with $\kappa \gtrsim 0.7$, whereas the bandwidth is the key barrier at $\kappa \lesssim 0.7$. As mentioned above, the precise value of $\kappa$ is unknown; \textit{ab initio} and experimental estimate range from $0.5$ to $0.8$, and the true value may be sample-dependent. Values above $0.8$ seem unlikely, as they disfavor the gapped quantum anomalous hall state, in tension with its experimental appearance at $\nu=3$ \cite{PierceCDW,xie2021fractional}. It is therefore plausible that bandwidth the primary barrier to realizing FCIs. Section \ref{sec:finite_field} demonstrated there is at least one way to reduce the bandwidth: reduce the Hartree dip by applying a magnetic field. We now suggest a possible route to realize FCIs at zero field \emph{if the limiting factor to their experimental realization is the bandwidth}. A complementary route relying on reducing $\kappa$ in multilayer systems was recently discussed by some of us in Ref.~\cite{TBorNotTB}

Our route is based on the following observation. The large bandwidth at $\nu = 3$ is caused by a strong \emph{dip} at the $\Gamma$ point coming from the Hartree-Fock dispersion. On the other hand, the bare dispersion has a \emph{peak} at $\Gamma$. (See Fig. \ref{fig:overview}c.) At the magic angle, this peak is too small to compensate for the large interaction generated dip. However, moving away from the magic angle increases the peak in the BM dispersion, partially cancelling out the Hartree dip. Therefore off-magic angles can reduce the total quasiparticle dispersion at $\nu=3$, while leaving the favorable quantum geometry largely intact. This reasoning is based on the observation that the interaction-generated quasiparticle dispersion $\hat{h}_T$, rather than the `bare' BM dispersion, is what controls the stability of FCI phases. In particular, we test these ideas using DMRG numerics and find that for $\kappa = 0.6$, hBN-aligned \tbg{} with twist angle $\theta \approx 1.17^\circ$ does indeed support an FCI phase at $\nu=3+\tfrac{2}{3}$ \emph{without external magnetic field.}

Fig. \ref{fig:larger_theta}a shows that increasing the twist reduces the bandwidth for the quasiparticle dispersion at $\nu = 3$ by raising the BM peak substantially while decreasing the Hartree-Fock dip. At $\kappa=0.6$, the magic angle --- i.e. the angle that minimizes the BM bandwidth --- is about $1.06^\circ$. However, one can see that the bandwidth for the quasiparticle bands at $\nu = 3$ is actually a \textit{maximum} near $1.06^\circ$; the total bandwidth is minimized off-magic angles. Fig. \ref{fig:larger_theta}b shows that increasing $\theta$ also improves the Berry deviation, at the cost of a larger violation of the trace condition. The optimal conditions for FCIs are a balance between small angles, where the trace violation is minimized, and larger angles where the dispersion and Berry deviation are small. (We note the behavior of the trace condition is itself angle-dependent.) This leads to a concrete hypothesis: \tbg{} at angles around $\theta \approx 1.17^\circ$ should support FCIs without external field,  
assuming  isospin polarization is retained. 

This hypothesis is borne out in DMRG. Fig. \ref{fig:larger_theta}c shows a phase diagram in the $(\theta,\kappa)$ plane at $\nu =3+\tfrac{2}{3}$. For small $\kappa$ and $\theta \approx  1.06^\circ$ near the magic angle, there is a CDW phase, detected through a large value of the translation order parameter $O_T$ of translations along $\v{a}_1$. (At larger $\kappa$, a metallic phase may appear.) An FCI phase appears at larger $\theta$. As in the previous section, it is diagnosed by its entanglement spectrum. The FCI order appears to be strongest directly next to the transition; at larger $\theta$, the larger trace violation $T[\eta]$ may decrease the gap in the FCI phase \cite{jacksonGeometricStabilityTopological2015}. Encouragingly, the FCI phase persists for all $\kappa \in [0.5,0.8]$, which suggests this phase should be {\bf experimentally observable without external magnetic field}. Caveats are noted in the conclusions.

\section{Experimental Comparison}
\label{sec:expt_comparison}

 Fractional Chern insulator states have recently been observed in Ref.~\cite{xie2021fractional} through compressibility measurements. We discuss the interpetation of these states as fractional Chern insulators and also point to four states that don't fit our expectations for conventional fractional Chern insulators. We offer potential explanations for these four states, but more work is needed to pin down their character.

Gapped states in a magnetic field have two important parameters: their zero field filling $s$ and their dimensionless Hall conductance $t = \sigma_{xy}h/e^2$ such that the Streda relation $\nu = t \phi + s$ holds. Non-integer $t$ implies intrinsic topological order whereas fractional $s$ may arise from topological order or translation symmetry breaking. In this work we have focused on fractional Chern insulators that are continuously connected to states that may form if the TBG Chern band is replaced by the lowest Landau level. Consider, for example, continuously interpolating between the TBG form factors $\Lambda_\bq$ and those of the LLL while also continuously turning off band dispersion. We emphasize that this adiabatic continuity is \emph{not} between an FCI and a FQH state --- these are sharply distinct by translation symmetry, we discuss this more below. In these states, as a result of the Galilean invariance of the LLL, we have
\begin{equation}
  (t,s) = (t_p,s_p) + (\delta\nu, C_b \delta\nu)
  \label{parent_chern_division}
\end{equation}
and $(t_p, s_p)$ are the integer Chern number and $\phi=0$ filling of the ``parent state'', respectively. The state $(t,s)$ is formed on top of the parent state by fractionally filling a $C_b = \pm 1$ band by $\delta \nu$. If the fractional Chern insulator state in the Chern $C_b$ band is continuously connected to a state that can form in the lowest Landau level, by Galilean invariance the additional Chern number on top of the parent state must be $C_b \delta \nu$. Note that the parent state is generally stabilized by the presence of the lattice and is not Galilean invariant; the deformation to the LLL is only applied to the Chern band(s) that are fractionally filled. The states \eqref{parent_chern_division} are characterized by $-C_b t + s - s_p$ equal to an integer. At first we consider translation-symmetric parent states with integer $s_p$ but we will later comment on potential parent states that break translation symmetry.
Also the Landau fans consist mostly of states pointing away from charge neutrality with near-maximal Chern number. Then, on the electron (hole) side we are typically filling (emptying) $C_b=-1$ bands such that $t+s$ is an integer.

We expect that the states \eqref{parent_chern_division} will be favored in finite magnetic field as opposed to more exotic fractional Chern insulators; our work in previous sections shows that nonzero magnetic field quickly causes the Chern bands of twisted bilayer graphene to resemble the lowest Landau level. More exotic fractional Chern insulators are necessarily de-stabilized under adiabatic evolution towards lowest Landau level band geometry. 

We emphasize that these FCIs are sharply distinct from ordinary fractional quantum Hall states that form in actual Landau levels on top of the zero-field parent state. Indeed, for a fractional quantum Hall state in a Landau level we have $(t,s) - (t_p,s_p) = (\pm \delta \nu_{LL},0)$ because the Landau levels contain no electrons as $\phi \to 0$. Fractional quantum Hall states therefore have fractional $t$ but integer $s = s_p$ and are sharply distinct from fractional Chern insulators which have fractional $t$ and fractional $s$ as long as the parent state is translation-symmetric. However, translation symmetry breaking in the parent state may lead to fractional quantum Hall states with fractional $s$.

Instead, for FCIs, the Landau level limit is $\phi \to 1$, where the $C_b = -1$ bands become Landau levels with effective magnetic flux $(1-\phi)$ on top of putative insulating states at integer filling $\nu_{\phi = 1} =  t + s$. Here, we assume that the entire group of $C_b = -1$ subbands is fractionally filled.
Thus, these ``conventional'' FCIs \eqref{parent_chern_division} at close to zero field may adiabatically continue into fractional quantum Hall states on top of insulating states at $\phi = 1$. 

\renewcommand{\arraystretch}{2.5}
\begin{table}
    \centering
    \begin{ruledtabular}
    \begin{tabular}{lccc}
    \toprule
  {\bf FCI} & $\mathbf{(t,\,s)}$ & $\mathbf{t+s}$ & \textbf{Remarks} \\
 \hline
 \hline
  {\bf 1} & $\left(\dfrac13,\dfrac{11}{3}\right)$ & $4$ & VP and SP \\
 \hline
 {\bf 2} &  $\left(\dfrac23,\dfrac{10}{3}\right)$ & $4$ & VP \\
 \hline
  {\bf 3} & $\left(-\dfrac43,-\dfrac{5}{3}\right)$ & $-3$ & \\
 \hline
  {\bf 4} & $\left(-\dfrac{13}{5},-\dfrac{2}{5}\right)$ & $-3$ \\
 \hline
  {\bf 5} & $\left(-\dfrac85,\dfrac{11}{10}\right)$ & $-\dfrac12$ \\
 \hline
  {\bf 6} & $\left(-\dfrac73,\dfrac{2}{9}\right)$ & $-2-\dfrac19$ \\
 \hline
  {\bf 7} & $\left( {-\dfrac52},-\dfrac{1}{5}\right)$ & $-3+\dfrac3{10}$ & Even Denominator\\
 \hline
  {\bf 8} & $\left({-\dfrac12},-\dfrac{8}{3}\right)$ & $-3-\dfrac16$ & Even Denominator \\ 
\end{tabular}
\end{ruledtabular}
    \caption{Fractional Chern insulator states observed in Ref. \cite{xie2021fractional}. The state \textbf{1} is expected to be spin polarized (SP) and valley polarized (VP) and is our focus for our DMRG study. The state \textbf{2} may be spin polarized or a spin singlet \cite{repellin2020ferromagnetism}. States \textbf{1}--\textbf{4} can be described as conventional FCIs on top of translation symmetric parent states \eqref{parent_chern_division} whereas states \textbf{5}--\textbf{8} cannot.
    }
    \label{tab:FCI_Expt}
\end{table}

\renewcommand{\arraystretch}{1.0}

There are four states in Ref.~\cite{xie2021fractional} that are straightforwardly interpreted as  FCIs \eqref{parent_chern_division} with translationally-symmetric parent states:
\begin{equation}
  (t,s) = \left(  \frac{1}{3},\frac{11}{3}
  \right),  \left( \frac{2}{3}, \frac{10}{3} \right), \left( -\frac{4}{3}, - \frac{5}{3}\right), \left( - \frac{13}{5}, -\frac{2}{5} \right) .
\label{conventionalFCIs}
\end{equation}
These states have $t+s = 4,4,-3,-3$ respectively and are shown in Table \ref{tab:FCI_Expt} as states \textbf{1}--\textbf{4} respectively. The state \textbf{1} corresponds to hole filling the fully filled $\nu = 4$ state by $\frac{1}{3}$ and is expected to be a spin and valley polarized Laughlin state of holes; this is the state we have focused on in our DMRG analysis above. The state \textbf{2} is probably valley polarized but may be either a (112) spin singlet Halperin state or a spin polarized Laughlin state of electrons; see Ref. \cite{repellin2020ferromagnetism} for an exact diagonalization study of the $\kappa$-dependent competition between these spin structures. The multicomponent nature of the other six FCIs is an interesting question for future study.

There are also four other states that do not have integer values of $t+s$:
\begin{equation}
 \left(-\frac{8}{5}, \frac{11}{10}  \right),  \left( -\frac{7}{3}, \frac{2}{9} \right),\left( - \frac{5}{2}, -\frac{1}{5} \right),\, \text{\&} \,  \left( - \frac{1}{2}, -\frac{8}{3} \right).
  \label{weirdFCIs}
\end{equation}
These states have $t + s = -\frac{1}{2},-2 - \frac{1}{9}, -3 + \frac{3}{10},  \text{and } -3 -\frac{1}{6} $ and are shown in Table \ref{tab:FCI_Expt} as states \textbf{5}--\textbf{8} respectively. The states \eqref{weirdFCIs} have other unusual properties. The last two states have even denominators, whereas the first two states point \emph{towards} charge neutrality unlike all of the other FCIs. While we will largely leave the study of these states for future work, we make a few comments on plausible explanations for these states. First, we note that if the fractional Chern insulator forms on top of a translation breaking state, i.e. a parent state with fractional $s_p$, then $t+s = s_p \mod 1$. Charge density waves at fillings $\frac{1}{6}, \frac{1}{9},$ and $\frac{1}{2}$ are not implausible.

The even denominator states, states \textbf{7} and \textbf{8} in Table \ref{tab:FCI_Expt}, are particularly striking. Note that in previous sections we have argued that the band geometry and dispersion of twisted bilayer graphene quickly approaches that of the \emph{lowest} Landau level upon addition of flux. A priori, one may not expect the presence of non-Abelian Pfaffian states since such states are most often proposed in higher Landau levels, which have a  different quantum geometry. Furthermore the multicomponent nature of twisted bilayer graphene enables the construction of Halperin states such as the $(331)$ state which also has half-integer filling. The $(331)$ Halperin state is a popular candidate state for the $\nu = 1/2$ state observed\cite{SuenExptBilayerQH1992,EisensteinExptBilayerQH1992,SuenExptWideWell1992,ShabaniExptWideWell2013} in quantum Hall bilayers and wide wells, where the pseudospin is layer and the lowest two subbands of the well respectively. The $\mathrm{SU}(2)$ pseudospin rotation symmetry in these systems is broken both by easy-plane anisotropy and an in-plane ``Zeeman field" associated with either layer-tunneling or the splitting between the two lowest subbands of the well. A long thread of work \cite{ZhaoWideWell2021,ZhuBilayerTunneling2016,PapicTunneling2010,FaugnoBilayerCF2020,ParkCFspin1998,DasSarmaPhaseDiagram2010,BarkeshliBilayerQH2011,ScarolaBilayerCF2001,ScarolaPairing2002,SeidelThinCylinder2008,RegnaultAbNonAbBridge2008,CappelliParafermion2001,WenTransitions2000,ReadGreen2000,NaudBilayer2000,FradkinLandauGinzburg1999,Nayak1995,HoTwoCptQH1995,NomuraGapEvolutionBilayer2004,GreiterPairingBilayer1992,ChakrabortyMultilayerQH1987,YoshiokaQHBilayer1989,HeQHBilayer1991,HeQHBilayer1993,PetersonQHBilayer2010,StorniMooreRead2010} has shown that the $(331)$ state may transition to the Moore Read state when the pseudospin Zeeman field is sufficiently strong, but not so strong as to reproduce the single component case where a composite Fermi liquid is the ground state. The competition between these states is very close and sensitive to microscopic details. 

Field induced FCIs in TBG could be in a similar scenario where the pseudospin could be actual spin or valley, though we highlight some differences. For actual spin, there is a a small Zeeman field from the magnetic field - much smaller than the Zeeman field of the ordinary FQHE or that of Ref. \cite{spanton2018observation}.  Note however that $\mathrm{SU}(2)$ rotation symmetry is near exact before breaking by the Zeeman field in contrast to the quantum Hall problem which has easy-plane anisotropy. For the valley pseudospin, the $\mathrm{SU}(2)$ rotation symmetry is broken weakly by $\kappa > 0$ with either an easy-plane or easy-axis anisotropy \cite{khalafChargedSkyrmionsTopological2020} and the hBN substrate acts as a Zeeman field. Note however that the Zeeman field is ``out-of-plane" with respect to the easy-plane or easy-axis anisotropy unlike the quantum Hall case. Additionally, small deviations away from LLL quantum geometry may also to alter the close competition between various candidate states at $\nu = \frac{1}{2}$.

Note that the state $\left( - \frac{5}{2}, -\frac{1}{5} \right)$, which has the somewhat puzzling value of $t+s = \frac{3}{10}$, may actually be a fractional quantum Hall state on top the nearby translation breaking state $\left(-2,-\frac{1}{5}\right)$. We do not call this an FCI because the fractional $s$ comes from translation breaking. Note however that the $\left( - \frac{5}{2}, -\frac{1}{5} \right)$ state has a larger $d\mu/dn$ peak than the putative $\left(-2,-\frac{1}{5}\right)$ parent state \cite{xie2021fractional}, which is atypical for parent-child relationships

\section{Conclusions}
\label{sec:conclusions}

This work has undertaken a detailed study of fractional Chern insulating phases in \tbg{} both with and without magnetic fields. Combining DMRG numerics on a microscopic model with the FCI indicators of bandwidth and geometry has led to a coherent physical picture.
\begin{itemize}
	\item As previously noted, the Chern insulating state at $\nu=3$ in hBN-aligned magic angle \tbg{} is a good ``parent state" for FCIs, with strong interactions and relatively large spectral gaps.  
	
	\item A DMRG study on a microscopic model of \tbg{} at the magic angle and filling $\nu = 3 + \frac{2}{3}$ finds an FCI phase of Laughlin type at parameters only slightly outside the experimentally realistic range.
	
	\item From the perspective of the FCI indicators, the main limiting factor for FCIs at the magic angle is excessive bandwidth for $\kappa \lesssim 0.7$ and non-ideal quantum geometry for $\kappa \gtrsim 0.7$. The large bandwidth is the sum of the Bistritzer-MacDonald bands and the dispersion generated by the background charge density through interactions, for a total of ${\sim} \SI{30}{meV}$. 
	
	\item Experiment \cite{xie2021fractional} finds applying small magnetic fields to \tbg{} drives a transition to an FCI at the magic angle. To understand when the fractional quantum hall effect might appear on top of Hofstadter minibands (as opposed to a single Chern band), we generalized the FCI indicators to the multiband case.
	
	\item We then extended the Hartree-Fock theory of magic-angle graphene to finite field and computed the bandwidth and quantum geometry. Remarkably, we found that even small fields dramatically reduce the bandwidth by collapsing the `Hartree dip' and improve the quantum geometry, implying a critical magnetic flux of $\phi \sim 0.1$, in good accord with experiment.

	\item We predict that \tbg{} at angles around $\theta \approx 1.17^\circ$ can host an FCI phase {\em without external magnetic field}. This is supported by a physical argument from the quantum geometry and bandwidth, and borne out in DMRG numerics for a range of chiral ratios $\kappa = 0.5-0.8$.
	
	Let us mention a few possible caveats.
	
    \smallskip 
    \textbf{Parent state.} Our approach to FCIs assumes that doping \tbg{} from $\nu=3$ to $4$ adds electrons to the spin- and valley-polarized Chern insulator at $\nu=3$. We note that several other states are known to be competitive, such as charge density waves \cite{PierceCDW}, and, in the presence of strain, the intervalley Kekul\'e state \cite{kwan2021kekul}. Our assumption of flavor polarization at larger angles to realize zero field FCIs  can be tested in future work.
    
	\smallskip
	\textbf{Strain.} Heterostrain values as small as $0.2\%$ enlarge the bandwidth by many \si{meV} and likely increase the violation of the trace condition, strongly disfavoring the FCI phase.	Experiments have previously measured strain in the range $0.1\% - 0.7\%$ \cite{kerelsky2019maximized,choi2019electronic,xie2019spectroscopic}, so it is crucial to have a low-strain sample.
	
    \smallskip 
	\textbf{Particle-Hole Asymmetry.} Experiments on \tbg{} show particle-hole asymmetry, whose microscopic origin remains to be established and is not incorporated in our model. Ref. \cite{MacdonaldPH} suggested that it leads to less (more) dispersive bands on the positive (negative) $\nu$ side by acting against the Hartree term. This implies that when FCIs are bandwidth-limited, particle-hole asymmetry could be favorable for FCIs on the electron-doped side. This also seems compatible with experiment \cite{xie2021fractional}.
    
    \smallskip 
    \textbf{Interaction strength.} FCIs require the ratio of bandwidth to interactions to be relatively small. However, the most of the bandwidth of \tbg{} is induced by interactions through the background charge density, and is therefore proportional to the interaction strength. Therefore, FCIs should depend only weakly on the interaction strength.
    
	\item Finally, we divided the experimentally observed FCI states into those that have ``conventional" explanations and those that are either exotic, or form on top of a translation breaking pairing state. We also discussed aspects of the two even-denominator states and proposed a scenario in which one or both could be non-Abelian.
\end{itemize}

\begin{acknowledgments}

We thank A. Stern, Y. Sheffer, J. Wang, and B. Mera for enlightening related discussions.
AV was supported by a Simons Investigator award and by the Simons Collaboration on Ultra-Quantum Matter, which is a grant from the Simons Foundation (651440, AV). EK was supported by a Simons Investigator
Fellowship and by the German National Academy of Sciences Leopoldina through grant LPDS 2018-02 Leopoldina fellowship. 
 This research is funded in part by the Gordon and Betty Moore Foundation’s EPiQS
Initiative, Grant GBMF8683 to D.E.P.
JH was funded by the U.S. Department of Energy, Office of Science, Office of Basic Energy Sciences, Materials Sciences and Engineering Division under Contract No. DE-AC02-05- CH11231 through the Scientific Discovery through Advanced Computing (SciDAC) program (KC23DAC Topological and Correlated Matter via Tensor Networks and Quantum Monte Carlo).
A.T.P. and P.L. acknowledge support from the Department of Defense through the National Defense Science and Engineering Graduate Fellowship Program.
Y.X. acknowledges partial support from the
Harvard Quantum Initiative in Science and Engineering.
A.T.P., Y.X and A.Y. acknowledge
support from the Harvard Quantum Initiative Seed Fund.
T.S. is funded by the Masason foundation. MZ was supported
by the ARO through the MURI program (grant number W911NF-17-1-0323) and the Alfred P Sloan Foundation.
\end{acknowledgments}

\bibliography{references}

\begin{thebibliography}{125}%
\makeatletter
\providecommand \@ifxundefined [1]{%
 \@ifx{#1\undefined}
}%
\providecommand \@ifnum [1]{%
 \ifnum #1\expandafter \@firstoftwo
 \else \expandafter \@secondoftwo
 \fi
}%
\providecommand \@ifx [1]{%
 \ifx #1\expandafter \@firstoftwo
 \else \expandafter \@secondoftwo
 \fi
}%
\providecommand \natexlab [1]{#1}%
\providecommand \enquote  [1]{``#1''}%
\providecommand \bibnamefont  [1]{#1}%
\providecommand \bibfnamefont [1]{#1}%
\providecommand \citenamefont [1]{#1}%
\providecommand \href@noop [0]{\@secondoftwo}%
\providecommand \href [0]{\begingroup \@sanitize@url \@href}%
\providecommand \@href[1]{\@@startlink{#1}\@@href}%
\providecommand \@@href[1]{\endgroup#1\@@endlink}%
\providecommand \@sanitize@url [0]{\catcode `\\12\catcode `\$12\catcode
  `\&12\catcode `\#12\catcode `\^12\catcode `\_12\catcode `\%12\relax}%
\providecommand \@@startlink[1]{}%
\providecommand \@@endlink[0]{}%
\providecommand \url  [0]{\begingroup\@sanitize@url \@url }%
\providecommand \@url [1]{\endgroup\@href {#1}{\urlprefix }}%
\providecommand \urlprefix  [0]{URL }%
\providecommand \Eprint [0]{\href }%
\providecommand \doibase [0]{http://dx.doi.org/}%
\providecommand \selectlanguage [0]{\@gobble}%
\providecommand \bibinfo  [0]{\@secondoftwo}%
\providecommand \bibfield  [0]{\@secondoftwo}%
\providecommand \translation [1]{[#1]}%
\providecommand \BibitemOpen [0]{}%
\providecommand \bibitemStop [0]{}%
\providecommand \bibitemNoStop [0]{.\EOS\space}%
\providecommand \EOS [0]{\spacefactor3000\relax}%
\providecommand \BibitemShut  [1]{\csname bibitem#1\endcsname}%
\let\auto@bib@innerbib\@empty
\bibitem [{\citenamefont {Halperin}\ and\ \citenamefont
  {Jain}(2020)}]{FQHbook}%
  \BibitemOpen
  \bibfield  {author} {\bibinfo {author} {\bibfnamefont {B.~I.}\ \bibnamefont
  {Halperin}}\ and\ \bibinfo {author} {\bibfnamefont {J.~K.}\ \bibnamefont
  {Jain}},\ }\href {\doibase 10.1142/11751} {\emph {\bibinfo {title}
  {Fractional Quantum Hall Effects}}}\ (\bibinfo  {publisher} {WORLD
  SCIENTIFIC},\ \bibinfo {year} {2020})\ \Eprint
  {http://arxiv.org/abs/https://worldscientific.com/doi/pdf/10.1142/11751}
  {https://worldscientific.com/doi/pdf/10.1142/11751} \BibitemShut {NoStop}%
\bibitem [{\citenamefont {Neupert}\ \emph {et~al.}(2011)\citenamefont
  {Neupert}, \citenamefont {Santos}, \citenamefont {Chamon},\ and\
  \citenamefont {Mudry}}]{FCINeupertChamon}%
  \BibitemOpen
  \bibfield  {author} {\bibinfo {author} {\bibfnamefont {T.}~\bibnamefont
  {Neupert}}, \bibinfo {author} {\bibfnamefont {L.}~\bibnamefont {Santos}},
  \bibinfo {author} {\bibfnamefont {C.}~\bibnamefont {Chamon}}, \ and\ \bibinfo
  {author} {\bibfnamefont {C.}~\bibnamefont {Mudry}},\ }\href {\doibase
  10.1103/PhysRevLett.106.236804} {\bibfield  {journal} {\bibinfo  {journal}
  {Phys. Rev. Lett.}\ }\textbf {\bibinfo {volume} {106}},\ \bibinfo {pages}
  {236804} (\bibinfo {year} {2011})}\BibitemShut {NoStop}%
\bibitem [{\citenamefont {Sheng}\ \emph {et~al.}(2011)\citenamefont {Sheng},
  \citenamefont {Gu}, \citenamefont {Sun},\ and\ \citenamefont
  {Sheng}}]{FCIDonnaSheng}%
  \BibitemOpen
  \bibfield  {author} {\bibinfo {author} {\bibfnamefont {D.~N.}\ \bibnamefont
  {Sheng}}, \bibinfo {author} {\bibfnamefont {Z.-C.}\ \bibnamefont {Gu}},
  \bibinfo {author} {\bibfnamefont {K.}~\bibnamefont {Sun}}, \ and\ \bibinfo
  {author} {\bibfnamefont {L.}~\bibnamefont {Sheng}},\ }\href {\doibase
  10.1038/ncomms1380} {\bibfield  {journal} {\bibinfo  {journal} {Nature
  Communications}\ }\textbf {\bibinfo {volume} {2}},\ \bibinfo {pages} {389}
  (\bibinfo {year} {2011})}\BibitemShut {NoStop}%
\bibitem [{\citenamefont {Regnault}\ and\ \citenamefont
  {Bernevig}(2011)}]{FCIBernevigRegnault}%
  \BibitemOpen
  \bibfield  {author} {\bibinfo {author} {\bibfnamefont {N.}~\bibnamefont
  {Regnault}}\ and\ \bibinfo {author} {\bibfnamefont {B.~A.}\ \bibnamefont
  {Bernevig}},\ }\href {\doibase 10.1103/PhysRevX.1.021014} {\bibfield
  {journal} {\bibinfo  {journal} {Phys. Rev. X}\ }\textbf {\bibinfo {volume}
  {1}},\ \bibinfo {pages} {021014} (\bibinfo {year} {2011})}\BibitemShut
  {NoStop}%
\bibitem [{\citenamefont {Spanton}\ \emph {et~al.}(2018)\citenamefont
  {Spanton}, \citenamefont {Zibrov}, \citenamefont {Zhou}, \citenamefont
  {Taniguchi}, \citenamefont {Watanabe}, \citenamefont {Zaletel},\ and\
  \citenamefont {Young}}]{spanton2018observation}%
  \BibitemOpen
  \bibfield  {author} {\bibinfo {author} {\bibfnamefont {E.~M.}\ \bibnamefont
  {Spanton}}, \bibinfo {author} {\bibfnamefont {A.~A.}\ \bibnamefont {Zibrov}},
  \bibinfo {author} {\bibfnamefont {H.}~\bibnamefont {Zhou}}, \bibinfo {author}
  {\bibfnamefont {T.}~\bibnamefont {Taniguchi}}, \bibinfo {author}
  {\bibfnamefont {K.}~\bibnamefont {Watanabe}}, \bibinfo {author}
  {\bibfnamefont {M.~P.}\ \bibnamefont {Zaletel}}, \ and\ \bibinfo {author}
  {\bibfnamefont {A.~F.}\ \bibnamefont {Young}},\ }\href@noop {} {\bibfield
  {journal} {\bibinfo  {journal} {Science}\ }\textbf {\bibinfo {volume}
  {360}},\ \bibinfo {pages} {62} (\bibinfo {year} {2018})}\BibitemShut
  {NoStop}%
\bibitem [{\citenamefont {Haldane}(1988)}]{HaldaneModel}%
  \BibitemOpen
  \bibfield  {author} {\bibinfo {author} {\bibfnamefont {F.~D.~M.}\
  \bibnamefont {Haldane}},\ }\href {\doibase 10.1103/PhysRevLett.61.2015}
  {\bibfield  {journal} {\bibinfo  {journal} {Phys. Rev. Lett.}\ }\textbf
  {\bibinfo {volume} {61}},\ \bibinfo {pages} {2015} (\bibinfo {year}
  {1988})}\BibitemShut {NoStop}%
\bibitem [{\citenamefont {Liu}\ \emph {et~al.}(2012)\citenamefont {Liu},
  \citenamefont {Bergholtz}, \citenamefont {Fan},\ and\ \citenamefont
  {L\"auchli}}]{FCIBergholtzLauchli}%
  \BibitemOpen
  \bibfield  {author} {\bibinfo {author} {\bibfnamefont {Z.}~\bibnamefont
  {Liu}}, \bibinfo {author} {\bibfnamefont {E.~J.}\ \bibnamefont {Bergholtz}},
  \bibinfo {author} {\bibfnamefont {H.}~\bibnamefont {Fan}}, \ and\ \bibinfo
  {author} {\bibfnamefont {A.~M.}\ \bibnamefont {L\"auchli}},\ }\href {\doibase
  10.1103/PhysRevLett.109.186805} {\bibfield  {journal} {\bibinfo  {journal}
  {Phys. Rev. Lett.}\ }\textbf {\bibinfo {volume} {109}},\ \bibinfo {pages}
  {186805} (\bibinfo {year} {2012})}\BibitemShut {NoStop}%
\bibitem [{\citenamefont {Wu}\ \emph {et~al.}(2012)\citenamefont {Wu},
  \citenamefont {Bernevig},\ and\ \citenamefont {Regnault}}]{FCIZoology}%
  \BibitemOpen
  \bibfield  {author} {\bibinfo {author} {\bibfnamefont {Y.-L.}\ \bibnamefont
  {Wu}}, \bibinfo {author} {\bibfnamefont {B.~A.}\ \bibnamefont {Bernevig}}, \
  and\ \bibinfo {author} {\bibfnamefont {N.}~\bibnamefont {Regnault}},\ }\href
  {\doibase 10.1103/PhysRevB.85.075116} {\bibfield  {journal} {\bibinfo
  {journal} {Phys. Rev. B}\ }\textbf {\bibinfo {volume} {85}},\ \bibinfo
  {pages} {075116} (\bibinfo {year} {2012})}\BibitemShut {NoStop}%
\bibitem [{\citenamefont {Roy}()}]{royBandGeometryFractional2014}%
  \BibitemOpen
  \bibfield  {author} {\bibinfo {author} {\bibfnamefont {R.}~\bibnamefont
  {Roy}},\ }\href {\doibase 10.1103/PhysRevB.90.165139} {\ \textbf {\bibinfo
  {volume} {90}},\ \bibinfo {pages} {165139}},\ \Eprint
  {http://arxiv.org/abs/1208.2055} {1208.2055} \BibitemShut {NoStop}%
\bibitem [{\citenamefont {Jackson}\ \emph {et~al.}()\citenamefont {Jackson},
  \citenamefont {Möller},\ and\ \citenamefont
  {Roy}}]{jacksonGeometricStabilityTopological2015}%
  \BibitemOpen
  \bibfield  {author} {\bibinfo {author} {\bibfnamefont {T.~S.}\ \bibnamefont
  {Jackson}}, \bibinfo {author} {\bibfnamefont {G.}~\bibnamefont {Möller}}, \
  and\ \bibinfo {author} {\bibfnamefont {R.}~\bibnamefont {Roy}},\ }\href
  {\doibase 10.1038/ncomms9629} {\ \textbf {\bibinfo {volume} {6}},\ \bibinfo
  {pages} {8629}}\BibitemShut {NoStop}%
\bibitem [{\citenamefont {Parameswaran}\ \emph {et~al.}(2013)\citenamefont
  {Parameswaran}, \citenamefont {Roy},\ and\ \citenamefont
  {Sondhi}}]{parameswaran2013fractional}%
  \BibitemOpen
  \bibfield  {author} {\bibinfo {author} {\bibfnamefont {S.~A.}\ \bibnamefont
  {Parameswaran}}, \bibinfo {author} {\bibfnamefont {R.}~\bibnamefont {Roy}}, \
  and\ \bibinfo {author} {\bibfnamefont {S.~L.}\ \bibnamefont {Sondhi}},\
  }\href@noop {} {\bibfield  {journal} {\bibinfo  {journal} {Comptes Rendus
  Physique}\ }\textbf {\bibinfo {volume} {14}},\ \bibinfo {pages} {816}
  (\bibinfo {year} {2013})}\BibitemShut {NoStop}%
\bibitem [{\citenamefont {Bergholtz}\ and\ \citenamefont
  {Liu}(2013)}]{bergholtz2013topological}%
  \BibitemOpen
  \bibfield  {author} {\bibinfo {author} {\bibfnamefont {E.~J.}\ \bibnamefont
  {Bergholtz}}\ and\ \bibinfo {author} {\bibfnamefont {Z.}~\bibnamefont
  {Liu}},\ }\href@noop {} {\bibfield  {journal} {\bibinfo  {journal}
  {International Journal of Modern Physics B}\ }\textbf {\bibinfo {volume}
  {27}},\ \bibinfo {pages} {1330017} (\bibinfo {year} {2013})}\BibitemShut
  {NoStop}%
\bibitem [{\citenamefont {Claassen}\ \emph {et~al.}(2015)\citenamefont
  {Claassen}, \citenamefont {Lee}, \citenamefont {Thomale}, \citenamefont
  {Qi},\ and\ \citenamefont {Devereaux}}]{claassen2015position}%
  \BibitemOpen
  \bibfield  {author} {\bibinfo {author} {\bibfnamefont {M.}~\bibnamefont
  {Claassen}}, \bibinfo {author} {\bibfnamefont {C.~H.}\ \bibnamefont {Lee}},
  \bibinfo {author} {\bibfnamefont {R.}~\bibnamefont {Thomale}}, \bibinfo
  {author} {\bibfnamefont {X.-L.}\ \bibnamefont {Qi}}, \ and\ \bibinfo {author}
  {\bibfnamefont {T.~P.}\ \bibnamefont {Devereaux}},\ }\href@noop {} {\bibfield
   {journal} {\bibinfo  {journal} {Physical review letters}\ }\textbf {\bibinfo
  {volume} {114}},\ \bibinfo {pages} {236802} (\bibinfo {year}
  {2015})}\BibitemShut {NoStop}%
\bibitem [{\citenamefont {Lee}\ \emph {et~al.}(2017)\citenamefont {Lee},
  \citenamefont {Claassen},\ and\ \citenamefont {Thomale}}]{Lee2017}%
  \BibitemOpen
  \bibfield  {author} {\bibinfo {author} {\bibfnamefont {C.~H.}\ \bibnamefont
  {Lee}}, \bibinfo {author} {\bibfnamefont {M.}~\bibnamefont {Claassen}}, \
  and\ \bibinfo {author} {\bibfnamefont {R.}~\bibnamefont {Thomale}},\ }\href
  {\doibase 10.1103/PhysRevB.96.165150} {\bibfield  {journal} {\bibinfo
  {journal} {Phys. Rev. B}\ }\textbf {\bibinfo {volume} {96}},\ \bibinfo
  {pages} {165150} (\bibinfo {year} {2017})}\BibitemShut {NoStop}%
\bibitem [{\citenamefont {Simon}\ and\ \citenamefont
  {Rudner}(2020{\natexlab{a}})}]{Simon2020}%
  \BibitemOpen
  \bibfield  {author} {\bibinfo {author} {\bibfnamefont {S.~H.}\ \bibnamefont
  {Simon}}\ and\ \bibinfo {author} {\bibfnamefont {M.~S.}\ \bibnamefont
  {Rudner}},\ }\href {\doibase 10.1103/PhysRevB.102.165148} {\bibfield
  {journal} {\bibinfo  {journal} {Phys. Rev. B}\ }\textbf {\bibinfo {volume}
  {102}},\ \bibinfo {pages} {165148} (\bibinfo {year}
  {2020}{\natexlab{a}})}\BibitemShut {NoStop}%
\bibitem [{\citenamefont {Mera}\ and\ \citenamefont
  {Ozawa}()}]{meraKahlerGeometryChern2021}%
  \BibitemOpen
  \bibfield  {author} {\bibinfo {author} {\bibfnamefont {B.}~\bibnamefont
  {Mera}}\ and\ \bibinfo {author} {\bibfnamefont {T.}~\bibnamefont {Ozawa}},\
  }\href {http://arxiv.org/abs/2103.11583} {\ }\Eprint
  {http://arxiv.org/abs/2103.11583} {2103.11583} \BibitemShut {NoStop}%
\bibitem [{\citenamefont {Ozawa}\ and\ \citenamefont
  {Mera}()}]{ozawaRelationsTopologyQuantum2021}%
  \BibitemOpen
  \bibfield  {author} {\bibinfo {author} {\bibfnamefont {T.}~\bibnamefont
  {Ozawa}}\ and\ \bibinfo {author} {\bibfnamefont {B.}~\bibnamefont {Mera}},\
  }\href {http://arxiv.org/abs/2103.11582} {\ }\Eprint
  {http://arxiv.org/abs/2103.11582} {2103.11582} \BibitemShut {NoStop}%
\bibitem [{\citenamefont {Zhang}(2021)}]{Zhang2021}%
  \BibitemOpen
  \bibfield  {author} {\bibinfo {author} {\bibfnamefont {A.}~\bibnamefont
  {Zhang}},\ }\href
  {http://iopscience.iop.org/article/10.1088/1674-1056/ac2f2c} {\bibfield
  {journal} {\bibinfo  {journal} {Chinese Physics B}\ } (\bibinfo {year}
  {2021})}\BibitemShut {NoStop}%
\bibitem [{\citenamefont {Mera}\ \emph {et~al.}(2021)\citenamefont {Mera},
  \citenamefont {Zhang},\ and\ \citenamefont {Goldman}}]{MeraDirac}%
  \BibitemOpen
  \bibfield  {author} {\bibinfo {author} {\bibfnamefont {B.}~\bibnamefont
  {Mera}}, \bibinfo {author} {\bibfnamefont {A.}~\bibnamefont {Zhang}}, \ and\
  \bibinfo {author} {\bibfnamefont {N.}~\bibnamefont {Goldman}},\ }\href@noop
  {} {\bibfield  {journal} {\bibinfo  {journal} {arXiv preprint
  arXiv:2106.00800}\ } (\bibinfo {year} {2021})}\BibitemShut {NoStop}%
\bibitem [{\citenamefont {Mera}\ and\ \citenamefont
  {Ozawa}(2021)}]{MeraOzawaEngineering2021}%
  \BibitemOpen
  \bibfield  {author} {\bibinfo {author} {\bibfnamefont {B.}~\bibnamefont
  {Mera}}\ and\ \bibinfo {author} {\bibfnamefont {T.}~\bibnamefont {Ozawa}},\
  }\href {\doibase 10.1103/PhysRevB.104.115160} {\bibfield  {journal} {\bibinfo
   {journal} {Phys. Rev. B}\ }\textbf {\bibinfo {volume} {104}},\ \bibinfo
  {pages} {115160} (\bibinfo {year} {2021})}\BibitemShut {NoStop}%
\bibitem [{\citenamefont {Varjas}\ \emph {et~al.}(2021)\citenamefont {Varjas},
  \citenamefont {Abouelkomsan}, \citenamefont {Yang},\ and\ \citenamefont
  {Bergholtz}}]{varjas2021topological}%
  \BibitemOpen
  \bibfield  {author} {\bibinfo {author} {\bibfnamefont {D.}~\bibnamefont
  {Varjas}}, \bibinfo {author} {\bibfnamefont {A.}~\bibnamefont
  {Abouelkomsan}}, \bibinfo {author} {\bibfnamefont {K.}~\bibnamefont {Yang}},
  \ and\ \bibinfo {author} {\bibfnamefont {E.~J.}\ \bibnamefont {Bergholtz}},\
  }\href@noop {} {\enquote {\bibinfo {title} {Topological lattice models with
  constant berry curvature},}\ } (\bibinfo {year} {2021}),\ \Eprint
  {http://arxiv.org/abs/2107.06902} {arXiv:2107.06902 [cond-mat.str-el]}
  \BibitemShut {NoStop}%
\bibitem [{\citenamefont {Yang}\ \emph {et~al.}(2011)\citenamefont {Yang},
  \citenamefont {Zhu}, \citenamefont {Xiao}, \citenamefont {Okamoto},
  \citenamefont {Wang},\ and\ \citenamefont {Ran}}]{okamoto}%
  \BibitemOpen
  \bibfield  {author} {\bibinfo {author} {\bibfnamefont {K.-Y.}\ \bibnamefont
  {Yang}}, \bibinfo {author} {\bibfnamefont {W.}~\bibnamefont {Zhu}}, \bibinfo
  {author} {\bibfnamefont {D.}~\bibnamefont {Xiao}}, \bibinfo {author}
  {\bibfnamefont {S.}~\bibnamefont {Okamoto}}, \bibinfo {author} {\bibfnamefont
  {Z.}~\bibnamefont {Wang}}, \ and\ \bibinfo {author} {\bibfnamefont
  {Y.}~\bibnamefont {Ran}},\ }\href {\doibase 10.1103/physrevb.84.201104}
  {\bibfield  {journal} {\bibinfo  {journal} {Physical Review B}\ }\textbf
  {\bibinfo {volume} {84}} (\bibinfo {year} {2011}),\
  10.1103/physrevb.84.201104}\BibitemShut {NoStop}%
\bibitem [{\citenamefont {Xie}\ \emph {et~al.}(2021)\citenamefont {Xie},
  \citenamefont {Pierce}, \citenamefont {Park}, \citenamefont {Parker},
  \citenamefont {Khalaf}, \citenamefont {Ledwith}, \citenamefont {Cao},
  \citenamefont {Lee}, \citenamefont {Chen}, \citenamefont {Forrester} \emph
  {et~al.}}]{xie2021fractional}%
  \BibitemOpen
  \bibfield  {author} {\bibinfo {author} {\bibfnamefont {Y.}~\bibnamefont
  {Xie}}, \bibinfo {author} {\bibfnamefont {A.~T.}\ \bibnamefont {Pierce}},
  \bibinfo {author} {\bibfnamefont {J.~M.}\ \bibnamefont {Park}}, \bibinfo
  {author} {\bibfnamefont {D.~E.}\ \bibnamefont {Parker}}, \bibinfo {author}
  {\bibfnamefont {E.}~\bibnamefont {Khalaf}}, \bibinfo {author} {\bibfnamefont
  {P.}~\bibnamefont {Ledwith}}, \bibinfo {author} {\bibfnamefont
  {Y.}~\bibnamefont {Cao}}, \bibinfo {author} {\bibfnamefont {S.~H.}\
  \bibnamefont {Lee}}, \bibinfo {author} {\bibfnamefont {S.}~\bibnamefont
  {Chen}}, \bibinfo {author} {\bibfnamefont {P.~R.}\ \bibnamefont {Forrester}},
   \emph {et~al.},\ }\href@noop {} {\bibfield  {journal} {\bibinfo  {journal}
  {arXiv preprint arXiv:2107.10854}\ } (\bibinfo {year} {2021})}\BibitemShut
  {NoStop}%
\bibitem [{\citenamefont {Cao}\ \emph {et~al.}(2018{\natexlab{a}})\citenamefont
  {Cao}, \citenamefont {Fatemi}, \citenamefont {Demir}, \citenamefont {Fang},
  \citenamefont {Tomarken}, \citenamefont {Luo}, \citenamefont
  {Sanchez-Yamagishi}, \citenamefont {Watanabe}, \citenamefont {Taniguchi},
  \citenamefont {Kaxiras} \emph {et~al.}}]{PabloMott}%
  \BibitemOpen
  \bibfield  {author} {\bibinfo {author} {\bibfnamefont {Y.}~\bibnamefont
  {Cao}}, \bibinfo {author} {\bibfnamefont {V.}~\bibnamefont {Fatemi}},
  \bibinfo {author} {\bibfnamefont {A.}~\bibnamefont {Demir}}, \bibinfo
  {author} {\bibfnamefont {S.}~\bibnamefont {Fang}}, \bibinfo {author}
  {\bibfnamefont {S.~L.}\ \bibnamefont {Tomarken}}, \bibinfo {author}
  {\bibfnamefont {J.~Y.}\ \bibnamefont {Luo}}, \bibinfo {author} {\bibfnamefont
  {J.~D.}\ \bibnamefont {Sanchez-Yamagishi}}, \bibinfo {author} {\bibfnamefont
  {K.}~\bibnamefont {Watanabe}}, \bibinfo {author} {\bibfnamefont
  {T.}~\bibnamefont {Taniguchi}}, \bibinfo {author} {\bibfnamefont
  {E.}~\bibnamefont {Kaxiras}},  \emph {et~al.},\ }\href@noop {} {\bibfield
  {journal} {\bibinfo  {journal} {Nature}\ }\textbf {\bibinfo {volume} {556}},\
  \bibinfo {pages} {80} (\bibinfo {year} {2018}{\natexlab{a}})}\BibitemShut
  {NoStop}%
\bibitem [{\citenamefont {Cao}\ \emph {et~al.}(2018{\natexlab{b}})\citenamefont
  {Cao}, \citenamefont {Fatemi}, \citenamefont {Fang}, \citenamefont
  {Watanabe}, \citenamefont {Taniguchi}, \citenamefont {Kaxiras},\ and\
  \citenamefont {Jarillo-Herrero}}]{PabloSC}%
  \BibitemOpen
  \bibfield  {author} {\bibinfo {author} {\bibfnamefont {Y.}~\bibnamefont
  {Cao}}, \bibinfo {author} {\bibfnamefont {V.}~\bibnamefont {Fatemi}},
  \bibinfo {author} {\bibfnamefont {S.}~\bibnamefont {Fang}}, \bibinfo {author}
  {\bibfnamefont {K.}~\bibnamefont {Watanabe}}, \bibinfo {author}
  {\bibfnamefont {T.}~\bibnamefont {Taniguchi}}, \bibinfo {author}
  {\bibfnamefont {E.}~\bibnamefont {Kaxiras}}, \ and\ \bibinfo {author}
  {\bibfnamefont {P.}~\bibnamefont {Jarillo-Herrero}},\ }\href@noop {}
  {\bibfield  {journal} {\bibinfo  {journal} {Nature}\ }\textbf {\bibinfo
  {volume} {556}},\ \bibinfo {pages} {43} (\bibinfo {year}
  {2018}{\natexlab{b}})}\BibitemShut {NoStop}%
\bibitem [{\citenamefont {Po}\ \emph {et~al.}(2018)\citenamefont {Po},
  \citenamefont {Zou}, \citenamefont {Vishwanath},\ and\ \citenamefont
  {Senthil}}]{PoTBG}%
  \BibitemOpen
  \bibfield  {author} {\bibinfo {author} {\bibfnamefont {H.~C.}\ \bibnamefont
  {Po}}, \bibinfo {author} {\bibfnamefont {L.}~\bibnamefont {Zou}}, \bibinfo
  {author} {\bibfnamefont {A.}~\bibnamefont {Vishwanath}}, \ and\ \bibinfo
  {author} {\bibfnamefont {T.}~\bibnamefont {Senthil}},\ }\href {\doibase
  10.1103/PhysRevX.8.031089} {\bibfield  {journal} {\bibinfo  {journal} {Phys.
  Rev. X}\ }\textbf {\bibinfo {volume} {8}},\ \bibinfo {pages} {031089}
  (\bibinfo {year} {2018})}\BibitemShut {NoStop}%
\bibitem [{\citenamefont {Song}\ \emph {et~al.}()\citenamefont {Song},
  \citenamefont {Wang}, \citenamefont {Shi}, \citenamefont {Li}, \citenamefont
  {Fang},\ and\ \citenamefont {Bernevig}}]{songAllMagicAngles2019}%
  \BibitemOpen
  \bibfield  {author} {\bibinfo {author} {\bibfnamefont {Z.}~\bibnamefont
  {Song}}, \bibinfo {author} {\bibfnamefont {Z.}~\bibnamefont {Wang}}, \bibinfo
  {author} {\bibfnamefont {W.}~\bibnamefont {Shi}}, \bibinfo {author}
  {\bibfnamefont {G.}~\bibnamefont {Li}}, \bibinfo {author} {\bibfnamefont
  {C.}~\bibnamefont {Fang}}, \ and\ \bibinfo {author} {\bibfnamefont {B.~A.}\
  \bibnamefont {Bernevig}},\ }\href {\doibase 10.1103/PhysRevLett.123.036401}
  {\ \textbf {\bibinfo {volume} {123}},\ \bibinfo {pages} {036401}},\ \Eprint
  {http://arxiv.org/abs/1807.10676} {1807.10676} \BibitemShut {NoStop}%
\bibitem [{\citenamefont {Hejazi}\ \emph {et~al.}(2019)\citenamefont {Hejazi},
  \citenamefont {Liu}, \citenamefont {Shapourian}, \citenamefont {Chen},\ and\
  \citenamefont {Balents}}]{HejaziTopological}%
  \BibitemOpen
  \bibfield  {author} {\bibinfo {author} {\bibfnamefont {K.}~\bibnamefont
  {Hejazi}}, \bibinfo {author} {\bibfnamefont {C.}~\bibnamefont {Liu}},
  \bibinfo {author} {\bibfnamefont {H.}~\bibnamefont {Shapourian}}, \bibinfo
  {author} {\bibfnamefont {X.}~\bibnamefont {Chen}}, \ and\ \bibinfo {author}
  {\bibfnamefont {L.}~\bibnamefont {Balents}},\ }\href {\doibase
  10.1103/PhysRevB.99.035111} {\bibfield  {journal} {\bibinfo  {journal} {Phys.
  Rev. B}\ }\textbf {\bibinfo {volume} {99}},\ \bibinfo {pages} {035111}
  (\bibinfo {year} {2019})}\BibitemShut {NoStop}%
\bibitem [{\citenamefont {{Ahn}}\ \emph {et~al.}(2019)\citenamefont {{Ahn}},
  \citenamefont {{Park}},\ and\ \citenamefont {{Yang}}}]{AhnStiefelWhitney}%
  \BibitemOpen
  \bibfield  {author} {\bibinfo {author} {\bibfnamefont {J.}~\bibnamefont
  {{Ahn}}}, \bibinfo {author} {\bibfnamefont {S.}~\bibnamefont {{Park}}}, \
  and\ \bibinfo {author} {\bibfnamefont {B.-J.}\ \bibnamefont {{Yang}}},\
  }\href {\doibase 10.1103/PhysRevX.9.021013} {\bibfield  {journal} {\bibinfo
  {journal} {Physical Review X}\ }\textbf {\bibinfo {volume} {9}},\ \bibinfo
  {eid} {021013} (\bibinfo {year} {2019})},\ \Eprint
  {http://arxiv.org/abs/1808.05375} {arXiv:1808.05375 [cond-mat.mes-hall]}
  \BibitemShut {NoStop}%
\bibitem [{\citenamefont {Bultinck}\ \emph
  {et~al.}(2020{\natexlab{a}})\citenamefont {Bultinck}, \citenamefont
  {Chatterjee},\ and\ \citenamefont {Zaletel}}]{BultinckAnomalousHall}%
  \BibitemOpen
  \bibfield  {author} {\bibinfo {author} {\bibfnamefont {N.}~\bibnamefont
  {Bultinck}}, \bibinfo {author} {\bibfnamefont {S.}~\bibnamefont
  {Chatterjee}}, \ and\ \bibinfo {author} {\bibfnamefont {M.~P.}\ \bibnamefont
  {Zaletel}},\ }\href {\doibase 10.1103/PhysRevLett.124.166601} {\bibfield
  {journal} {\bibinfo  {journal} {Phys. Rev. Lett.}\ }\textbf {\bibinfo
  {volume} {124}},\ \bibinfo {pages} {166601} (\bibinfo {year}
  {2020}{\natexlab{a}})}\BibitemShut {NoStop}%
\bibitem [{\citenamefont {{Zhang}}\ \emph {et~al.}(2019)\citenamefont
  {{Zhang}}, \citenamefont {{Mao}},\ and\ \citenamefont
  {{Senthil}}}]{ZhangAnomalousHall}%
  \BibitemOpen
  \bibfield  {author} {\bibinfo {author} {\bibfnamefont {Y.-H.}\ \bibnamefont
  {{Zhang}}}, \bibinfo {author} {\bibfnamefont {D.}~\bibnamefont {{Mao}}}, \
  and\ \bibinfo {author} {\bibfnamefont {T.}~\bibnamefont {{Senthil}}},\
  }\href@noop {} {\bibfield  {journal} {\bibinfo  {journal} {arXiv e-prints}\
  ,\ \bibinfo {eid} {arXiv:1901.08209}} (\bibinfo {year} {2019})}\BibitemShut
  {NoStop}%
\bibitem [{\citenamefont {{Sharpe}}\ \emph {et~al.}(2019)\citenamefont
  {{Sharpe}}, \citenamefont {{Fox}}, \citenamefont {{Barnard}}, \citenamefont
  {{Finney}}, \citenamefont {{Watanabe}}, \citenamefont {{Taniguchi}},
  \citenamefont {{Kastner}},\ and\ \citenamefont
  {{Goldhaber-Gordon}}}]{SharpeQAH}%
  \BibitemOpen
  \bibfield  {author} {\bibinfo {author} {\bibfnamefont {A.~L.}\ \bibnamefont
  {{Sharpe}}}, \bibinfo {author} {\bibfnamefont {E.~J.}\ \bibnamefont {{Fox}}},
  \bibinfo {author} {\bibfnamefont {A.~W.}\ \bibnamefont {{Barnard}}}, \bibinfo
  {author} {\bibfnamefont {J.}~\bibnamefont {{Finney}}}, \bibinfo {author}
  {\bibfnamefont {K.}~\bibnamefont {{Watanabe}}}, \bibinfo {author}
  {\bibfnamefont {T.}~\bibnamefont {{Taniguchi}}}, \bibinfo {author}
  {\bibfnamefont {M.~A.}\ \bibnamefont {{Kastner}}}, \ and\ \bibinfo {author}
  {\bibfnamefont {D.}~\bibnamefont {{Goldhaber-Gordon}}},\ }\href@noop {}
  {\bibfield  {journal} {\bibinfo  {journal} {arXiv e-prints}\ ,\ \bibinfo
  {eid} {arXiv:1901.03520}} (\bibinfo {year} {2019})}\BibitemShut {NoStop}%
\bibitem [{\citenamefont {Serlin}\ \emph {et~al.}(2020)\citenamefont {Serlin},
  \citenamefont {Tschirhart}, \citenamefont {Polshyn}, \citenamefont {Zhang},
  \citenamefont {Zhu}, \citenamefont {Watanabe}, \citenamefont {Taniguchi},
  \citenamefont {Balents},\ and\ \citenamefont {Young}}]{YoungQAH}%
  \BibitemOpen
  \bibfield  {author} {\bibinfo {author} {\bibfnamefont {M.}~\bibnamefont
  {Serlin}}, \bibinfo {author} {\bibfnamefont {C.~L.}\ \bibnamefont
  {Tschirhart}}, \bibinfo {author} {\bibfnamefont {H.}~\bibnamefont {Polshyn}},
  \bibinfo {author} {\bibfnamefont {Y.}~\bibnamefont {Zhang}}, \bibinfo
  {author} {\bibfnamefont {J.}~\bibnamefont {Zhu}}, \bibinfo {author}
  {\bibfnamefont {K.}~\bibnamefont {Watanabe}}, \bibinfo {author}
  {\bibfnamefont {T.}~\bibnamefont {Taniguchi}}, \bibinfo {author}
  {\bibfnamefont {L.}~\bibnamefont {Balents}}, \ and\ \bibinfo {author}
  {\bibfnamefont {A.~F.}\ \bibnamefont {Young}},\ }\href {\doibase
  10.1126/science.aay5533} {\bibfield  {journal} {\bibinfo  {journal}
  {Science}\ }\textbf {\bibinfo {volume} {367}},\ \bibinfo {pages} {900}
  (\bibinfo {year} {2020})}\BibitemShut {NoStop}%
\bibitem [{\citenamefont {Tarnopolsky}\ \emph {et~al.}(2019)\citenamefont
  {Tarnopolsky}, \citenamefont {Kruchkov},\ and\ \citenamefont
  {Vishwanath}}]{TarnopolskyChiralModel}%
  \BibitemOpen
  \bibfield  {author} {\bibinfo {author} {\bibfnamefont {G.}~\bibnamefont
  {Tarnopolsky}}, \bibinfo {author} {\bibfnamefont {A.~J.}\ \bibnamefont
  {Kruchkov}}, \ and\ \bibinfo {author} {\bibfnamefont {A.}~\bibnamefont
  {Vishwanath}},\ }\href {\doibase 10.1103/PhysRevLett.122.106405} {\bibfield
  {journal} {\bibinfo  {journal} {Phys. Rev. Lett.}\ }\textbf {\bibinfo
  {volume} {122}},\ \bibinfo {pages} {106405} (\bibinfo {year}
  {2019})}\BibitemShut {NoStop}%
\bibitem [{\citenamefont {Ledwith}\ \emph {et~al.}()\citenamefont {Ledwith},
  \citenamefont {Tarnopolsky}, \citenamefont {Khalaf},\ and\ \citenamefont
  {Vishwanath}}]{ledwithFractionalChernInsulator2020}%
  \BibitemOpen
  \bibfield  {author} {\bibinfo {author} {\bibfnamefont {P.~J.}\ \bibnamefont
  {Ledwith}}, \bibinfo {author} {\bibfnamefont {G.}~\bibnamefont
  {Tarnopolsky}}, \bibinfo {author} {\bibfnamefont {E.}~\bibnamefont {Khalaf}},
  \ and\ \bibinfo {author} {\bibfnamefont {A.}~\bibnamefont {Vishwanath}},\
  }\href {\doibase 10.1103/PhysRevResearch.2.023237} {\ \textbf {\bibinfo
  {volume} {2}},\ \bibinfo {pages} {023237}}\BibitemShut {NoStop}%
\bibitem [{\citenamefont {Wang}\ \emph {et~al.}(2021)\citenamefont {Wang},
  \citenamefont {Cano}, \citenamefont {Millis}, \citenamefont {Liu},\ and\
  \citenamefont {Yang}}]{WangExactLandauLevel}%
  \BibitemOpen
  \bibfield  {author} {\bibinfo {author} {\bibfnamefont {J.}~\bibnamefont
  {Wang}}, \bibinfo {author} {\bibfnamefont {J.}~\bibnamefont {Cano}}, \bibinfo
  {author} {\bibfnamefont {A.~J.}\ \bibnamefont {Millis}}, \bibinfo {author}
  {\bibfnamefont {Z.}~\bibnamefont {Liu}}, \ and\ \bibinfo {author}
  {\bibfnamefont {B.}~\bibnamefont {Yang}},\ }\href@noop {} {\bibfield
  {journal} {\bibinfo  {journal} {arXiv preprint arXiv:2105.07491}\ } (\bibinfo
  {year} {2021})}\BibitemShut {NoStop}%
\bibitem [{\citenamefont {Repellin}\ and\ \citenamefont
  {Senthil}()}]{repellinChernBandsTwisted2019}%
  \BibitemOpen
  \bibfield  {author} {\bibinfo {author} {\bibfnamefont {C.}~\bibnamefont
  {Repellin}}\ and\ \bibinfo {author} {\bibfnamefont {T.}~\bibnamefont
  {Senthil}},\ }\href {http://arxiv.org/abs/1912.11469} {\ }\Eprint
  {http://arxiv.org/abs/1912.11469} {1912.11469} \BibitemShut {NoStop}%
\bibitem [{\citenamefont {Abouelkomsan}\ \emph {et~al.}(2020)\citenamefont
  {Abouelkomsan}, \citenamefont {Liu},\ and\ \citenamefont
  {Bergholtz}}]{abouelkomsan2020particle}%
  \BibitemOpen
  \bibfield  {author} {\bibinfo {author} {\bibfnamefont {A.}~\bibnamefont
  {Abouelkomsan}}, \bibinfo {author} {\bibfnamefont {Z.}~\bibnamefont {Liu}}, \
  and\ \bibinfo {author} {\bibfnamefont {E.~J.}\ \bibnamefont {Bergholtz}},\
  }\href@noop {} {\bibfield  {journal} {\bibinfo  {journal} {Physical review
  letters}\ }\textbf {\bibinfo {volume} {124}},\ \bibinfo {pages} {106803}
  (\bibinfo {year} {2020})}\BibitemShut {NoStop}%
\bibitem [{\citenamefont {Wilhelm}\ \emph {et~al.}(2021)\citenamefont
  {Wilhelm}, \citenamefont {Lang},\ and\ \citenamefont
  {L{\"a}uchli}}]{wilhelm2021interplay}%
  \BibitemOpen
  \bibfield  {author} {\bibinfo {author} {\bibfnamefont {P.}~\bibnamefont
  {Wilhelm}}, \bibinfo {author} {\bibfnamefont {T.~C.}\ \bibnamefont {Lang}}, \
  and\ \bibinfo {author} {\bibfnamefont {A.~M.}\ \bibnamefont {L{\"a}uchli}},\
  }\href@noop {} {\bibfield  {journal} {\bibinfo  {journal} {Physical Review
  B}\ }\textbf {\bibinfo {volume} {103}},\ \bibinfo {pages} {125406} (\bibinfo
  {year} {2021})}\BibitemShut {NoStop}%
\bibitem [{\citenamefont {Xie}\ and\ \citenamefont
  {MacDonald}(2020)}]{XieMacdonald}%
  \BibitemOpen
  \bibfield  {author} {\bibinfo {author} {\bibfnamefont {M.}~\bibnamefont
  {Xie}}\ and\ \bibinfo {author} {\bibfnamefont {A.~H.}\ \bibnamefont
  {MacDonald}},\ }\href {\doibase 10.1103/PhysRevLett.124.097601} {\bibfield
  {journal} {\bibinfo  {journal} {Phys. Rev. Lett.}\ }\textbf {\bibinfo
  {volume} {124}},\ \bibinfo {pages} {097601} (\bibinfo {year}
  {2020})}\BibitemShut {NoStop}%
\bibitem [{\citenamefont {Repellin}\ \emph {et~al.}(2020)\citenamefont
  {Repellin}, \citenamefont {Dong}, \citenamefont {Zhang},\ and\ \citenamefont
  {Senthil}}]{repellin2020ferromagnetism}%
  \BibitemOpen
  \bibfield  {author} {\bibinfo {author} {\bibfnamefont {C.}~\bibnamefont
  {Repellin}}, \bibinfo {author} {\bibfnamefont {Z.}~\bibnamefont {Dong}},
  \bibinfo {author} {\bibfnamefont {Y.-H.}\ \bibnamefont {Zhang}}, \ and\
  \bibinfo {author} {\bibfnamefont {T.}~\bibnamefont {Senthil}},\ }\href@noop
  {} {\bibfield  {journal} {\bibinfo  {journal} {Physical Review Letters}\
  }\textbf {\bibinfo {volume} {124}},\ \bibinfo {pages} {187601} (\bibinfo
  {year} {2020})}\BibitemShut {NoStop}%
\bibitem [{\citenamefont {Liu}\ \emph {et~al.}(2021)\citenamefont {Liu},
  \citenamefont {Khalaf}, \citenamefont {Lee},\ and\ \citenamefont
  {Vishwanath}}]{ShangNematic}%
  \BibitemOpen
  \bibfield  {author} {\bibinfo {author} {\bibfnamefont {S.}~\bibnamefont
  {Liu}}, \bibinfo {author} {\bibfnamefont {E.}~\bibnamefont {Khalaf}},
  \bibinfo {author} {\bibfnamefont {J.~Y.}\ \bibnamefont {Lee}}, \ and\
  \bibinfo {author} {\bibfnamefont {A.}~\bibnamefont {Vishwanath}},\ }\href
  {\doibase 10.1103/PhysRevResearch.3.013033} {\bibfield  {journal} {\bibinfo
  {journal} {Phys. Rev. Research}\ }\textbf {\bibinfo {volume} {3}},\ \bibinfo
  {pages} {013033} (\bibinfo {year} {2021})}\BibitemShut {NoStop}%
\bibitem [{\citenamefont {Bultinck}\ \emph
  {et~al.}(2020{\natexlab{b}})\citenamefont {Bultinck}, \citenamefont {Khalaf},
  \citenamefont {Liu}, \citenamefont {Chatterjee}, \citenamefont {Vishwanath},\
  and\ \citenamefont {Zaletel}}]{bultinck2020ground}%
  \BibitemOpen
  \bibfield  {author} {\bibinfo {author} {\bibfnamefont {N.}~\bibnamefont
  {Bultinck}}, \bibinfo {author} {\bibfnamefont {E.}~\bibnamefont {Khalaf}},
  \bibinfo {author} {\bibfnamefont {S.}~\bibnamefont {Liu}}, \bibinfo {author}
  {\bibfnamefont {S.}~\bibnamefont {Chatterjee}}, \bibinfo {author}
  {\bibfnamefont {A.}~\bibnamefont {Vishwanath}}, \ and\ \bibinfo {author}
  {\bibfnamefont {M.~P.}\ \bibnamefont {Zaletel}},\ }\href@noop {} {\bibfield
  {journal} {\bibinfo  {journal} {Physical Review X}\ }\textbf {\bibinfo
  {volume} {10}},\ \bibinfo {pages} {031034} (\bibinfo {year}
  {2020}{\natexlab{b}})}\BibitemShut {NoStop}%
\bibitem [{\citenamefont {Bi}\ \emph {et~al.}(2019)\citenamefont {Bi},
  \citenamefont {Yuan},\ and\ \citenamefont {Fu}}]{BiFuStrain}%
  \BibitemOpen
  \bibfield  {author} {\bibinfo {author} {\bibfnamefont {Z.}~\bibnamefont
  {Bi}}, \bibinfo {author} {\bibfnamefont {N.~F.~Q.}\ \bibnamefont {Yuan}}, \
  and\ \bibinfo {author} {\bibfnamefont {L.}~\bibnamefont {Fu}},\ }\href
  {\doibase 10.1103/PhysRevB.100.035448} {\bibfield  {journal} {\bibinfo
  {journal} {Phys. Rev. B}\ }\textbf {\bibinfo {volume} {100}},\ \bibinfo
  {pages} {035448} (\bibinfo {year} {2019})}\BibitemShut {NoStop}%
\bibitem [{\citenamefont {Parker}\ \emph
  {et~al.}(2020{\natexlab{a}})\citenamefont {Parker}, \citenamefont {Soejima},
  \citenamefont {Hauschild}, \citenamefont {Zaletel},\ and\ \citenamefont
  {Bultinck}}]{parker2020strain}%
  \BibitemOpen
  \bibfield  {author} {\bibinfo {author} {\bibfnamefont {D.~E.}\ \bibnamefont
  {Parker}}, \bibinfo {author} {\bibfnamefont {T.}~\bibnamefont {Soejima}},
  \bibinfo {author} {\bibfnamefont {J.}~\bibnamefont {Hauschild}}, \bibinfo
  {author} {\bibfnamefont {M.~P.}\ \bibnamefont {Zaletel}}, \ and\ \bibinfo
  {author} {\bibfnamefont {N.}~\bibnamefont {Bultinck}},\ }\href@noop {}
  {\bibfield  {journal} {\bibinfo  {journal} {arXiv preprint arXiv:2012.09885}\
  } (\bibinfo {year} {2020}{\natexlab{a}})}\BibitemShut {NoStop}%
\bibitem [{\citenamefont {Cea}\ and\ \citenamefont {Guinea}(2020)}]{GuineaHF}%
  \BibitemOpen
  \bibfield  {author} {\bibinfo {author} {\bibfnamefont {T.}~\bibnamefont
  {Cea}}\ and\ \bibinfo {author} {\bibfnamefont {F.}~\bibnamefont {Guinea}},\
  }\href {\doibase 10.1103/PhysRevB.102.045107} {\bibfield  {journal} {\bibinfo
   {journal} {Phys. Rev. B}\ }\textbf {\bibinfo {volume} {102}},\ \bibinfo
  {pages} {045107} (\bibinfo {year} {2020})}\BibitemShut {NoStop}%
\bibitem [{\citenamefont {Bernevig}\ \emph {et~al.}(2021)\citenamefont
  {Bernevig}, \citenamefont {Lian}, \citenamefont {Cowsik}, \citenamefont
  {Xie}, \citenamefont {Regnault},\ and\ \citenamefont {Song}}]{TBGV}%
  \BibitemOpen
  \bibfield  {author} {\bibinfo {author} {\bibfnamefont {B.~A.}\ \bibnamefont
  {Bernevig}}, \bibinfo {author} {\bibfnamefont {B.}~\bibnamefont {Lian}},
  \bibinfo {author} {\bibfnamefont {A.}~\bibnamefont {Cowsik}}, \bibinfo
  {author} {\bibfnamefont {F.}~\bibnamefont {Xie}}, \bibinfo {author}
  {\bibfnamefont {N.}~\bibnamefont {Regnault}}, \ and\ \bibinfo {author}
  {\bibfnamefont {Z.-D.}\ \bibnamefont {Song}},\ }\href {\doibase
  10.1103/PhysRevB.103.205415} {\bibfield  {journal} {\bibinfo  {journal}
  {Phys. Rev. B}\ }\textbf {\bibinfo {volume} {103}},\ \bibinfo {pages}
  {205415} (\bibinfo {year} {2021})}\BibitemShut {NoStop}%
\bibitem [{\citenamefont {Pierce}\ \emph {et~al.}(2021)\citenamefont {Pierce},
  \citenamefont {Xie}, \citenamefont {Park}, \citenamefont {Khalaf},
  \citenamefont {Lee}, \citenamefont {Cao}, \citenamefont {Parker},
  \citenamefont {Forrester}, \citenamefont {Chen}, \citenamefont {Watanabe},
  \citenamefont {Taniguchi}, \citenamefont {Vishwanath}, \citenamefont
  {Jarillo-Herrero},\ and\ \citenamefont {Yacoby}}]{PierceCDW}%
  \BibitemOpen
  \bibfield  {author} {\bibinfo {author} {\bibfnamefont {A.~T.}\ \bibnamefont
  {Pierce}}, \bibinfo {author} {\bibfnamefont {Y.}~\bibnamefont {Xie}},
  \bibinfo {author} {\bibfnamefont {J.~M.}\ \bibnamefont {Park}}, \bibinfo
  {author} {\bibfnamefont {E.}~\bibnamefont {Khalaf}}, \bibinfo {author}
  {\bibfnamefont {S.~H.}\ \bibnamefont {Lee}}, \bibinfo {author} {\bibfnamefont
  {Y.}~\bibnamefont {Cao}}, \bibinfo {author} {\bibfnamefont {D.~E.}\
  \bibnamefont {Parker}}, \bibinfo {author} {\bibfnamefont {P.~R.}\
  \bibnamefont {Forrester}}, \bibinfo {author} {\bibfnamefont {S.}~\bibnamefont
  {Chen}}, \bibinfo {author} {\bibfnamefont {K.}~\bibnamefont {Watanabe}},
  \bibinfo {author} {\bibfnamefont {T.}~\bibnamefont {Taniguchi}}, \bibinfo
  {author} {\bibfnamefont {A.}~\bibnamefont {Vishwanath}}, \bibinfo {author}
  {\bibfnamefont {P.}~\bibnamefont {Jarillo-Herrero}}, \ and\ \bibinfo {author}
  {\bibfnamefont {A.}~\bibnamefont {Yacoby}},\ }\href {\doibase
  10.1038/s41567-021-01347-4} {\bibfield  {journal} {\bibinfo  {journal}
  {Nature Physics}\ }\textbf {\bibinfo {volume} {17}},\ \bibinfo {pages} {1210}
  (\bibinfo {year} {2021})}\BibitemShut {NoStop}%
\bibitem [{\citenamefont {Kang}\ \emph {et~al.}(2021)\citenamefont {Kang},
  \citenamefont {Bernevig},\ and\ \citenamefont
  {Vafek}}]{VafekBernevigCascade}%
  \BibitemOpen
  \bibfield  {author} {\bibinfo {author} {\bibfnamefont {J.}~\bibnamefont
  {Kang}}, \bibinfo {author} {\bibfnamefont {B.~A.}\ \bibnamefont {Bernevig}},
  \ and\ \bibinfo {author} {\bibfnamefont {O.}~\bibnamefont {Vafek}},\ }\href
  {\doibase 10.1103/PhysRevLett.127.266402} {\bibfield  {journal} {\bibinfo
  {journal} {Phys. Rev. Lett.}\ }\textbf {\bibinfo {volume} {127}},\ \bibinfo
  {pages} {266402} (\bibinfo {year} {2021})}\BibitemShut {NoStop}%
\bibitem [{\citenamefont {Soejima}\ \emph {et~al.}(2020)\citenamefont
  {Soejima}, \citenamefont {Parker}, \citenamefont {Bultinck}, \citenamefont
  {Hauschild}, \citenamefont {Zaletel} \emph {et~al.}}]{soejima2020efficient}%
  \BibitemOpen
  \bibfield  {author} {\bibinfo {author} {\bibfnamefont {T.}~\bibnamefont
  {Soejima}}, \bibinfo {author} {\bibfnamefont {D.~E.}\ \bibnamefont {Parker}},
  \bibinfo {author} {\bibfnamefont {N.}~\bibnamefont {Bultinck}}, \bibinfo
  {author} {\bibfnamefont {J.}~\bibnamefont {Hauschild}}, \bibinfo {author}
  {\bibfnamefont {M.~P.}\ \bibnamefont {Zaletel}},  \emph {et~al.},\
  }\href@noop {} {\bibfield  {journal} {\bibinfo  {journal} {Physical Review
  B}\ }\textbf {\bibinfo {volume} {102}},\ \bibinfo {pages} {205111} (\bibinfo
  {year} {2020})}\BibitemShut {NoStop}%
\bibitem [{\citenamefont {Ledwith}\ \emph
  {et~al.}(2021{\natexlab{a}})\citenamefont {Ledwith}, \citenamefont {Khalaf},\
  and\ \citenamefont {Vishwanath}}]{ledwith2021strong}%
  \BibitemOpen
  \bibfield  {author} {\bibinfo {author} {\bibfnamefont {P.~J.}\ \bibnamefont
  {Ledwith}}, \bibinfo {author} {\bibfnamefont {E.}~\bibnamefont {Khalaf}}, \
  and\ \bibinfo {author} {\bibfnamefont {A.}~\bibnamefont {Vishwanath}},\
  }\href@noop {} {\bibfield  {journal} {\bibinfo  {journal} {arXiv preprint
  arXiv:2105.08858}\ } (\bibinfo {year} {2021}{\natexlab{a}})}\BibitemShut
  {NoStop}%
\bibitem [{\citenamefont {Lian}\ \emph {et~al.}(2021)\citenamefont {Lian},
  \citenamefont {Song}, \citenamefont {Regnault}, \citenamefont {Efetov},
  \citenamefont {Yazdani},\ and\ \citenamefont {Bernevig}}]{TBG1V}%
  \BibitemOpen
  \bibfield  {author} {\bibinfo {author} {\bibfnamefont {B.}~\bibnamefont
  {Lian}}, \bibinfo {author} {\bibfnamefont {Z.-D.}\ \bibnamefont {Song}},
  \bibinfo {author} {\bibfnamefont {N.}~\bibnamefont {Regnault}}, \bibinfo
  {author} {\bibfnamefont {D.~K.}\ \bibnamefont {Efetov}}, \bibinfo {author}
  {\bibfnamefont {A.}~\bibnamefont {Yazdani}}, \ and\ \bibinfo {author}
  {\bibfnamefont {B.~A.}\ \bibnamefont {Bernevig}},\ }\href {\doibase
  10.1103/physrevb.103.205414} {\bibfield  {journal} {\bibinfo  {journal}
  {Physical Review B}\ }\textbf {\bibinfo {volume} {103}} (\bibinfo {year}
  {2021}),\ 10.1103/physrevb.103.205414}\BibitemShut {NoStop}%
\bibitem [{\citenamefont {Carr}\ \emph {et~al.}(2019)\citenamefont {Carr},
  \citenamefont {Fang}, \citenamefont {Zhu},\ and\ \citenamefont
  {Kaxiras}}]{CarrKaxiras}%
  \BibitemOpen
  \bibfield  {author} {\bibinfo {author} {\bibfnamefont {S.}~\bibnamefont
  {Carr}}, \bibinfo {author} {\bibfnamefont {S.}~\bibnamefont {Fang}}, \bibinfo
  {author} {\bibfnamefont {Z.}~\bibnamefont {Zhu}}, \ and\ \bibinfo {author}
  {\bibfnamefont {E.}~\bibnamefont {Kaxiras}},\ }\href {\doibase
  10.1103/PhysRevResearch.1.013001} {\bibfield  {journal} {\bibinfo  {journal}
  {Phys. Rev. Research}\ }\textbf {\bibinfo {volume} {1}},\ \bibinfo {pages}
  {013001} (\bibinfo {year} {2019})}\BibitemShut {NoStop}%
\bibitem [{\citenamefont {Nam}\ and\ \citenamefont
  {Koshino}(2017)}]{NamRelaxation}%
  \BibitemOpen
  \bibfield  {author} {\bibinfo {author} {\bibfnamefont {N.~N.~T.}\
  \bibnamefont {Nam}}\ and\ \bibinfo {author} {\bibfnamefont {M.}~\bibnamefont
  {Koshino}},\ }\href {\doibase 10.1103/PhysRevB.96.075311} {\bibfield
  {journal} {\bibinfo  {journal} {Phys. Rev. B}\ }\textbf {\bibinfo {volume}
  {96}},\ \bibinfo {pages} {075311} (\bibinfo {year} {2017})}\BibitemShut
  {NoStop}%
\bibitem [{\citenamefont {Koshino}\ \emph {et~al.}(2018)\citenamefont
  {Koshino}, \citenamefont {Yuan}, \citenamefont {Koretsune}, \citenamefont
  {Ochi}, \citenamefont {Kuroki},\ and\ \citenamefont
  {Fu}}]{Koshino2018outofplane}%
  \BibitemOpen
  \bibfield  {author} {\bibinfo {author} {\bibfnamefont {M.}~\bibnamefont
  {Koshino}}, \bibinfo {author} {\bibfnamefont {N.~F.~Q.}\ \bibnamefont
  {Yuan}}, \bibinfo {author} {\bibfnamefont {T.}~\bibnamefont {Koretsune}},
  \bibinfo {author} {\bibfnamefont {M.}~\bibnamefont {Ochi}}, \bibinfo {author}
  {\bibfnamefont {K.}~\bibnamefont {Kuroki}}, \ and\ \bibinfo {author}
  {\bibfnamefont {L.}~\bibnamefont {Fu}},\ }\href {\doibase
  10.1103/PhysRevX.8.031087} {\bibfield  {journal} {\bibinfo  {journal} {Phys.
  Rev. X}\ }\textbf {\bibinfo {volume} {8}},\ \bibinfo {pages} {031087}
  (\bibinfo {year} {2018})}\BibitemShut {NoStop}%
\bibitem [{\citenamefont {Carr}\ \emph
  {et~al.}(2018{\natexlab{a}})\citenamefont {Carr}, \citenamefont {Massatt},
  \citenamefont {Torrisi}, \citenamefont {Cazeaux}, \citenamefont {Luskin},\
  and\ \citenamefont {Kaxiras}}]{Carr2018relax}%
  \BibitemOpen
  \bibfield  {author} {\bibinfo {author} {\bibfnamefont {S.}~\bibnamefont
  {Carr}}, \bibinfo {author} {\bibfnamefont {D.}~\bibnamefont {Massatt}},
  \bibinfo {author} {\bibfnamefont {S.~B.}\ \bibnamefont {Torrisi}}, \bibinfo
  {author} {\bibfnamefont {P.}~\bibnamefont {Cazeaux}}, \bibinfo {author}
  {\bibfnamefont {M.}~\bibnamefont {Luskin}}, \ and\ \bibinfo {author}
  {\bibfnamefont {E.}~\bibnamefont {Kaxiras}},\ }\href {\doibase
  10.1103/PhysRevB.98.224102} {\bibfield  {journal} {\bibinfo  {journal} {Phys.
  Rev. B}\ }\textbf {\bibinfo {volume} {98}},\ \bibinfo {pages} {224102}
  (\bibinfo {year} {2018}{\natexlab{a}})}\BibitemShut {NoStop}%
\bibitem [{\citenamefont {Carr}\ \emph {et~al.}(2020)\citenamefont {Carr},
  \citenamefont {Fang},\ and\ \citenamefont {Kaxiras}}]{Carr2020review}%
  \BibitemOpen
  \bibfield  {author} {\bibinfo {author} {\bibfnamefont {S.}~\bibnamefont
  {Carr}}, \bibinfo {author} {\bibfnamefont {S.}~\bibnamefont {Fang}}, \ and\
  \bibinfo {author} {\bibfnamefont {E.}~\bibnamefont {Kaxiras}},\ }\href
  {\doibase 10.1038/s41578-020-0214-0} {\bibfield  {journal} {\bibinfo
  {journal} {Nature Reviews Materials}\ }\textbf {\bibinfo {volume} {5}},\
  \bibinfo {pages} {748} (\bibinfo {year} {2020})}\BibitemShut {NoStop}%
\bibitem [{\citenamefont {Guinea}\ and\ \citenamefont
  {Walet}(2019)}]{Guinea2019Relax}%
  \BibitemOpen
  \bibfield  {author} {\bibinfo {author} {\bibfnamefont {F.}~\bibnamefont
  {Guinea}}\ and\ \bibinfo {author} {\bibfnamefont {N.~R.}\ \bibnamefont
  {Walet}},\ }\href {\doibase 10.1103/PhysRevB.99.205134} {\bibfield  {journal}
  {\bibinfo  {journal} {Phys. Rev. B}\ }\textbf {\bibinfo {volume} {99}},\
  \bibinfo {pages} {205134} (\bibinfo {year} {2019})}\BibitemShut {NoStop}%
\bibitem [{\citenamefont {Ledwith}\ \emph
  {et~al.}(2021{\natexlab{b}})\citenamefont {Ledwith}, \citenamefont {Khalaf},
  \citenamefont {Zhu}, \citenamefont {Carr}, \citenamefont {Kaxiras},\ and\
  \citenamefont {Vishwanath}}]{TBorNotTB}%
  \BibitemOpen
  \bibfield  {author} {\bibinfo {author} {\bibfnamefont {P.~J.}\ \bibnamefont
  {Ledwith}}, \bibinfo {author} {\bibfnamefont {E.}~\bibnamefont {Khalaf}},
  \bibinfo {author} {\bibfnamefont {Z.}~\bibnamefont {Zhu}}, \bibinfo {author}
  {\bibfnamefont {S.}~\bibnamefont {Carr}}, \bibinfo {author} {\bibfnamefont
  {E.}~\bibnamefont {Kaxiras}}, \ and\ \bibinfo {author} {\bibfnamefont
  {A.}~\bibnamefont {Vishwanath}},\ }\href@noop {} {\bibfield  {journal}
  {\bibinfo  {journal} {arXiv preprint arXiv:2111.11060}\ } (\bibinfo {year}
  {2021}{\natexlab{b}})}\BibitemShut {NoStop}%
\bibitem [{\citenamefont {Fang}\ and\ \citenamefont
  {Kaxiras}(2016)}]{Fang2016}%
  \BibitemOpen
  \bibfield  {author} {\bibinfo {author} {\bibfnamefont {S.}~\bibnamefont
  {Fang}}\ and\ \bibinfo {author} {\bibfnamefont {E.}~\bibnamefont {Kaxiras}},\
  }\href {\doibase 10.1103/PhysRevB.93.235153} {\bibfield  {journal} {\bibinfo
  {journal} {Phys. Rev. B}\ }\textbf {\bibinfo {volume} {93}},\ \bibinfo
  {pages} {235153} (\bibinfo {year} {2016})}\BibitemShut {NoStop}%
\bibitem [{\citenamefont {Carr}\ \emph
  {et~al.}(2018{\natexlab{b}})\citenamefont {Carr}, \citenamefont {Fang},
  \citenamefont {Jarillo-Herrero},\ and\ \citenamefont
  {Kaxiras}}]{Carr2018PressureDep}%
  \BibitemOpen
  \bibfield  {author} {\bibinfo {author} {\bibfnamefont {S.}~\bibnamefont
  {Carr}}, \bibinfo {author} {\bibfnamefont {S.}~\bibnamefont {Fang}}, \bibinfo
  {author} {\bibfnamefont {P.}~\bibnamefont {Jarillo-Herrero}}, \ and\ \bibinfo
  {author} {\bibfnamefont {E.}~\bibnamefont {Kaxiras}},\ }\href {\doibase
  10.1103/PhysRevB.98.085144} {\bibfield  {journal} {\bibinfo  {journal} {Phys.
  Rev. B}\ }\textbf {\bibinfo {volume} {98}},\ \bibinfo {pages} {085144}
  (\bibinfo {year} {2018}{\natexlab{b}})}\BibitemShut {NoStop}%
\bibitem [{\citenamefont {Xie}\ \emph {et~al.}(2019)\citenamefont {Xie},
  \citenamefont {Lian}, \citenamefont {J{\"a}ck}, \citenamefont {Liu},
  \citenamefont {Chiu}, \citenamefont {Watanabe}, \citenamefont {Taniguchi},
  \citenamefont {Bernevig},\ and\ \citenamefont
  {Yazdani}}]{xie2019spectroscopic}%
  \BibitemOpen
  \bibfield  {author} {\bibinfo {author} {\bibfnamefont {Y.}~\bibnamefont
  {Xie}}, \bibinfo {author} {\bibfnamefont {B.}~\bibnamefont {Lian}}, \bibinfo
  {author} {\bibfnamefont {B.}~\bibnamefont {J{\"a}ck}}, \bibinfo {author}
  {\bibfnamefont {X.}~\bibnamefont {Liu}}, \bibinfo {author} {\bibfnamefont
  {C.-L.}\ \bibnamefont {Chiu}}, \bibinfo {author} {\bibfnamefont
  {K.}~\bibnamefont {Watanabe}}, \bibinfo {author} {\bibfnamefont
  {T.}~\bibnamefont {Taniguchi}}, \bibinfo {author} {\bibfnamefont {B.~A.}\
  \bibnamefont {Bernevig}}, \ and\ \bibinfo {author} {\bibfnamefont
  {A.}~\bibnamefont {Yazdani}},\ }\href@noop {} {\bibfield  {journal} {\bibinfo
   {journal} {Nature}\ }\textbf {\bibinfo {volume} {572}},\ \bibinfo {pages}
  {101} (\bibinfo {year} {2019})}\BibitemShut {NoStop}%
\bibitem [{\citenamefont {Das}\ \emph {et~al.}(2021)\citenamefont {Das},
  \citenamefont {Lu}, \citenamefont {Herzog-Arbeitman}, \citenamefont {Song},
  \citenamefont {Watanabe}, \citenamefont {Taniguchi}, \citenamefont
  {Bernevig},\ and\ \citenamefont {Efetov}}]{Das2021}%
  \BibitemOpen
  \bibfield  {author} {\bibinfo {author} {\bibfnamefont {I.}~\bibnamefont
  {Das}}, \bibinfo {author} {\bibfnamefont {X.}~\bibnamefont {Lu}}, \bibinfo
  {author} {\bibfnamefont {J.}~\bibnamefont {Herzog-Arbeitman}}, \bibinfo
  {author} {\bibfnamefont {Z.-D.}\ \bibnamefont {Song}}, \bibinfo {author}
  {\bibfnamefont {K.}~\bibnamefont {Watanabe}}, \bibinfo {author}
  {\bibfnamefont {T.}~\bibnamefont {Taniguchi}}, \bibinfo {author}
  {\bibfnamefont {B.~A.}\ \bibnamefont {Bernevig}}, \ and\ \bibinfo {author}
  {\bibfnamefont {D.~K.}\ \bibnamefont {Efetov}},\ }\href {\doibase
  10.1038/s41567-021-01186-3} {\bibfield  {journal} {\bibinfo  {journal}
  {Nature Physics}\ }\textbf {\bibinfo {volume} {17}},\ \bibinfo {pages} {710}
  (\bibinfo {year} {2021})}\BibitemShut {NoStop}%
\bibitem [{\citenamefont {Vafek}\ and\ \citenamefont
  {Kang}(2020)}]{vafek2020renormalization}%
  \BibitemOpen
  \bibfield  {author} {\bibinfo {author} {\bibfnamefont {O.}~\bibnamefont
  {Vafek}}\ and\ \bibinfo {author} {\bibfnamefont {J.}~\bibnamefont {Kang}},\
  }\href@noop {} {\bibfield  {journal} {\bibinfo  {journal} {Physical Review
  Letters}\ }\textbf {\bibinfo {volume} {125}},\ \bibinfo {pages} {257602}
  (\bibinfo {year} {2020})}\BibitemShut {NoStop}%
\bibitem [{\citenamefont {Parker}\ \emph
  {et~al.}(2020{\natexlab{b}})\citenamefont {Parker}, \citenamefont {Cao},\
  and\ \citenamefont {Zaletel}}]{parker2020local}%
  \BibitemOpen
  \bibfield  {author} {\bibinfo {author} {\bibfnamefont {D.~E.}\ \bibnamefont
  {Parker}}, \bibinfo {author} {\bibfnamefont {X.}~\bibnamefont {Cao}}, \ and\
  \bibinfo {author} {\bibfnamefont {M.~P.}\ \bibnamefont {Zaletel}},\
  }\href@noop {} {\bibfield  {journal} {\bibinfo  {journal} {Physical Review
  B}\ }\textbf {\bibinfo {volume} {102}},\ \bibinfo {pages} {035147} (\bibinfo
  {year} {2020}{\natexlab{b}})}\BibitemShut {NoStop}%
\bibitem [{\citenamefont {Grushin}\ \emph {et~al.}(2015)\citenamefont
  {Grushin}, \citenamefont {Motruk}, \citenamefont {Zaletel},\ and\
  \citenamefont {Pollmann}}]{grushin2015characterization}%
  \BibitemOpen
  \bibfield  {author} {\bibinfo {author} {\bibfnamefont {A.~G.}\ \bibnamefont
  {Grushin}}, \bibinfo {author} {\bibfnamefont {J.}~\bibnamefont {Motruk}},
  \bibinfo {author} {\bibfnamefont {M.~P.}\ \bibnamefont {Zaletel}}, \ and\
  \bibinfo {author} {\bibfnamefont {F.}~\bibnamefont {Pollmann}},\ }\href@noop
  {} {\bibfield  {journal} {\bibinfo  {journal} {Physical Review B}\ }\textbf
  {\bibinfo {volume} {91}},\ \bibinfo {pages} {035136} (\bibinfo {year}
  {2015})}\BibitemShut {NoStop}%
\bibitem [{\citenamefont {Motruk}\ \emph {et~al.}(2016)\citenamefont {Motruk},
  \citenamefont {Zaletel}, \citenamefont {Mong},\ and\ \citenamefont
  {Pollmann}}]{motruk2016density}%
  \BibitemOpen
  \bibfield  {author} {\bibinfo {author} {\bibfnamefont {J.}~\bibnamefont
  {Motruk}}, \bibinfo {author} {\bibfnamefont {M.~P.}\ \bibnamefont {Zaletel}},
  \bibinfo {author} {\bibfnamefont {R.~S.}\ \bibnamefont {Mong}}, \ and\
  \bibinfo {author} {\bibfnamefont {F.}~\bibnamefont {Pollmann}},\ }\href@noop
  {} {\bibfield  {journal} {\bibinfo  {journal} {Physical Review B}\ }\textbf
  {\bibinfo {volume} {93}},\ \bibinfo {pages} {155139} (\bibinfo {year}
  {2016})}\BibitemShut {NoStop}%
\bibitem [{\citenamefont {Zaletel}\ \emph {et~al.}(2013)\citenamefont
  {Zaletel}, \citenamefont {Mong},\ and\ \citenamefont
  {Pollmann}}]{zaletel2013topological}%
  \BibitemOpen
  \bibfield  {author} {\bibinfo {author} {\bibfnamefont {M.~P.}\ \bibnamefont
  {Zaletel}}, \bibinfo {author} {\bibfnamefont {R.~S.}\ \bibnamefont {Mong}}, \
  and\ \bibinfo {author} {\bibfnamefont {F.}~\bibnamefont {Pollmann}},\
  }\href@noop {} {\bibfield  {journal} {\bibinfo  {journal} {Physical review
  letters}\ }\textbf {\bibinfo {volume} {110}},\ \bibinfo {pages} {236801}
  (\bibinfo {year} {2013})}\BibitemShut {NoStop}%
\bibitem [{\citenamefont {Zaletel}\ \emph {et~al.}(2015)\citenamefont
  {Zaletel}, \citenamefont {Mong}, \citenamefont {Pollmann},\ and\
  \citenamefont {Rezayi}}]{zaletel2015infinite}%
  \BibitemOpen
  \bibfield  {author} {\bibinfo {author} {\bibfnamefont {M.~P.}\ \bibnamefont
  {Zaletel}}, \bibinfo {author} {\bibfnamefont {R.~S.}\ \bibnamefont {Mong}},
  \bibinfo {author} {\bibfnamefont {F.}~\bibnamefont {Pollmann}}, \ and\
  \bibinfo {author} {\bibfnamefont {E.~H.}\ \bibnamefont {Rezayi}},\
  }\href@noop {} {\bibfield  {journal} {\bibinfo  {journal} {Physical Review
  B}\ }\textbf {\bibinfo {volume} {91}},\ \bibinfo {pages} {045115} (\bibinfo
  {year} {2015})}\BibitemShut {NoStop}%
\bibitem [{\citenamefont {Qi}(2011)}]{qi2011generic}%
  \BibitemOpen
  \bibfield  {author} {\bibinfo {author} {\bibfnamefont {X.-L.}\ \bibnamefont
  {Qi}},\ }\href@noop {} {\bibfield  {journal} {\bibinfo  {journal} {Physical
  review letters}\ }\textbf {\bibinfo {volume} {107}},\ \bibinfo {pages}
  {126803} (\bibinfo {year} {2011})}\BibitemShut {NoStop}%
\bibitem [{\citenamefont {Kang}\ and\ \citenamefont
  {Vafek}(2020)}]{VafekBraiding}%
  \BibitemOpen
  \bibfield  {author} {\bibinfo {author} {\bibfnamefont {J.}~\bibnamefont
  {Kang}}\ and\ \bibinfo {author} {\bibfnamefont {O.}~\bibnamefont {Vafek}},\
  }\href {\doibase 10.1103/PhysRevB.102.035161} {\bibfield  {journal} {\bibinfo
   {journal} {Phys. Rev. B}\ }\textbf {\bibinfo {volume} {102}},\ \bibinfo
  {pages} {035161} (\bibinfo {year} {2020})}\BibitemShut {NoStop}%
\bibitem [{\citenamefont {Li}\ and\ \citenamefont
  {Haldane}(2008)}]{li2008entanglement}%
  \BibitemOpen
  \bibfield  {author} {\bibinfo {author} {\bibfnamefont {H.}~\bibnamefont
  {Li}}\ and\ \bibinfo {author} {\bibfnamefont {F.~D.~M.}\ \bibnamefont
  {Haldane}},\ }\href@noop {} {\bibfield  {journal} {\bibinfo  {journal}
  {Physical review letters}\ }\textbf {\bibinfo {volume} {101}},\ \bibinfo
  {pages} {010504} (\bibinfo {year} {2008})}\BibitemShut {NoStop}%
\bibitem [{\citenamefont {Seidel}\ \emph {et~al.}(2005)\citenamefont {Seidel},
  \citenamefont {Fu}, \citenamefont {Lee}, \citenamefont {Leinaas},\ and\
  \citenamefont {Moore}}]{seidel2005incompressible}%
  \BibitemOpen
  \bibfield  {author} {\bibinfo {author} {\bibfnamefont {A.}~\bibnamefont
  {Seidel}}, \bibinfo {author} {\bibfnamefont {H.}~\bibnamefont {Fu}}, \bibinfo
  {author} {\bibfnamefont {D.-H.}\ \bibnamefont {Lee}}, \bibinfo {author}
  {\bibfnamefont {J.~M.}\ \bibnamefont {Leinaas}}, \ and\ \bibinfo {author}
  {\bibfnamefont {J.}~\bibnamefont {Moore}},\ }\href@noop {} {\bibfield
  {journal} {\bibinfo  {journal} {Physical review letters}\ }\textbf {\bibinfo
  {volume} {95}},\ \bibinfo {pages} {266405} (\bibinfo {year}
  {2005})}\BibitemShut {NoStop}%
\bibitem [{\citenamefont {Simon}\ and\ \citenamefont
  {Rudner}(2020{\natexlab{b}})}]{simon2020contrasting}%
  \BibitemOpen
  \bibfield  {author} {\bibinfo {author} {\bibfnamefont {S.~H.}\ \bibnamefont
  {Simon}}\ and\ \bibinfo {author} {\bibfnamefont {M.~S.}\ \bibnamefont
  {Rudner}},\ }\href@noop {} {\bibfield  {journal} {\bibinfo  {journal}
  {Physical Review B}\ }\textbf {\bibinfo {volume} {102}},\ \bibinfo {pages}
  {165148} (\bibinfo {year} {2020}{\natexlab{b}})}\BibitemShut {NoStop}%
\bibitem [{\citenamefont {Girvin}\ \emph {et~al.}({\natexlab{a}})\citenamefont
  {Girvin}, \citenamefont {{MacDonald}},\ and\ \citenamefont
  {Platzman}}]{girvinCollectiveExcitationGapFractional1985}%
  \BibitemOpen
  \bibfield  {author} {\bibinfo {author} {\bibfnamefont {S.~M.}\ \bibnamefont
  {Girvin}}, \bibinfo {author} {\bibfnamefont {A.~H.}\ \bibnamefont
  {{MacDonald}}}, \ and\ \bibinfo {author} {\bibfnamefont {P.~M.}\ \bibnamefont
  {Platzman}},\ }\href {\doibase 10.1103/PhysRevLett.54.581} {\ \textbf
  {\bibinfo {volume} {54}},\ \bibinfo {pages} {581}
  ({\natexlab{a}})}\BibitemShut {NoStop}%
\bibitem [{\citenamefont {Hofstadter}()}]{hofstadterEnergyLevelsWave1976}%
  \BibitemOpen
  \bibfield  {author} {\bibinfo {author} {\bibfnamefont {D.~R.}\ \bibnamefont
  {Hofstadter}},\ }\href {\doibase 10.1103/PhysRevB.14.2239} {\ \textbf
  {\bibinfo {volume} {14}},\ \bibinfo {pages} {2239}}\BibitemShut {NoStop}%
\bibitem [{\citenamefont {Bistritzer}\ and\ \citenamefont
  {{MacDonald}}()}]{bistritzerMoireBandsTwisted2011}%
  \BibitemOpen
  \bibfield  {author} {\bibinfo {author} {\bibfnamefont {R.}~\bibnamefont
  {Bistritzer}}\ and\ \bibinfo {author} {\bibfnamefont {A.~H.}\ \bibnamefont
  {{MacDonald}}},\ }\href {\doibase 10.1073/pnas.1108174108} {\ \textbf
  {\bibinfo {volume} {108}},\ \bibinfo {pages} {12233}},\ \bibinfo {note}
  {{ISBN}: 9781108174107 Publisher: National Academy of Sciences Section:
  Physical Sciences}\BibitemShut {NoStop}%
\bibitem [{\citenamefont {Zhang}\ \emph {et~al.}()\citenamefont {Zhang},
  \citenamefont {Po},\ and\ \citenamefont
  {Senthil}}]{zhangLandauLevelDegeneracy2019}%
  \BibitemOpen
  \bibfield  {author} {\bibinfo {author} {\bibfnamefont {Y.-H.}\ \bibnamefont
  {Zhang}}, \bibinfo {author} {\bibfnamefont {H.~C.}\ \bibnamefont {Po}}, \
  and\ \bibinfo {author} {\bibfnamefont {T.}~\bibnamefont {Senthil}},\ }\href
  {\doibase 10.1103/PhysRevB.100.125104} {\ \textbf {\bibinfo {volume} {100}},\
  \bibinfo {pages} {125104}},\ \Eprint {http://arxiv.org/abs/1904.10452}
  {1904.10452} \BibitemShut {NoStop}%
\bibitem [{\citenamefont {Hejazi}\ \emph {et~al.}()\citenamefont {Hejazi},
  \citenamefont {Liu},\ and\ \citenamefont
  {Balents}}]{hejaziLandauLevelsTwisted2019}%
  \BibitemOpen
  \bibfield  {author} {\bibinfo {author} {\bibfnamefont {K.}~\bibnamefont
  {Hejazi}}, \bibinfo {author} {\bibfnamefont {C.}~\bibnamefont {Liu}}, \ and\
  \bibinfo {author} {\bibfnamefont {L.}~\bibnamefont {Balents}},\ }\href
  {\doibase 10.1103/PhysRevB.100.035115} {\ \textbf {\bibinfo {volume} {100}},\
  \bibinfo {pages} {035115}}\BibitemShut {NoStop}%
\bibitem [{\citenamefont {Wang}\ and\ \citenamefont
  {Vafek}(2021)}]{wang2021narrow}%
  \BibitemOpen
  \bibfield  {author} {\bibinfo {author} {\bibfnamefont {X.}~\bibnamefont
  {Wang}}\ and\ \bibinfo {author} {\bibfnamefont {O.}~\bibnamefont {Vafek}},\
  }\href@noop {} {\bibfield  {journal} {\bibinfo  {journal} {arXiv preprint
  arXiv:2112.08620}\ } (\bibinfo {year} {2021})}\BibitemShut {NoStop}%
\bibitem [{\citenamefont {Onsager}(1952)}]{onsager1952interpretation}%
  \BibitemOpen
  \bibfield  {author} {\bibinfo {author} {\bibfnamefont {L.}~\bibnamefont
  {Onsager}},\ }\href@noop {} {\bibfield  {journal} {\bibinfo  {journal} {The
  London, Edinburgh, and Dublin Philosophical Magazine and Journal of Science}\
  }\textbf {\bibinfo {volume} {43}},\ \bibinfo {pages} {1006} (\bibinfo {year}
  {1952})}\BibitemShut {NoStop}%
\bibitem [{\citenamefont {Lifshitz}\ and\ \citenamefont
  {Kosevich}(1956)}]{lifshitz1956theory}%
  \BibitemOpen
  \bibfield  {author} {\bibinfo {author} {\bibfnamefont {I.}~\bibnamefont
  {Lifshitz}}\ and\ \bibinfo {author} {\bibfnamefont {A.}~\bibnamefont
  {Kosevich}},\ }\href@noop {} {\bibfield  {journal} {\bibinfo  {journal} {Sov.
  Phys. JETP}\ }\textbf {\bibinfo {volume} {2}},\ \bibinfo {pages} {636}
  (\bibinfo {year} {1956})}\BibitemShut {NoStop}%
\bibitem [{\citenamefont {Mikitik}\ and\ \citenamefont
  {Sharlai}(1999)}]{mikitik1999manifestation}%
  \BibitemOpen
  \bibfield  {author} {\bibinfo {author} {\bibfnamefont {G.}~\bibnamefont
  {Mikitik}}\ and\ \bibinfo {author} {\bibfnamefont {Y.~V.}\ \bibnamefont
  {Sharlai}},\ }\href@noop {} {\bibfield  {journal} {\bibinfo  {journal}
  {Physical review letters}\ }\textbf {\bibinfo {volume} {82}},\ \bibinfo
  {pages} {2147} (\bibinfo {year} {1999})}\BibitemShut {NoStop}%
\bibitem [{\citenamefont {Suen}\ \emph {et~al.}(1992)\citenamefont {Suen},
  \citenamefont {Engel}, \citenamefont {Santos}, \citenamefont {Shayegan},\
  and\ \citenamefont {Tsui}}]{SuenExptBilayerQH1992}%
  \BibitemOpen
  \bibfield  {author} {\bibinfo {author} {\bibfnamefont {Y.~W.}\ \bibnamefont
  {Suen}}, \bibinfo {author} {\bibfnamefont {L.~W.}\ \bibnamefont {Engel}},
  \bibinfo {author} {\bibfnamefont {M.~B.}\ \bibnamefont {Santos}}, \bibinfo
  {author} {\bibfnamefont {M.}~\bibnamefont {Shayegan}}, \ and\ \bibinfo
  {author} {\bibfnamefont {D.~C.}\ \bibnamefont {Tsui}},\ }\href {\doibase
  10.1103/PhysRevLett.68.1379} {\bibfield  {journal} {\bibinfo  {journal}
  {Phys. Rev. Lett.}\ }\textbf {\bibinfo {volume} {68}},\ \bibinfo {pages}
  {1379} (\bibinfo {year} {1992})}\BibitemShut {NoStop}%
\bibitem [{\citenamefont {Eisenstein}\ \emph {et~al.}(1992)\citenamefont
  {Eisenstein}, \citenamefont {Boebinger}, \citenamefont {Pfeiffer},
  \citenamefont {West},\ and\ \citenamefont
  {He}}]{EisensteinExptBilayerQH1992}%
  \BibitemOpen
  \bibfield  {author} {\bibinfo {author} {\bibfnamefont {J.~P.}\ \bibnamefont
  {Eisenstein}}, \bibinfo {author} {\bibfnamefont {G.~S.}\ \bibnamefont
  {Boebinger}}, \bibinfo {author} {\bibfnamefont {L.~N.}\ \bibnamefont
  {Pfeiffer}}, \bibinfo {author} {\bibfnamefont {K.~W.}\ \bibnamefont {West}},
  \ and\ \bibinfo {author} {\bibfnamefont {S.}~\bibnamefont {He}},\ }\href
  {\doibase 10.1103/PhysRevLett.68.1383} {\bibfield  {journal} {\bibinfo
  {journal} {Phys. Rev. Lett.}\ }\textbf {\bibinfo {volume} {68}},\ \bibinfo
  {pages} {1383} (\bibinfo {year} {1992})}\BibitemShut {NoStop}%
\bibitem [{\citenamefont {Suen}\ \emph {et~al.}(1994)\citenamefont {Suen},
  \citenamefont {Manoharan}, \citenamefont {Ying}, \citenamefont {Santos},\
  and\ \citenamefont {Shayegan}}]{SuenExptWideWell1992}%
  \BibitemOpen
  \bibfield  {author} {\bibinfo {author} {\bibfnamefont {Y.~W.}\ \bibnamefont
  {Suen}}, \bibinfo {author} {\bibfnamefont {H.~C.}\ \bibnamefont {Manoharan}},
  \bibinfo {author} {\bibfnamefont {X.}~\bibnamefont {Ying}}, \bibinfo {author}
  {\bibfnamefont {M.~B.}\ \bibnamefont {Santos}}, \ and\ \bibinfo {author}
  {\bibfnamefont {M.}~\bibnamefont {Shayegan}},\ }\href {\doibase
  10.1103/PhysRevLett.72.3405} {\bibfield  {journal} {\bibinfo  {journal}
  {Phys. Rev. Lett.}\ }\textbf {\bibinfo {volume} {72}},\ \bibinfo {pages}
  {3405} (\bibinfo {year} {1994})}\BibitemShut {NoStop}%
\bibitem [{\citenamefont {Shabani}\ \emph {et~al.}(2013)\citenamefont
  {Shabani}, \citenamefont {Liu}, \citenamefont {Shayegan}, \citenamefont
  {Pfeiffer}, \citenamefont {West},\ and\ \citenamefont
  {Baldwin}}]{ShabaniExptWideWell2013}%
  \BibitemOpen
  \bibfield  {author} {\bibinfo {author} {\bibfnamefont {J.}~\bibnamefont
  {Shabani}}, \bibinfo {author} {\bibfnamefont {Y.}~\bibnamefont {Liu}},
  \bibinfo {author} {\bibfnamefont {M.}~\bibnamefont {Shayegan}}, \bibinfo
  {author} {\bibfnamefont {L.~N.}\ \bibnamefont {Pfeiffer}}, \bibinfo {author}
  {\bibfnamefont {K.~W.}\ \bibnamefont {West}}, \ and\ \bibinfo {author}
  {\bibfnamefont {K.~W.}\ \bibnamefont {Baldwin}},\ }\href {\doibase
  10.1103/PhysRevB.88.245413} {\bibfield  {journal} {\bibinfo  {journal} {Phys.
  Rev. B}\ }\textbf {\bibinfo {volume} {88}},\ \bibinfo {pages} {245413}
  (\bibinfo {year} {2013})}\BibitemShut {NoStop}%
\bibitem [{\citenamefont {Zhao}\ \emph {et~al.}(2021)\citenamefont {Zhao},
  \citenamefont {Faugno}, \citenamefont {Pu}, \citenamefont {Balram},\ and\
  \citenamefont {Jain}}]{ZhaoWideWell2021}%
  \BibitemOpen
  \bibfield  {author} {\bibinfo {author} {\bibfnamefont {T.}~\bibnamefont
  {Zhao}}, \bibinfo {author} {\bibfnamefont {W.~N.}\ \bibnamefont {Faugno}},
  \bibinfo {author} {\bibfnamefont {S.}~\bibnamefont {Pu}}, \bibinfo {author}
  {\bibfnamefont {A.~C.}\ \bibnamefont {Balram}}, \ and\ \bibinfo {author}
  {\bibfnamefont {J.~K.}\ \bibnamefont {Jain}},\ }\href {\doibase
  10.1103/PhysRevB.103.155306} {\bibfield  {journal} {\bibinfo  {journal}
  {Phys. Rev. B}\ }\textbf {\bibinfo {volume} {103}},\ \bibinfo {pages}
  {155306} (\bibinfo {year} {2021})}\BibitemShut {NoStop}%
\bibitem [{\citenamefont {Zhu}\ \emph {et~al.}(2016)\citenamefont {Zhu},
  \citenamefont {Liu}, \citenamefont {Haldane},\ and\ \citenamefont
  {Sheng}}]{ZhuBilayerTunneling2016}%
  \BibitemOpen
  \bibfield  {author} {\bibinfo {author} {\bibfnamefont {W.}~\bibnamefont
  {Zhu}}, \bibinfo {author} {\bibfnamefont {Z.}~\bibnamefont {Liu}}, \bibinfo
  {author} {\bibfnamefont {F.~D.~M.}\ \bibnamefont {Haldane}}, \ and\ \bibinfo
  {author} {\bibfnamefont {D.~N.}\ \bibnamefont {Sheng}},\ }\href {\doibase
  10.1103/PhysRevB.94.245147} {\bibfield  {journal} {\bibinfo  {journal} {Phys.
  Rev. B}\ }\textbf {\bibinfo {volume} {94}},\ \bibinfo {pages} {245147}
  (\bibinfo {year} {2016})}\BibitemShut {NoStop}%
\bibitem [{\citenamefont {Papi\ifmmode~\acute{c}\else \'{c}\fi{}}\ \emph
  {et~al.}(2010)\citenamefont {Papi\ifmmode~\acute{c}\else \'{c}\fi{}},
  \citenamefont {Goerbig}, \citenamefont {Regnault},\ and\ \citenamefont
  {Milovanovi\ifmmode~\acute{c}\else \'{c}\fi{}}}]{PapicTunneling2010}%
  \BibitemOpen
  \bibfield  {author} {\bibinfo {author} {\bibfnamefont {Z.}~\bibnamefont
  {Papi\ifmmode~\acute{c}\else \'{c}\fi{}}}, \bibinfo {author} {\bibfnamefont
  {M.~O.}\ \bibnamefont {Goerbig}}, \bibinfo {author} {\bibfnamefont
  {N.}~\bibnamefont {Regnault}}, \ and\ \bibinfo {author} {\bibfnamefont
  {M.~V.}\ \bibnamefont {Milovanovi\ifmmode~\acute{c}\else \'{c}\fi{}}},\
  }\href {\doibase 10.1103/PhysRevB.82.075302} {\bibfield  {journal} {\bibinfo
  {journal} {Phys. Rev. B}\ }\textbf {\bibinfo {volume} {82}},\ \bibinfo
  {pages} {075302} (\bibinfo {year} {2010})}\BibitemShut {NoStop}%
\bibitem [{\citenamefont {Faugno}\ \emph {et~al.}(2020)\citenamefont {Faugno},
  \citenamefont {Balram}, \citenamefont {W\'ojs},\ and\ \citenamefont
  {Jain}}]{FaugnoBilayerCF2020}%
  \BibitemOpen
  \bibfield  {author} {\bibinfo {author} {\bibfnamefont {W.~N.}\ \bibnamefont
  {Faugno}}, \bibinfo {author} {\bibfnamefont {A.~C.}\ \bibnamefont {Balram}},
  \bibinfo {author} {\bibfnamefont {A.}~\bibnamefont {W\'ojs}}, \ and\ \bibinfo
  {author} {\bibfnamefont {J.~K.}\ \bibnamefont {Jain}},\ }\href {\doibase
  10.1103/PhysRevB.101.085412} {\bibfield  {journal} {\bibinfo  {journal}
  {Phys. Rev. B}\ }\textbf {\bibinfo {volume} {101}},\ \bibinfo {pages}
  {085412} (\bibinfo {year} {2020})}\BibitemShut {NoStop}%
\bibitem [{\citenamefont {Park}\ and\ \citenamefont
  {Jain}(1998)}]{ParkCFspin1998}%
  \BibitemOpen
  \bibfield  {author} {\bibinfo {author} {\bibfnamefont {K.}~\bibnamefont
  {Park}}\ and\ \bibinfo {author} {\bibfnamefont {J.~K.}\ \bibnamefont
  {Jain}},\ }\href {\doibase 10.1103/PhysRevLett.80.4237} {\bibfield  {journal}
  {\bibinfo  {journal} {Phys. Rev. Lett.}\ }\textbf {\bibinfo {volume} {80}},\
  \bibinfo {pages} {4237} (\bibinfo {year} {1998})}\BibitemShut {NoStop}%
\bibitem [{\citenamefont {Peterson}\ and\ \citenamefont
  {Das~Sarma}(2010)}]{DasSarmaPhaseDiagram2010}%
  \BibitemOpen
  \bibfield  {author} {\bibinfo {author} {\bibfnamefont {M.~R.}\ \bibnamefont
  {Peterson}}\ and\ \bibinfo {author} {\bibfnamefont {S.}~\bibnamefont
  {Das~Sarma}},\ }\href {\doibase 10.1103/PhysRevB.81.165304} {\bibfield
  {journal} {\bibinfo  {journal} {Phys. Rev. B}\ }\textbf {\bibinfo {volume}
  {81}},\ \bibinfo {pages} {165304} (\bibinfo {year} {2010})}\BibitemShut
  {NoStop}%
\bibitem [{\citenamefont {Barkeshli}\ and\ \citenamefont
  {Wen}(2011)}]{BarkeshliBilayerQH2011}%
  \BibitemOpen
  \bibfield  {author} {\bibinfo {author} {\bibfnamefont {M.}~\bibnamefont
  {Barkeshli}}\ and\ \bibinfo {author} {\bibfnamefont {X.-G.}\ \bibnamefont
  {Wen}},\ }\href {\doibase 10.1103/PhysRevB.84.115121} {\bibfield  {journal}
  {\bibinfo  {journal} {Phys. Rev. B}\ }\textbf {\bibinfo {volume} {84}},\
  \bibinfo {pages} {115121} (\bibinfo {year} {2011})}\BibitemShut {NoStop}%
\bibitem [{\citenamefont {Scarola}\ and\ \citenamefont
  {Jain}(2001)}]{ScarolaBilayerCF2001}%
  \BibitemOpen
  \bibfield  {author} {\bibinfo {author} {\bibfnamefont {V.~W.}\ \bibnamefont
  {Scarola}}\ and\ \bibinfo {author} {\bibfnamefont {J.~K.}\ \bibnamefont
  {Jain}},\ }\href {\doibase 10.1103/PhysRevB.64.085313} {\bibfield  {journal}
  {\bibinfo  {journal} {Phys. Rev. B}\ }\textbf {\bibinfo {volume} {64}},\
  \bibinfo {pages} {085313} (\bibinfo {year} {2001})}\BibitemShut {NoStop}%
\bibitem [{\citenamefont {Scarola}\ \emph {et~al.}(2002)\citenamefont
  {Scarola}, \citenamefont {Jain},\ and\ \citenamefont
  {Rezayi}}]{ScarolaPairing2002}%
  \BibitemOpen
  \bibfield  {author} {\bibinfo {author} {\bibfnamefont {V.~W.}\ \bibnamefont
  {Scarola}}, \bibinfo {author} {\bibfnamefont {J.~K.}\ \bibnamefont {Jain}}, \
  and\ \bibinfo {author} {\bibfnamefont {E.~H.}\ \bibnamefont {Rezayi}},\
  }\href {\doibase 10.1103/PhysRevLett.88.216804} {\bibfield  {journal}
  {\bibinfo  {journal} {Phys. Rev. Lett.}\ }\textbf {\bibinfo {volume} {88}},\
  \bibinfo {pages} {216804} (\bibinfo {year} {2002})}\BibitemShut {NoStop}%
\bibitem [{\citenamefont {Seidel}\ and\ \citenamefont
  {Yang}(2008)}]{SeidelThinCylinder2008}%
  \BibitemOpen
  \bibfield  {author} {\bibinfo {author} {\bibfnamefont {A.}~\bibnamefont
  {Seidel}}\ and\ \bibinfo {author} {\bibfnamefont {K.}~\bibnamefont {Yang}},\
  }\href {\doibase 10.1103/PhysRevLett.101.036804} {\bibfield  {journal}
  {\bibinfo  {journal} {Phys. Rev. Lett.}\ }\textbf {\bibinfo {volume} {101}},\
  \bibinfo {pages} {036804} (\bibinfo {year} {2008})}\BibitemShut {NoStop}%
\bibitem [{\citenamefont {Regnault}\ \emph {et~al.}(2008)\citenamefont
  {Regnault}, \citenamefont {Goerbig},\ and\ \citenamefont
  {Jolicoeur}}]{RegnaultAbNonAbBridge2008}%
  \BibitemOpen
  \bibfield  {author} {\bibinfo {author} {\bibfnamefont {N.}~\bibnamefont
  {Regnault}}, \bibinfo {author} {\bibfnamefont {M.~O.}\ \bibnamefont
  {Goerbig}}, \ and\ \bibinfo {author} {\bibfnamefont {T.}~\bibnamefont
  {Jolicoeur}},\ }\href {\doibase 10.1103/PhysRevLett.101.066803} {\bibfield
  {journal} {\bibinfo  {journal} {Phys. Rev. Lett.}\ }\textbf {\bibinfo
  {volume} {101}},\ \bibinfo {pages} {066803} (\bibinfo {year}
  {2008})}\BibitemShut {NoStop}%
\bibitem [{\citenamefont {Cappelli}\ \emph {et~al.}(2001)\citenamefont
  {Cappelli}, \citenamefont {Georgiev},\ and\ \citenamefont
  {Todorov}}]{CappelliParafermion2001}%
  \BibitemOpen
  \bibfield  {author} {\bibinfo {author} {\bibfnamefont {A.}~\bibnamefont
  {Cappelli}}, \bibinfo {author} {\bibfnamefont {L.~S.}\ \bibnamefont
  {Georgiev}}, \ and\ \bibinfo {author} {\bibfnamefont {I.~T.}\ \bibnamefont
  {Todorov}},\ }\href {\doibase https://doi.org/10.1016/S0550-3213(00)00774-4}
  {\bibfield  {journal} {\bibinfo  {journal} {Nuclear Physics B}\ }\textbf
  {\bibinfo {volume} {599}},\ \bibinfo {pages} {499} (\bibinfo {year}
  {2001})}\BibitemShut {NoStop}%
\bibitem [{\citenamefont {Wen}(2000)}]{WenTransitions2000}%
  \BibitemOpen
  \bibfield  {author} {\bibinfo {author} {\bibfnamefont {X.-G.}\ \bibnamefont
  {Wen}},\ }\href {\doibase 10.1103/PhysRevLett.84.3950} {\bibfield  {journal}
  {\bibinfo  {journal} {Phys. Rev. Lett.}\ }\textbf {\bibinfo {volume} {84}},\
  \bibinfo {pages} {3950} (\bibinfo {year} {2000})}\BibitemShut {NoStop}%
\bibitem [{\citenamefont {Read}\ and\ \citenamefont
  {Green}(2000)}]{ReadGreen2000}%
  \BibitemOpen
  \bibfield  {author} {\bibinfo {author} {\bibfnamefont {N.}~\bibnamefont
  {Read}}\ and\ \bibinfo {author} {\bibfnamefont {D.}~\bibnamefont {Green}},\
  }\href {\doibase 10.1103/PhysRevB.61.10267} {\bibfield  {journal} {\bibinfo
  {journal} {Phys. Rev. B}\ }\textbf {\bibinfo {volume} {61}},\ \bibinfo
  {pages} {10267} (\bibinfo {year} {2000})}\BibitemShut {NoStop}%
\bibitem [{\citenamefont {Naud}\ \emph {et~al.}(2000)\citenamefont {Naud},
  \citenamefont {Pryadko},\ and\ \citenamefont {Sondhi}}]{NaudBilayer2000}%
  \BibitemOpen
  \bibfield  {author} {\bibinfo {author} {\bibfnamefont {J.}~\bibnamefont
  {Naud}}, \bibinfo {author} {\bibfnamefont {L.~P.}\ \bibnamefont {Pryadko}}, \
  and\ \bibinfo {author} {\bibfnamefont {S.}~\bibnamefont {Sondhi}},\ }\href
  {\doibase https://doi.org/10.1016/S0550-3213(99)00658-6} {\bibfield
  {journal} {\bibinfo  {journal} {Nuclear Physics B}\ }\textbf {\bibinfo
  {volume} {565}},\ \bibinfo {pages} {572} (\bibinfo {year}
  {2000})}\BibitemShut {NoStop}%
\bibitem [{\citenamefont {Fradkin}\ \emph {et~al.}(1999)\citenamefont
  {Fradkin}, \citenamefont {Nayak},\ and\ \citenamefont
  {Schoutens}}]{FradkinLandauGinzburg1999}%
  \BibitemOpen
  \bibfield  {author} {\bibinfo {author} {\bibfnamefont {E.}~\bibnamefont
  {Fradkin}}, \bibinfo {author} {\bibfnamefont {C.}~\bibnamefont {Nayak}}, \
  and\ \bibinfo {author} {\bibfnamefont {K.}~\bibnamefont {Schoutens}},\ }\href
  {\doibase https://doi.org/10.1016/S0550-3213(99)00039-5} {\bibfield
  {journal} {\bibinfo  {journal} {Nuclear Physics B}\ }\textbf {\bibinfo
  {volume} {546}},\ \bibinfo {pages} {711} (\bibinfo {year}
  {1999})}\BibitemShut {NoStop}%
\bibitem [{\citenamefont {Nayak}\ and\ \citenamefont
  {Wilczek}(1996)}]{Nayak1995}%
  \BibitemOpen
  \bibfield  {author} {\bibinfo {author} {\bibfnamefont {C.}~\bibnamefont
  {Nayak}}\ and\ \bibinfo {author} {\bibfnamefont {F.}~\bibnamefont
  {Wilczek}},\ }\href {\doibase https://doi.org/10.1016/0550-3213(96)00430-0}
  {\bibfield  {journal} {\bibinfo  {journal} {Nuclear Physics B}\ }\textbf
  {\bibinfo {volume} {479}},\ \bibinfo {pages} {529} (\bibinfo {year}
  {1996})}\BibitemShut {NoStop}%
\bibitem [{\citenamefont {Ho}(1995)}]{HoTwoCptQH1995}%
  \BibitemOpen
  \bibfield  {author} {\bibinfo {author} {\bibfnamefont {T.-L.}\ \bibnamefont
  {Ho}},\ }\href {\doibase 10.1103/PhysRevLett.75.1186} {\bibfield  {journal}
  {\bibinfo  {journal} {Phys. Rev. Lett.}\ }\textbf {\bibinfo {volume} {75}},\
  \bibinfo {pages} {1186} (\bibinfo {year} {1995})}\BibitemShut {NoStop}%
\bibitem [{\citenamefont {Nomura}\ and\ \citenamefont
  {Yoshioka}(2004)}]{NomuraGapEvolutionBilayer2004}%
  \BibitemOpen
  \bibfield  {author} {\bibinfo {author} {\bibfnamefont {K.}~\bibnamefont
  {Nomura}}\ and\ \bibinfo {author} {\bibfnamefont {D.}~\bibnamefont
  {Yoshioka}},\ }\href {\doibase 10.1143/JPSJ.73.2612} {\bibfield  {journal}
  {\bibinfo  {journal} {Journal of the Physical Society of Japan}\ }\textbf
  {\bibinfo {volume} {73}},\ \bibinfo {pages} {2612} (\bibinfo {year}
  {2004})},\ \Eprint
  {http://arxiv.org/abs/https://doi.org/10.1143/JPSJ.73.2612}
  {https://doi.org/10.1143/JPSJ.73.2612} \BibitemShut {NoStop}%
\bibitem [{\citenamefont {Greiter}\ \emph {et~al.}(1992)\citenamefont
  {Greiter}, \citenamefont {Wen},\ and\ \citenamefont
  {Wilczek}}]{GreiterPairingBilayer1992}%
  \BibitemOpen
  \bibfield  {author} {\bibinfo {author} {\bibfnamefont {M.}~\bibnamefont
  {Greiter}}, \bibinfo {author} {\bibfnamefont {X.~G.}\ \bibnamefont {Wen}}, \
  and\ \bibinfo {author} {\bibfnamefont {F.}~\bibnamefont {Wilczek}},\ }\href
  {\doibase 10.1103/PhysRevB.46.9586} {\bibfield  {journal} {\bibinfo
  {journal} {Phys. Rev. B}\ }\textbf {\bibinfo {volume} {46}},\ \bibinfo
  {pages} {9586} (\bibinfo {year} {1992})}\BibitemShut {NoStop}%
\bibitem [{\citenamefont {Chakraborty}\ and\ \citenamefont
  {Pietil\"ainen}(1987)}]{ChakrabortyMultilayerQH1987}%
  \BibitemOpen
  \bibfield  {author} {\bibinfo {author} {\bibfnamefont {T.}~\bibnamefont
  {Chakraborty}}\ and\ \bibinfo {author} {\bibfnamefont {P.}~\bibnamefont
  {Pietil\"ainen}},\ }\href {\doibase 10.1103/PhysRevLett.59.2784} {\bibfield
  {journal} {\bibinfo  {journal} {Phys. Rev. Lett.}\ }\textbf {\bibinfo
  {volume} {59}},\ \bibinfo {pages} {2784} (\bibinfo {year}
  {1987})}\BibitemShut {NoStop}%
\bibitem [{\citenamefont {Yoshioka}\ \emph {et~al.}(1989)\citenamefont
  {Yoshioka}, \citenamefont {MacDonald},\ and\ \citenamefont
  {Girvin}}]{YoshiokaQHBilayer1989}%
  \BibitemOpen
  \bibfield  {author} {\bibinfo {author} {\bibfnamefont {D.}~\bibnamefont
  {Yoshioka}}, \bibinfo {author} {\bibfnamefont {A.~H.}\ \bibnamefont
  {MacDonald}}, \ and\ \bibinfo {author} {\bibfnamefont {S.~M.}\ \bibnamefont
  {Girvin}},\ }\href {\doibase 10.1103/PhysRevB.39.1932} {\bibfield  {journal}
  {\bibinfo  {journal} {Phys. Rev. B}\ }\textbf {\bibinfo {volume} {39}},\
  \bibinfo {pages} {1932} (\bibinfo {year} {1989})}\BibitemShut {NoStop}%
\bibitem [{\citenamefont {He}\ \emph {et~al.}(1991)\citenamefont {He},
  \citenamefont {Xie}, \citenamefont {Das~Sarma},\ and\ \citenamefont
  {Zhang}}]{HeQHBilayer1991}%
  \BibitemOpen
  \bibfield  {author} {\bibinfo {author} {\bibfnamefont {S.}~\bibnamefont
  {He}}, \bibinfo {author} {\bibfnamefont {X.~C.}\ \bibnamefont {Xie}},
  \bibinfo {author} {\bibfnamefont {S.}~\bibnamefont {Das~Sarma}}, \ and\
  \bibinfo {author} {\bibfnamefont {F.~C.}\ \bibnamefont {Zhang}},\ }\href
  {\doibase 10.1103/PhysRevB.43.9339} {\bibfield  {journal} {\bibinfo
  {journal} {Phys. Rev. B}\ }\textbf {\bibinfo {volume} {43}},\ \bibinfo
  {pages} {9339} (\bibinfo {year} {1991})}\BibitemShut {NoStop}%
\bibitem [{\citenamefont {He}\ \emph {et~al.}(1993)\citenamefont {He},
  \citenamefont {Das~Sarma},\ and\ \citenamefont {Xie}}]{HeQHBilayer1993}%
  \BibitemOpen
  \bibfield  {author} {\bibinfo {author} {\bibfnamefont {S.}~\bibnamefont
  {He}}, \bibinfo {author} {\bibfnamefont {S.}~\bibnamefont {Das~Sarma}}, \
  and\ \bibinfo {author} {\bibfnamefont {X.~C.}\ \bibnamefont {Xie}},\ }\href
  {\doibase 10.1103/PhysRevB.47.4394} {\bibfield  {journal} {\bibinfo
  {journal} {Phys. Rev. B}\ }\textbf {\bibinfo {volume} {47}},\ \bibinfo
  {pages} {4394} (\bibinfo {year} {1993})}\BibitemShut {NoStop}%
\bibitem [{\citenamefont {Peterson}\ \emph {et~al.}(2010)\citenamefont
  {Peterson}, \citenamefont {Papi\ifmmode~\acute{c}\else \'{c}\fi{}},\ and\
  \citenamefont {Das~Sarma}}]{PetersonQHBilayer2010}%
  \BibitemOpen
  \bibfield  {author} {\bibinfo {author} {\bibfnamefont {M.~R.}\ \bibnamefont
  {Peterson}}, \bibinfo {author} {\bibfnamefont {Z.}~\bibnamefont
  {Papi\ifmmode~\acute{c}\else \'{c}\fi{}}}, \ and\ \bibinfo {author}
  {\bibfnamefont {S.}~\bibnamefont {Das~Sarma}},\ }\href {\doibase
  10.1103/PhysRevB.82.235312} {\bibfield  {journal} {\bibinfo  {journal} {Phys.
  Rev. B}\ }\textbf {\bibinfo {volume} {82}},\ \bibinfo {pages} {235312}
  (\bibinfo {year} {2010})}\BibitemShut {NoStop}%
\bibitem [{\citenamefont {Storni}\ \emph {et~al.}(2010)\citenamefont {Storni},
  \citenamefont {Morf},\ and\ \citenamefont {Das~Sarma}}]{StorniMooreRead2010}%
  \BibitemOpen
  \bibfield  {author} {\bibinfo {author} {\bibfnamefont {M.}~\bibnamefont
  {Storni}}, \bibinfo {author} {\bibfnamefont {R.~H.}\ \bibnamefont {Morf}}, \
  and\ \bibinfo {author} {\bibfnamefont {S.}~\bibnamefont {Das~Sarma}},\ }\href
  {\doibase 10.1103/PhysRevLett.104.076803} {\bibfield  {journal} {\bibinfo
  {journal} {Phys. Rev. Lett.}\ }\textbf {\bibinfo {volume} {104}},\ \bibinfo
  {pages} {076803} (\bibinfo {year} {2010})}\BibitemShut {NoStop}%
\bibitem [{\citenamefont {Khalaf}\ \emph {et~al.}()\citenamefont {Khalaf},
  \citenamefont {Chatterjee}, \citenamefont {Bultinck}, \citenamefont
  {Zaletel},\ and\ \citenamefont
  {Vishwanath}}]{khalafChargedSkyrmionsTopological2020}%
  \BibitemOpen
  \bibfield  {author} {\bibinfo {author} {\bibfnamefont {E.}~\bibnamefont
  {Khalaf}}, \bibinfo {author} {\bibfnamefont {S.}~\bibnamefont {Chatterjee}},
  \bibinfo {author} {\bibfnamefont {N.}~\bibnamefont {Bultinck}}, \bibinfo
  {author} {\bibfnamefont {M.~P.}\ \bibnamefont {Zaletel}}, \ and\ \bibinfo
  {author} {\bibfnamefont {A.}~\bibnamefont {Vishwanath}},\ }\href
  {http://arxiv.org/abs/2004.00638} {\ }\Eprint
  {http://arxiv.org/abs/2004.00638} {2004.00638} \BibitemShut {NoStop}%
\bibitem [{\citenamefont {Kwan}\ \emph {et~al.}(2021)\citenamefont {Kwan},
  \citenamefont {Wagner}, \citenamefont {Soejima}, \citenamefont {Zaletel},
  \citenamefont {Simon}, \citenamefont {Parameswaran},\ and\ \citenamefont
  {Bultinck}}]{kwan2021kekul}%
  \BibitemOpen
  \bibfield  {author} {\bibinfo {author} {\bibfnamefont {Y.~H.}\ \bibnamefont
  {Kwan}}, \bibinfo {author} {\bibfnamefont {G.}~\bibnamefont {Wagner}},
  \bibinfo {author} {\bibfnamefont {T.}~\bibnamefont {Soejima}}, \bibinfo
  {author} {\bibfnamefont {M.~P.}\ \bibnamefont {Zaletel}}, \bibinfo {author}
  {\bibfnamefont {S.~H.}\ \bibnamefont {Simon}}, \bibinfo {author}
  {\bibfnamefont {S.~A.}\ \bibnamefont {Parameswaran}}, \ and\ \bibinfo
  {author} {\bibfnamefont {N.}~\bibnamefont {Bultinck}},\ }\href@noop {}
  {\bibfield  {journal} {\bibinfo  {journal} {arXiv preprint arXiv:2105.05857}\
  } (\bibinfo {year} {2021})}\BibitemShut {NoStop}%
\bibitem [{\citenamefont {Kerelsky}\ \emph {et~al.}(2019)\citenamefont
  {Kerelsky}, \citenamefont {McGilly}, \citenamefont {Kennes}, \citenamefont
  {Xian}, \citenamefont {Yankowitz}, \citenamefont {Chen}, \citenamefont
  {Watanabe}, \citenamefont {Taniguchi}, \citenamefont {Hone}, \citenamefont
  {Dean} \emph {et~al.}}]{kerelsky2019maximized}%
  \BibitemOpen
  \bibfield  {author} {\bibinfo {author} {\bibfnamefont {A.}~\bibnamefont
  {Kerelsky}}, \bibinfo {author} {\bibfnamefont {L.~J.}\ \bibnamefont
  {McGilly}}, \bibinfo {author} {\bibfnamefont {D.~M.}\ \bibnamefont {Kennes}},
  \bibinfo {author} {\bibfnamefont {L.}~\bibnamefont {Xian}}, \bibinfo {author}
  {\bibfnamefont {M.}~\bibnamefont {Yankowitz}}, \bibinfo {author}
  {\bibfnamefont {S.}~\bibnamefont {Chen}}, \bibinfo {author} {\bibfnamefont
  {K.}~\bibnamefont {Watanabe}}, \bibinfo {author} {\bibfnamefont
  {T.}~\bibnamefont {Taniguchi}}, \bibinfo {author} {\bibfnamefont
  {J.}~\bibnamefont {Hone}}, \bibinfo {author} {\bibfnamefont {C.}~\bibnamefont
  {Dean}},  \emph {et~al.},\ }\href@noop {} {\bibfield  {journal} {\bibinfo
  {journal} {Nature}\ }\textbf {\bibinfo {volume} {572}},\ \bibinfo {pages}
  {95} (\bibinfo {year} {2019})}\BibitemShut {NoStop}%
\bibitem [{\citenamefont {Choi}\ \emph {et~al.}(2019)\citenamefont {Choi},
  \citenamefont {Kemmer}, \citenamefont {Peng}, \citenamefont {Thomson},
  \citenamefont {Arora}, \citenamefont {Polski}, \citenamefont {Zhang},
  \citenamefont {Ren}, \citenamefont {Alicea}, \citenamefont {Refael} \emph
  {et~al.}}]{choi2019electronic}%
  \BibitemOpen
  \bibfield  {author} {\bibinfo {author} {\bibfnamefont {Y.}~\bibnamefont
  {Choi}}, \bibinfo {author} {\bibfnamefont {J.}~\bibnamefont {Kemmer}},
  \bibinfo {author} {\bibfnamefont {Y.}~\bibnamefont {Peng}}, \bibinfo {author}
  {\bibfnamefont {A.}~\bibnamefont {Thomson}}, \bibinfo {author} {\bibfnamefont
  {H.}~\bibnamefont {Arora}}, \bibinfo {author} {\bibfnamefont
  {R.}~\bibnamefont {Polski}}, \bibinfo {author} {\bibfnamefont
  {Y.}~\bibnamefont {Zhang}}, \bibinfo {author} {\bibfnamefont
  {H.}~\bibnamefont {Ren}}, \bibinfo {author} {\bibfnamefont {J.}~\bibnamefont
  {Alicea}}, \bibinfo {author} {\bibfnamefont {G.}~\bibnamefont {Refael}},
  \emph {et~al.},\ }\href@noop {} {\bibfield  {journal} {\bibinfo  {journal}
  {Nature Physics}\ }\textbf {\bibinfo {volume} {15}},\ \bibinfo {pages} {1174}
  (\bibinfo {year} {2019})}\BibitemShut {NoStop}%
\bibitem [{\citenamefont {Xie}\ and\ \citenamefont
  {MacDonald}(2021)}]{MacdonaldPH}%
  \BibitemOpen
  \bibfield  {author} {\bibinfo {author} {\bibfnamefont {M.}~\bibnamefont
  {Xie}}\ and\ \bibinfo {author} {\bibfnamefont {A.~H.}\ \bibnamefont
  {MacDonald}},\ }\href {\doibase 10.1103/PhysRevLett.127.196401} {\bibfield
  {journal} {\bibinfo  {journal} {Phys. Rev. Lett.}\ }\textbf {\bibinfo
  {volume} {127}},\ \bibinfo {pages} {196401} (\bibinfo {year}
  {2021})}\BibitemShut {NoStop}%
\bibitem [{\citenamefont
  {Zak}({\natexlab{a}})}]{zakMagneticTranslationGroup1964a}%
  \BibitemOpen
  \bibfield  {author} {\bibinfo {author} {\bibfnamefont {J.}~\bibnamefont
  {Zak}},\ }\href {\doibase 10.1103/PhysRev.134.A1602} {\ \textbf {\bibinfo
  {volume} {134}},\ \bibinfo {pages} {A1602} ({\natexlab{a}})}\BibitemShut
  {NoStop}%
\bibitem [{\citenamefont
  {Zak}({\natexlab{b}})}]{zakMagneticTranslationGroup1964}%
  \BibitemOpen
  \bibfield  {author} {\bibinfo {author} {\bibfnamefont {J.}~\bibnamefont
  {Zak}},\ }\href {\doibase 10.1103/PhysRev.134.A1607} {\ \textbf {\bibinfo
  {volume} {134}},\ \bibinfo {pages} {A1607} ({\natexlab{b}})}\BibitemShut
  {NoStop}%
\bibitem [{\citenamefont {Zak}(1964)}]{zak1964group}%
  \BibitemOpen
  \bibfield  {author} {\bibinfo {author} {\bibfnamefont {J.}~\bibnamefont
  {Zak}},\ }\href@noop {} {\bibfield  {journal} {\bibinfo  {journal} {Physical
  Review}\ }\textbf {\bibinfo {volume} {136}},\ \bibinfo {pages} {A776}
  (\bibinfo {year} {1964})}\BibitemShut {NoStop}%
\bibitem [{\citenamefont {Hofmann}\ \emph {et~al.}(2021)\citenamefont
  {Hofmann}, \citenamefont {Khalaf}, \citenamefont {Vishwanath}, \citenamefont
  {Berg},\ and\ \citenamefont {Lee}}]{hofmann2021fermionic}%
  \BibitemOpen
  \bibfield  {author} {\bibinfo {author} {\bibfnamefont {J.~S.}\ \bibnamefont
  {Hofmann}}, \bibinfo {author} {\bibfnamefont {E.}~\bibnamefont {Khalaf}},
  \bibinfo {author} {\bibfnamefont {A.}~\bibnamefont {Vishwanath}}, \bibinfo
  {author} {\bibfnamefont {E.}~\bibnamefont {Berg}}, \ and\ \bibinfo {author}
  {\bibfnamefont {J.~Y.}\ \bibnamefont {Lee}},\ }\href@noop {} {\bibfield
  {journal} {\bibinfo  {journal} {arXiv preprint arXiv:2105.12112}\ } (\bibinfo
  {year} {2021})}\BibitemShut {NoStop}%
\bibitem [{\citenamefont {Hauschild}\ and\ \citenamefont
  {Pollmann}(2018)}]{tenpy}%
  \BibitemOpen
  \bibfield  {author} {\bibinfo {author} {\bibfnamefont {J.}~\bibnamefont
  {Hauschild}}\ and\ \bibinfo {author} {\bibfnamefont {F.}~\bibnamefont
  {Pollmann}},\ }\href {\doibase 10.21468/SciPostPhysLectNotes.5} {\bibfield
  {journal} {\bibinfo  {journal} {SciPost Phys. Lect. Notes}\ ,\ \bibinfo
  {pages} {5}} (\bibinfo {year} {2018})},\ \bibinfo {note} {code available from
  \url{https://github.com/tenpy/tenpy}},\ \Eprint
  {http://arxiv.org/abs/1805.00055} {arXiv:1805.00055} \BibitemShut {NoStop}%
\bibitem [{\citenamefont {Girvin}\ \emph {et~al.}({\natexlab{b}})\citenamefont
  {Girvin}, \citenamefont {{MacDonald}},\ and\ \citenamefont
  {Platzman}}]{girvinMagnetorotonTheoryCollective1986}%
  \BibitemOpen
  \bibfield  {author} {\bibinfo {author} {\bibfnamefont {S.~M.}\ \bibnamefont
  {Girvin}}, \bibinfo {author} {\bibfnamefont {A.~H.}\ \bibnamefont
  {{MacDonald}}}, \ and\ \bibinfo {author} {\bibfnamefont {P.~M.}\ \bibnamefont
  {Platzman}},\ }\href {\doibase 10.1103/PhysRevB.33.2481} {\ \textbf {\bibinfo
  {volume} {33}},\ \bibinfo {pages} {2481} ({\natexlab{b}})}\BibitemShut
  {NoStop}%
\bibitem [{\citenamefont {Blount}(1962)}]{blount1962formalisms}%
  \BibitemOpen
  \bibfield  {author} {\bibinfo {author} {\bibfnamefont {E.}~\bibnamefont
  {Blount}},\ }in\ \href@noop {} {\emph {\bibinfo {booktitle} {Solid state
  physics}}},\ Vol.~\bibinfo {volume} {13}\ (\bibinfo  {publisher} {Elsevier},\
  \bibinfo {year} {1962})\ pp.\ \bibinfo {pages} {305--373}\BibitemShut
  {NoStop}%
\end{thebibliography}%

\newpage

\appendix
\begin{widetext}

This paper contains five technical Appendices. The first three Appendices construct our model for \tbg{} and define Hartree-Fock at finite fields. 
\begin{description}
    \item[Appendix \ref{app:continuum_formalism}] this section provides notation and definitions for working with magnetic and non-magnetic continuum models on equal footing in first and second quantization, and derives the Hartree-Fock contributions. The treatment is general and does not use the details of \tbg{}.
    \item[Appendix \ref{app:TBG_in_B_field}] this section gives the details of the BM model for non-interacting \tbg{} and its generalization to finite magnetic field.
    \item[Appendix \ref{app:Hamiltonian_construction}] this section constructs a standard model of interacting \tbg{} using ``Hartree-Fock subtraction", with particular focus on the case of $\nu=3$ and how the Hamiltonian is split into interaction and dispersion parts.
    \item[Appendix \ref{app:numerical_details}] provides numerical details on the DMRG calculations in the main text. 
    \item[Appendix \ref{app:GMP}] this section proves the relationship between the multiband FCI indicators and the GMP algebra . 
\end{description}
Insofar as is possible, each Appendix is self-contained. 

\section{Continuum Models, Magnetic Translations, and Hartree-Fock}
\label{app:continuum_formalism}

This Appendix provides a self-contained formalism for working with continuum models with and without magnetic field on an equal footing, allowing us to work with the Bistritzer-MacDonald model and its finite field extensions. \cite{bistritzerMoireBandsTwisted2011,zhangLandauLevelDegeneracy2019,hejaziLandauLevelsTwisted2019,ozawaRelationsTopologyQuantum2021} on equal footing. The crux of our approach is that all model-dependent information is reduced to three operators $\hat{\lambda}, \hat{\mathcal{T}}$, and $\hat{V}_{\v{G}}$, whose matrix elements must be computed. We start by reviewing some salient facts about the magnetic translation algebra.   Then we describe continuum models the single-particle level to establish definitions, and consider the Coulomb interaction in second quantization. Finally re-derive the Hartree-Fock Hamiltonian as a partial trace operation that will be used to integrate out the remote bands in App. \ref{app:Hamiltonian_construction}. 

\subsection{Magnetic Translation Algebra}
\label{app:subsection_on_magnetic_translation}

This section will briefly review the magnetic translation algebra \cite{zakMagneticTranslationGroup1964a,zakMagneticTranslationGroup1964,zak1964group,ozawaRelationsTopologyQuantum2021}. A standard model for electrons in a solid is $\hat{H} = \hat{p}^2/(2m) + \hat{V}(\v{r})$, where $\hat{V}(\v{r}+\v{R}) = \hat{V}(\v{r})$ is periodic under lattice translations $\v{R} \in \Lambda$. The corresponding discrete translation operators $\hat{T}(\v{R}) := e^{i\v{R}\cdot \hat{\v{p}}}$ commute both with the Hamiltonian and each other. That is, $[\hat{H},\hat{T}(\v{R})]= [\hat{T}(\v{R}), \hat{T}(\v{R}')] = 0$, which is the assumption needed for Bloch's theorem.

In a magnetic field $\v{B}(\br) = B(\v{r}) \hat{\v{z}} = \bnabla \times \bA$ (we restrict to the 2d case and with a periodic or flat magnetic field so that $B(\v{r} + \v{R}) = B(\v{R})$), the canonical momentum becomes $\v{\pi} = \v{p} + \v{A}$ and $\hat{H}_B = \hat{\pi}^2/(2m) + \hat{V}(\v{r})$, which fails to commute with ordinary translation operators. To rectify this, we define magnetic translation operators
\begin{equation}
    \hat{\mathcal{M}}(\v{R}) := e^{-i \xi_{\v{R}}(\v{r})} T(\v{R})
    \label{eq:magnetic_translation_algebra}
\end{equation}
. The magnetic twists $\xi_{\v{R}}$ are chosen to satisfy $\v{A}(\v{r}+\v{R}) - \v{A}(\v{r}) = \v{\nabla}\xi_{\v{R}}(\v{r})$. One can check that $[\hat{\v{\pi}}, \hat{\mathcal{M}}(\v{R})] = 0$, so $[\hat{H}_B,\hat{\mathcal{M}}(\v{R})]=0$. However, Eq. \eqref{eq:magnetic_translation_algebra} is a projective representation of the group of lattice translations, and commutators are modified by phase factors relative to the normal case. In particular, if $\v{R}_1$ and $\v{R}_2$ are the primitive vectors,
\begin{equation}
    \hat{\mathcal{M}}(\v{R}_1)\hat{\mathcal{M}}(\v{R}_2) = \hat{\mathcal{M}}(\v{R}_2)\hat{\mathcal{M}}(\v{R}_1)\zeta, \text{ where } \zeta := e^{i 2\pi \phi}, \quad \phi := \frac{B A}{h/e},
    \label{eq:magnetic_translation_non_commutation}
\end{equation}
where $A = \hat{z} \cdot (\v{R}_1 \times \v{R}_2)$ is the (signed) area of the unit cell and $\phi$ is the flux in units of the flux quantum.

To apply Bloch's theorem, the translation operators must mutually commute. If the flux is a rational number
\begin{equation}
    \phi = \frac{BA}{h/e} = \frac{p}{q} \in \mathbb{Q},
\end{equation}
then we may enlarge the unit cell by a factor of $q$, with new primitive vectors $\tilde{\v{R}}_1 = r\v{R}_1$ and $\tilde{\v{R}}_2 = s \v{R}_2$ where $rs = q$. Since $\zeta^{rs} = e^{i 2\pi (p/q) q} = 1$, one has $[\hat{\mathcal{M}}(\tilde{\v{R}}_1), \hat{\mathcal{M}}(\tilde{\v{R}}_2)] = 0$. We may then employ a (generalized) Bloch's theorem for the expanded magnetic unit cell. We note that the resulting Bloch wavefunctions have differ from the non-magnetic ones by their boundary conditions, which will be important below.

Henceforth, we assume that the unit cell has always been properly expanded so that the (magnetic) translation operators commute.

\subsection{First Quantization}

Suppose we have a sample geometry $M$ with a finite area (IR cutoff) $\Asample \gg 1$. Suppose we have a lattice $\Lambda	= \st{n \v{R}_1 + m \v{R}_2}$ with reciprocal lattice  $\Lambda^* = \st{n \v{G}_1 + m \v{G}_2}$. Let $A = \Asample/N$ be the area of a (possibly magnetic) unit cell. Let $\mathcal{M}(\v{R})$ be the magnetic translations operators described above, which mutually commute (after expanding the unit cell if necessary)\cite{zakMagneticTranslationGroup1964,zakMagneticTranslationGroup1964a}. The Bloch states are simultaneous eigenstates of the Hamiltonian and the translation algebra
\begin{align}
	\label{eq:non-interacting_bandstructure}
	\hat{H}_0 \ket{\phi_{\v{k}n\mu}} &= \epsilon_{\v{k}n\mu} \ket{\phi_{\v{k}n \mu}}\\
	\mathcal{M}(\v{R}) \ket{\phi_{\v{k}n\mu}} &= e^{i\v{k}\cdot\v{R}} \ket{\phi_{\v{k}n \mu}}
\end{align}
where $\v{k} \in \bz$, $n$ labels bands, $\mu$ labels flavors such as spin and valley, and $a = (\gamma, \sigma,\ldots) = (\text{layer}, \text{sublattice},\dots)$ runs over internal degrees of freedom. For convenience, we group $\alpha = (\mu, n)$, and suppress internal (``$a$") indices whenever possible. 

The Bloch states $\phi_{\v{k}\alpha}^a(\v{r}) := \braket{\v{r},a|\phi_{\v{k}\alpha}}$ comprise a unitary transformation between the $(\v{r},a,\mu)$ and $(\v{k},n,\mu)$ bases:\footnote{Due to the finite sample area, the $\v{k}$-points are discrete.}
\begin{equation}
	\braket{\phi_{\v{k}\alpha}|\phi_{\v{k}'\beta}} = \int_{M} \dr \; \overline{\phi}_{\v{k}\alpha}(\v{r}) \phi_{\v{k}'\beta}(\v{r}) = \delta_{\v{k}\v{k}'} \delta_{\alpha\beta}; 
	\hspace{2em}  
	\sum_{\alpha} \sum_{\v{k} \in \bz} \overline{\phi}_{\v{k}\alpha}(\v{r}) \phi_{\v{k} \alpha}(\v{r}') = \delta(\v{r}-\v{r}').
	\label{eq:Bloch_waves_unitary}
\end{equation}
Bloch waves, defined by $u_{\v{k}\alpha}(\v{r}) := N^{1/2} e^{-i \bk \cdot \br} \phi_{\v{k}\alpha}(\v{r})$, are the eigenvectors of $\hat{H}_0(\v{k}) := e^{-i\v{k} \cdot \v{r}} \hat{H}_0 e^{i \v{k} \cdot \v{r}}$,  and are normalized on a single unit cell as $\braket{u_{\v{k}\alpha}|u_{\v{k}\beta}} = \int_{uc} \dr \; \overline{u}_{\v{k}\alpha}(\v{r}) u_{\v{k}\beta}(\v{r}) = \delta_{\alpha\beta}$. 

The magnetic field imposes twisted boundary conditions on the real-space unit cell. Translation acts as $\mathcal{M}(\v{R}) \phi_{\v{k}\alpha}(\v{r}) = e^{-i \xi_{\v{R}}(\v{r})} \phi_{\v{k}\alpha}(\v{r}+\v{R})$, so
\begin{equation}
	\phi_{\v{k}\alpha}(\v{r} + \v{R}) = e^{i\v{k} \cdot \v{R} + i \xi_{\v{R}}(\v{r})} \phi_{\v{k}\alpha}(\v{r}) \implies u_{\v{k}\alpha}(\v{r} + \v{R}) = e^{i \xi_{\v{R}}(\v{r})} u_{\v{k}\alpha}(\v{r}).
	\label{eq:Bloch_real_space_translation}
\end{equation}

For any $\v{k}$, we define $\v{k} = [\v{k}] + \v{G}$ where $[\v{k}]$ is in the first Brillouin zone and $\v{G} \in \Lambda^*$. In this section we impose a periodic gauge by \textit{defining} 
\begin{equation}
	\ket{u_{\v{k}\alpha}} := \hat{V}_{\v{G}}(\v{k}) \ket{u_{[\v{k}]\alpha}}; \quad  \hat{V}_{\v{G}}(\v{k}) := e^{-i\v{G} \cdot \hat{\v{r}}}.
	\label{eq:period_gauge_fixing}
\end{equation}
for $\v{k}$ outside the first Brillouin zone. One can check that $\ket{u_{\v{k}\alpha}}$ is an eigenstate of $\hat{H}(\v{k}) =\hat{H}([\v{k}] + \v{G})$. 

We define form factors as matrices
\begin{equation}
	[\Lambda_{\v{q}}(\v{k})]_{\alpha\beta} := \braket{u_{\v{k},\alpha}|u_{\v{k}+\v{q},\beta}},
	\label{eq:form_factors}
\end{equation}
for any $\v{k},\v{k}+\v{q}$. They obey the relation $\Lambda_{\v{q}}(\v{k})^\dagger = \Lambda_{-\v{q}}(\v{k}+\v{q})$ and, due to the periodic gauge, $\Lambda_{\v{G}}(\v{k})^\dagger = \Lambda_{-\v{G}}(\v{k})$.

To work with a continuum Hamiltonian in practice, one must choose a computational basis $v_{\v{k} I}(\v{r}) = \braket{r|v_{\v{k}I}}$ where the internal index $I$ indexes a function space on the (magnetic) Brillouin zone. By performing the unitary transformation
\begin{equation}
	\ket{u_{\v{k}\alpha}} = \sum_{I} \ket{v_{\v{k} I}} U^{I}_{\v{k}\alpha},
\end{equation}
\eqref{eq:non-interacting_bandstructure} is equivalent to 
\begin{equation}
	[\hat{H}_0(\v{k})]_J^I U_{\v{k} \alpha}^J = \epsilon_{\v{k}\alpha} U_{\v{k} \alpha}^I
	\label{eq:numerical_eigenproblem}
\end{equation}
where $[\hat{H}_0(\v{k})]_J^I = \braket{v_{\v{k} I}|\hat{H}_0 | v_{\v{k}J}}$ and the $U$'s are often found numerically in practice.

The drawback of this representation is that quantities that are natural in real space have non-trivial matrix elements in the computational basis. We now define three of the most common matrix elements for moire systems, which must be computed \textit{analytically} once a basis is chosen. Let $\v{k}$ be in the first Brillouin zone, let $\v{G} = r \v{G}_1 + s \v{G}_2$ be a reciprocal lattice vector, and let $\v{q}$ be an unconstrained momentum. Define: 
\begin{subequations}
	\begin{align}
			[\mathcal{T}_{\v{G}}(\v{k})]^{I'}_I &:= \braket{v_{\v{k} I'}|e^{-i\v{G} \cdot \hat{\v{r}}}| v_{\v{k}I}} & (\text{tunneling matrix element})\\
	[\lambda_{\v{q}}(\v{k})]^{I'}_I &:= \braket{v_{\v{k}I'} | v_{\v{k}+\v{q},I}} & (\text{computational basis form factor})\\
	[\mathcal{V}_{\v{G}}(\v{k})]^{I'}_{I} &:= \braket{v_{\v{k}+\v{G},I'}|e^{-i\v{G}\cdot \hat{\v{r}}}|v_{\v{k},I}} & (\text{reciprocal lattice translation})
\end{align}\label{eq:moire_matrix_elements}\end{subequations}
The $\mathcal{\mathcal{T}}$'s appear as tunnelling terms in \tbg{}. Form factors are matrices in the band basis $[\Lambda_{\v{q}}(\v{k})]^\alpha_\beta~=~\braket{u_{\v{k},\alpha}|u_{\v{k}+\v{q},\beta}}$, and may be computed in the computational basis as
\begin{equation}
	\Lambda_{\v{q}}(\v{k}) 
	= \braket{u_{\v{k}}|u_{\v{k}+\v{q}}}
	= U^\dagger_{\v{k}} \braket{v_{\v{k}}|v_{\v{k}+\v{q}}} U_{\v{k}+\v{q}}
	= U_{\v{k}}^\dagger \lambda_{\q}(\k) \mathcal{V}_{\v{G}}(\v{k}+\v{q}) U_{[\v{k}+\v{q}]},
	\label{eq:form_factor_equation}
\end{equation}
for any $\v{k}$ in the first Brillouin zone. In terms of these, the Berry curvature may be computed from an infinitesimal Wilson loop
\begin{equation}
	\mathcal{F}(\v{k}) = \lim_{\delta\to 0} \im \log [\Lambda_{\delta\v{q}_x} \Lambda_{\delta\v{q}_y} \Lambda_{-\delta\v{q}_x} \Lambda_{-\delta\v{q}_y}]
	\label{eq:Berry_curvature_form_factors_appendix}
\end{equation}
and similarly the quantum metric can be computed in terms of $\Lambda_{\delta \v{q}} \Lambda_{-\delta \v{q}}$ with $\delta\v{q}$ along the first, second, or diagonal axes respectively. These definitions are gauge invariant and numerically stable. Finally, the periodic gauge choice \eqref{eq:period_gauge_fixing} implies the relation $U_{\v{k}+\v{G}, \alpha}^I = [\mathcal{V}_{\v{G}}(\v{k})]^I_J U_{\v{k},\alpha}$. So, thanks to our choice of a periodic gauge, all physical quantities can be computed from the solutions to Eq. \eqref{eq:numerical_eigenproblem} within the first Brillouin zone alone ---  a result which applies in either the magnetic or non-magnetic cases.

\subsection{Second Quantization}

Suppose we have a microscopic creation operator $\hat{\psi}_{\mu a}^\dagger(\v{r})$ that obeys the standard anti-commutation relations $\{\hat{\psi}_{\mu a}^\dagger(\v{r}),\hat{\psi}_{\nu b}^\dagger(\v{r}') \} = \delta_{\mu\nu} \delta_{ab} \delta(\v{r} - \v{r}')$. For $\v{k} \in \bz$, i.e. only inside the first Brillouin zone, define band creation operators
\begin{equation}
    \hat{c}^\dagger_{\v{k} n \mu} := 
    \sum_a \int_{M} \dr \; \phi_{\v{k}n\mu}^a(\v{r}) \hat{\psi}_{\mu a}^\dagger(\v{r}).
    \label{eq:band_creation_operator}
\end{equation}
Employing the unitarity condition Eq. \eqref{eq:Bloch_waves_unitary}, we have anti-commutation relations $\{ \hat{c}_{\v{k}\alpha}, \hat{c}_{\v{k}'\beta}^\dagger \} = \delta_{\alpha\beta} \delta_{\v{k},\v{k}'}$ and the inverse transform
\begin{equation}
	\hat{\psi}_{\mu a}^\dagger(\v{r}) = \sum_{n} \sum_{\v{k} \in \bz} \hat{c}_{\v{k} n \mu}^\dagger \overline{\phi}_{\v{k} n \mu a}(\v{r}).
	\label{eq:microscopic_to_band_creation_operator}
\end{equation}

Our goal is to express the Coulomb interaction in this basis. The key ingredient we need is an expression for the density operator at finite wavevector. Charge density operators are defined as 
\begin{equation}
	\hat{\rho}(\v{q}) := \sum_{\mu} \sum_{a} \int_{M} \dr \; e^{-i\v{q}\cdot \v{r}} \hat{\psi}_{\mu a}^\dagger(\v{r}) \hat{\psi}_{\mu a}(\v{r}). 
\end{equation}
Transforming to the band operators and splitting $\int_M \dr = \sum_{\v{R} \in \Lambda} \int_{uc}$ yields 
\begin{equation}
	\hat{\rho}(\v{q}) = \sum_{\alpha\beta} \sum_{\v{k}, \v{k}' \in \bz} 
	\sum_{\v{R}\in \Lambda} \int_{uc} \dr \;
	e^{-i\v{q}\cdot (\v{r}+\v{R})} 
	\overline{\phi}_{\v{k}\alpha}(\v{r}+\v{R}) \phi_{\v{k}'\beta}(\v{r}+\v{R}) 
	\hat{c}^\dagger_{\v{k}\alpha} \hat{c}_{\v{k}'\beta}. 
\end{equation}
Using Eq. \eqref{eq:Bloch_real_space_translation}, we translate to the fundamental unit cell (the magnetic twists cancel),
\begin{equation}
	\hat{\rho}(\v{q}) = \sum_{\alpha\beta} \sum_{\v{k}, \v{k}' \in \bz} 
	\int_{uc} \dr \;
	\left( 
		\sum_{\v{R} \in \Lambda} e^{-i (\v{k}-\v{k}' + \v{q}) \cdot \v{R}} e^{-i\v{q} \cdot \v{r}} 
	\right)
	\overline{\phi}_{\v{k}\alpha}(\v{r}) \phi_{\v{k}'\beta}(\v{r})  e^{-i\v{q}\cdot \v{r}}
	\hat{c}^\dagger_{\v{k}\alpha} \hat{c}_{\v{k}'\beta}. 
\end{equation}
Using Poisson summation on the lattice sum (parentheses above) gives $N \sum_{\v{G} \in \Lambda^*} \delta(\v{k}'-\v{k}-q + \v{G})$. For a fixed $\v{k}$, there is unique non-zero term $N \delta(\v{k}'-[\v{k}+\v{q}])$ in this sum where $\v{k}+\v{q} = [\v{k}+\v{q}] + \v{G}$. Eliminating the $\v{k}'$ sum and switching to the Bloch waves gives
\begin{equation}
	\hat{\rho}(\v{q}) = \sum_{\v{k} \in \bz} 
	\hat{c}^\dagger_{\v{k}\alpha} 
	\left(
	 \sum_{\alpha\beta}
	\int_{uc} \dr \; 
	\overline{u}_{\v{k}\alpha}(\v{r}) e^{-i\v{k}\cdot \v{r}} e^{-i\v{q} \cdot \v{r}} e^{i [\v{k}+\v{q}] \cdot \v{r}} u_{[\v{k}+\v{q}],\beta}
	\right)
	\hat{c}_{[\v{k}+\v{q}]\beta}. 
\end{equation}
As $e^{-i\v{k}\cdot \hat{\v{r}}} e^{-i\v{q} \cdot \hat{\v{r}}} e^{i [\v{k}+\v{q}] \cdot \hat{\v{r}}} = e^{-i \v{G} \cdot \hat{\v{r}}} = \hat{V}_{\v{G}}(\v{k})$, the term in parentheses is $\braket{u_{\v{k}}| \hat{\mathcal{V}}_{\v{G}}(\v{k}) | u_{[\v{k}+\v{q}]}} = \braket{u_{\v{k}}|u_{\v{k}+\v{q}}} = \Lambda_{\v{q}}(\v{k})$. So, altogether,
\begin{equation}
	\hat{\rho}(\v{q}) = \sum_{\v{k} \in \bz} 
	\hat{c}^\dagger_{\v{k}}
	\Lambda_{\v{q}}(\v{k})
	\hat{c}_{[\v{k}+\v{q}]},
\end{equation}
where the operators and form factors are vectors and matrices in the band-flavor space respectively.

Coulomb and other density-density interactions are defined as
\begin{equation}
\hat{V} = \frac{1}{2} \sum_{\mu \nu} \sum_{a,b} \int_{M} \dr \,\dr' \; V(\v{r} - \v{r}') \hat{\psi}_{\mu a}^\dagger(\v{r}) \hat{\psi}_{\nu b}^\dagger(\v{r}') \hat{\psi}_{\nu b}(\v{r}') \hat{\psi}_{\mu a}(\v{r}).    
\end{equation}
The potential can be written as $V(\v{r}) = \frac{1}{\Asample} \sum_{\v{q} \in M^*} V(\v{q}) e^{i\v{q} \cdot \v{r}}$, where the sum runs over all of reciprocal space $M^*$. Hence
\begin{equation}
    \hat{V} =
	\frac{1}{2\Asample} \sum_{\v{q} \in \R^2} \; V(\v{q}) \hat{\rho}(\v{q}) \hat{\rho}(-\v{q}) - \frac{1}{2A} \sum_{\v{q} \in \R^2} \; V(\v{0}) \hat{\rho}(\v{0}) 
 :=	\frac{1}{2\Asample} \sum_{\v{q} \in \R^2} \; V(\v{q}) :\hat{\rho}(\v{q}) \hat{\rho}(-\v{q}):,
    \label{eq:Coulomb_interaction_band_basis}
\end{equation}
where the normal ordering is defined relative to the vacuum. The second term $V(\v{0}) \hat{\rho}(\v{0})$ is a chemical potential. Note that Eq. \eqref{eq:Coulomb_interaction_band_basis} takes the same form in magnetic and non-magnetic systems. Therefore the interaction is entirely specified by the Bloch waves $\ket{u_{\v{k}}}$ in the first Brillouin zone, together with the operators $\hat{V}_{\v{G}}(\v{k})$.

 \subsection{Derivation of Hartree-Fock}
 \label{app:HF_derivation}

 This section will derive Hartree-Fock using a partial-trace approach, wherein only some of the bands are integrated out at mean-field level. Standard Hartree-Fock is the case where all bands are integrated out. However, the partial version will prove useful below when integrating out the remote bands of the BM model. 

 Consider the four-body interaction
 \begin{equation}
	 \hat{V} := V:\hat{\rho} \hat{\rho}: = \frac{1}{2 A_s} \sum_{\v{q} \in \R^2} \sum_{\v{k},\v{k}' \in \bz} V(\v{q}) \,
	 [\Lambda_{-\q}(\k)]_{\alpha\delta} [\Lambda_{\q}(\k')]_{\beta\gamma} \d_{\k,\alpha} \d_{\k',\beta} \c_{[\k'+\q],\gamma} \c_{[\k-\q],\delta}
 \end{equation}
 where $\alpha,\beta,\gamma,\delta$ run over all species (i.e. bands and flavors). Suppose the species are partitioned into \textit{active} species $\mathcal{A}$ and \textit{remote} species $\mathcal{R}$. Physically, one normally considers the eight flat bands of \tbg{} to be active, and the rest remote. In Step 5 of App. \ref{app:Hamiltonian_construction} and the main text, we take the more restrictive perspective of a single active band.
 
 Our goal is to integrate out the remote bands, which is a partial trace
 \begin{equation}
	 \hat{V} \mapsto \tr_{\mathcal{R}}[\hat{\rho}_{\mathcal{R}} \hat{V}].
 \end{equation}
Here $\rho_{\mathcal{R}}$ is a density matrix for the remote bands, not a charge density operator. We treat remote bands at mean-field level such that $\rho_{\mathcal{R}}$ is quadratic, and any quadratic density matrix is fully characterized by its correlation matrix $\tr[ \hat{\rho}_\mathcal{R}\; \c_{\alpha}^\dagger \c_{\beta}] = (P_{\mathcal{R}})_{\beta\alpha}$. For convenience, define $\braket{\mathcal{O}}_{P_{\mathcal{R}}}~:=~\tr_{\mathcal{R}}[ \hat{\rho}[P_{\mathcal{R}}] \mathcal{O}]$, which is an \textit{operator} on the Hilbert space of the active bands. For brevity we drop the subscript $\mathcal{R}$ on $\rho$ when the $P_{\mathcal{R}}$ dependence is made explicit.

 Integrating out the remote bands of $\hat{V}$ --- an operation that changes the Hilbert space --- requires us to evaluate terms of the form
 \begin{equation}
	 \tr_{\mathcal{R}}[\hat{\rho}[P_{\mathcal{R}}] \d_A \d_B \c_C \c_D].
 \end{equation}
 There are six cases depending on which fermions are remote \cite{repellin2020ferromagnetism}:
 1. $A,B,C,D \in \mathcal{R}$,
 2. $B,D \in \mathcal{R}$,
 3. $B,C \in \mathcal{R}$,
 4. $A,D \in \mathcal{R}$,
 5. $A,C \in \mathcal{R}$,
 6. All in $\mathcal{A}$.
 These may all be evaluated straightforwardly from Wick's theorem, and the fact that one can pull fermions in $\mathcal{A}$ outside the partial trace, which yields:
 \begin{enumerate}
	 \item $\tr_{\mathcal{R}}[\hat{\rho}[P_{\mathcal{R}}] \d_A \d_B \c_C \c_D] = \phantom{-} \text{constant}$
	 \item $\tr_{\mathcal{R}}[\hat{\rho}[P_{\mathcal{R}}] \d_A \d_B \c_C \c_D] = - \braket{\d_B \c_D}_{P_{\mathcal{R}}} \d_A \c_C$
	 \item $\tr_{\mathcal{R}}[\hat{\rho}[P_{\mathcal{R}}] \d_A \d_B \c_C \c_D] = \phantom{-} \braket{\d_B \c_C}_{P_{\mathcal{R}}} \d_A \c_D$
	 \item $\tr_{\mathcal{R}}[\hat{\rho}[P_{\mathcal{R}}] \d_A \d_B \c_C \c_D] = \phantom{-} \braket{\d_A \c_D}_{P_{\mathcal{R}}} \d_B \c_C$
	 \item $\tr_{\mathcal{R}}[\hat{\rho}[P_{\mathcal{R}}] \d_A \d_B \c_C \c_D] =  -\braket{\d_A \c_C}_{P_{\mathcal{R}}} \d_B \c_D$
	 \item $\tr_{\mathcal{R}}[\hat{\rho}[P_{\mathcal{R}}] \d_A \d_B \c_C \c_D] = \phantom{-} \braket{I}_{P_{\mathcal{R}}} \d_A \d_B \c_C \c_D = \d_A \d_B \c_C \c_D$.
 \end{enumerate}
 We now restore momentum indices. We will find 3 \& 4 combine to give the Hartree Hamiltonian, while 2 \& 5 give the Fock term. For term 4, $\braket{\d_A \c_D}_P = \braket{\d_{\k,\alpha} \c_{[\k-\q],\delta}}_P = [P(\v{k})]_{\delta\alpha} \delta(\k - [\k'-\q])$, so 
 \begin{equation}
	 \hat{V}_4 = 
 \frac{1}{2 A_s} \sum_{\v{q} \in \R^2} \sum_{\v{k},\v{k}' \in \bz} V(\v{q}) \,
	 [\Lambda_{-\q}(\k)]_{\alpha\delta} [\Lambda_{\q}(\k')]_{\beta\gamma} \d_{\k',\beta} P(\v{k}) \delta(\k - [\k'-\q]) \c_{[\k'+\q],\gamma}
 \end{equation}
 The $\delta$-function is only satisfied if $\q = \v{G}$ for some $\v{G}$ in the reciprocal lattice, so 
 \begin{equation}
	 \hat{V}_4 
	 = \sum_{\k' \in \bz} \d_{\k', \beta} \left[
  \frac{1}{2A_s} 
  \sum_{\v{G}} V(\v{G}) [\Lambda_{\v{G}}(\v{k})]_{\beta \gamma} \left( \sum_{\k \in \bz} [P(\k)]_{\alpha\delta}  [\Lambda_{-\v{G}}(\k)]_{\alpha\delta} \right)
	 \right] \c_{\k',\gamma}.
 \end{equation}
 Due to the periodic gauge, $\Lambda_{-\v{G}}(\k) = \Lambda_{\v{G}}(\k)^\dagger$, which means $\hat{V}_4$ is half the Hartree term (and $\hat{V}_3$ is the other half).

 Meanwhile, term 5 involves the expectation $\braket{\d_A \c_C}_P = \braket{\d_{\k,\alpha} \c_{[\k'+\q],\gamma}}_P = [P(\k)]_{\gamma\alpha} \delta(\k - [\k'+\q])$, giving 
 \begin{equation}
	 \hat{V}_5 =  \frac{1}{2 A_s} \sum_{\v{q} \in \R^2} \sum_{\v{k},\v{k}' \in \bz} V(\v{q}) \,
	 [\Lambda_{-\q}(\k)]_{\alpha\delta} [\Lambda_{\q}(\k')]_{\beta\gamma}  [P(\k)]_{\gamma\alpha} \delta(\k - [\k'+\q])  \d_{\k',\beta} \c_{[\k-\q],\delta}
 \end{equation}
 The $\delta$-function implies $\k = \k'+\q + \v{G}$ for some reciprocal lattice vector $\v{G}$, so 
 \begin{equation}
	 \hat{V}_5 = \sum_{\k' \in \bz} \d_{\k',\beta} \left[
		 - \frac{1}{2 A_s} \sum_{\q \in \R^2} [\Lambda_{\q}(\k')]_{\beta \gamma} [P(\k' + \q)]_{\alpha\gamma} [\Lambda_{-\q}(\k'+\q)]_{\alpha \delta}
	 \right] \c_{\k',\delta} 
 \end{equation}
 Using the relation $\Lambda_{-\q}(\k'+\q) = \Lambda_{\q}(\k')^\dagger$, we recognize this as half the Fock contribution.

 Combining all these, we have
 \begin{subequations}
 \begin{align}
	 \label{eq:HF_derivation_final}
	 \tr[\hat{\rho}[P] \hat{V}] &= \sum_{i} \hat{V}_i = \hat{H}_{H}[P] + \hat{H}_F[P] + \hat{V}_{\mathcal{A}} + \text{ const}\\
	 \label{eq:Hartree_Hamiltonian}
	 \hat{H}_H[P] &= 
	 \sum_{\k \in \bz} \d_{\k} \left[
  \frac{1}{A_s} 
  \sum_{\v{G}} V(\v{G}) \Lambda_{\v{G}}(\v{k}) \Tr[P^T \Lambda_{\v{G}}^\dagger]
	 \right] \c_{\k}\\
	 \label{eq:Fock_Hamiltonian}
	 \hat{H}_F[P] &=
	 \sum_{\k \in \bz} \d_{\k} \left[
  -\frac{1}{A_s} 
  \sum_{\v{q}} V(\v{q}) \Lambda_{\v{q}}(\v{k}) P(\k + \q)^T \Lambda_{\v{q}}(\k')^\dagger]
	 \right] \c_{\k}\\
	 \hat{V}_{\mathcal{A}} &=
	 \frac{1}{2 A_s}\sum_{\alpha,\beta,\gamma,\delta \in \mathcal{A}} \sum_{\v{q} \in \R^2} \sum_{\v{k},\v{k}' \in \bz} V(\v{q}) \,
	 [\Lambda_{-\q}(\k)]_{\alpha\delta} [\Lambda_{\q}(\k')]_{\beta\gamma} \d_{\k,\alpha} \d_{\k',\beta} \c_{[\k'+\q],\gamma} \c_{[\k-\q],\delta}
 \end{align}
 \end{subequations}
 Here and below we use a convenient matrix notation for the form factors and correlation matrices (which runs over all bands of both $\mathcal{R}$ and $\mathcal{A}$), $\Tr$ gives the trace over both $\k'$ and all bands, and the fermions $\d,\c$ run over active bands $\mathcal{A}$. Note that the last term is the \textit{restriction} of $\hat{V}$ to the active bands.

If $\mathcal{R}$ runs over all bands, then $\hat{H}_{HF}$ is a quadratic operator on all the bands, and  $E_{HF} = \Tr\big[\hat{\rho}[P] \hat{H}_{HF}[P]\big]$ is simply a number. In this case, one can use this as a procedure for finding the ``best quadratic representation" of an operator with respect to a given correlation matrix: 
 \begin{equation}
	 \hat{V} \to \hat{V}\big|_{MF,P} := \hat{H}_{HF}[P].
	 \label{eq:mean_field_operator_projection}
 \end{equation}
 This formula is used below to implement the Hartree-Fock subtraction procedure to account for the background charge density of remote bands in \tbg{}.

This Appendix has now defined all the notation and Hamiltonian-independent formalism we shall use to consider the problem \tbg{} at finite field. We again highlight that the only model-specific information needed are the single particle energies and wavefunctions within the first Brillouin zone, and the matrix elements from Eq. \eqref{eq:moire_matrix_elements}.

\section{TBG in Magnetic Field}
\label{app:TBG_in_B_field}

In this section we review the construction of a model for MATBG at finite field and derive the form of Hartree-Fock interactions at finite field.

This Appendix will construct a model for \tbg{} at finite field. We start by recalling the BM model, then discuss its modifications from a magnetic field and compute the needed matrix elements. We then describe how the remote bands should be integrated out and the details of a ``Hartree-Fock subtraction" procedure employed to prevent double-counting various interaction effects. We conclude by discussing the Hamiltonian for the spin- and valley-polarized $\nu=3$ state, and the proper way to split it into a dispersion and interaction.

\subsection{BM Model}

To start, let us review the standard Bistritzer-MacDonald model.  The moir\'e unit cell and Brillouin zone are spanned by primitive and reciprocal lattice vectors
\begin{equation}
	\v{a}_{1,2} = \frac{4\pi}{3 k_\theta}(\pm \sqrt{3}/2, 1/2); \quad  \v{b}_{1,2} = k_\theta (\pm \sqrt{3}/2, 3/2) 
\end{equation}
so $\v{a}_i\cdot \v{b}_j = 2\pi \delta_{ij}$. Here $k_\theta = \frac{8\pi}{3\sqrt{3} a} \sin \theta/2$, where $\theta$ is the twist angle and $a \approx \SI{1.36}{\angstrom}$ is the carbon-carbon bond length of graphene. Let $A_0$ the unit cell area.

The BM model is defined as a real-space continuum model with four species: two layers and two sublattices, labeled by Pauli matrices $\gamma^\mu$ and $\sigma^\mu$ respectively. The Hamiltonian in the $K$ valley consists of two Dirac ones, one for each layer, and an interlayer tunneling term
\begin{equation}
	H^K_{BM}(\v{r}) = \begin{pmatrix} 
		h^+ & T(\v{r})\\
		T(\v{r})^\dagger & h^-
	\end{pmatrix},
	\quad
	h^{\pm} = - i\hbar v_F \v{\sigma}_{\pm \theta/2} \cdot \left[ \v{\nabla} \mp \frac{i}{2} \v{q}_{1} \right],\quad
	T(\v{r}) 
	= \sum_{s=0}^2 T_s e^{-i \v{b}_s \cdot \v{r}},
    \quad T_s	= \begin{pmatrix} 
	w_0  & \zeta^s w_1 \\
	\zeta^{-s} w_1 & w_0
	\end{pmatrix},
	\label{eq:BM_model_K_valley_appendix}
\end{equation}
where $\v{\sigma}_{\theta/2} = e^{-i \theta \sigma_z/4} \v{\sigma} e^{i \theta \sigma_z/4}$,  $\zeta = e^{i 2\pi/3}$, $\v{b}_0 = 0$, $\v{q}_1 = k_\theta(0,-1)$ is the offset of the Dirac point, and $v_F$ is the Fermi velocity of graphene. The moire Brillouin zone (mBZ) is shown in Fig. \ref{fig:moire_reciprocal_geometry}.

The other valley may be obtained under time-reversal symmetry, and we express it here for convenience. For simplicity, let us make the approximation $\v{\sigma}_{\theta/2} \approx \v{\sigma}$, which has the effect of making the model particle-hole symmetric. Let $\tau \in \st{K,K'}$ index the valley. Then
\begin{equation}
	H_{BM}(\v{r}) = \begin{pmatrix} 
		h_{t} & T(\v{r})\\
		T(\v{r})^\dagger & h_b
	\end{pmatrix},
	\quad
	h(\v{r}) = -v_F[\tau_z \sigma_x p_x + \sigma_y p_y] - v_F[\tau_z \sigma_x K_x - \tau_z \gamma_z \sigma_y K_y], 
	\label{eq:BM_model_both_valleys}
\end{equation}
where $p_\alpha = -i \partial_\alpha$ and $K_{\alpha}$ are the coordinates of $K^t = R(\theta/2) K_{\text{graphene}}$. The tunneling term is unchanged.

We take a computational basis of Fourier series in the real-space unit cell
\begin{equation}
	\braket{\v{r}|v_{\v{k}I}} = 	v_{\v{k},\gamma,\sigma,\tau,\v{G}}(\v{r}) =  A_0^{-1/2} e^{-i\v{G}\cdot\v{r}} \ket{\gamma,\sigma,\tau} 
\end{equation}
where $\v{G}$ runs over reciprocal lattice vectors and $I = (\gamma,\sigma,\tau,\v{G})$ is a multi-index. The matrix elements from Eq. \eqref{eq:moire_matrix_elements} are then relatively simple:
\begin{equation}
	[\mathcal{T}_{\v{G''}}(\v{k})]_{\gamma\sigma \v{G}}^{\gamma'\sigma' \v{G}'}
= 
[\mathcal{V}_{\v{G''}}(\v{k})]_{\gamma\sigma \v{G}}^{\gamma'\sigma' \v{G}'}
= \delta_{\gamma'\gamma} \delta_{\sigma'\sigma} \delta_{\tau\tau'} \delta_{\v{G}-\v{G}'',\v{G}'},\quad
[\lambda_{\v{q}}(\v{k})]^{I}_J = \delta^I_J. 
\end{equation}
In particular, the form factor in the computational basis, $\lambda_{\v{q}}$, is trivial. This will not be the case at finite field.

\begin{figure}
	\includegraphics[width=0.5\textwidth]{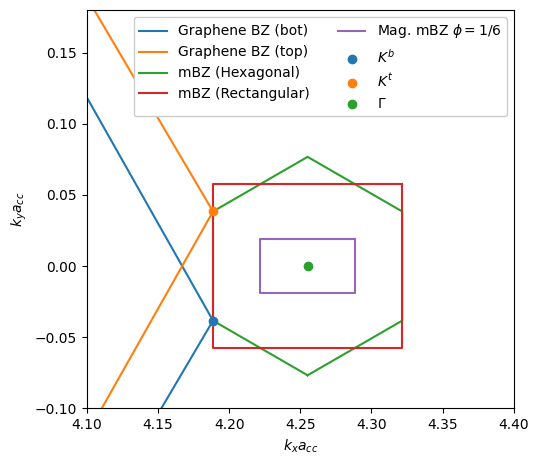}
	\caption{Reciprocal space geometry of MATBG. The magnetic Brillouin zone we employ is always rectangular, with width $Q_x$ (half the width of the rectangular mBZ) and height $Q_y/q$ ($1/q$ times the height of the mBZ) so that $A_{\text{mag. BZ}} = (1/2q) A_{mBZ}$.}
	\label{fig:moire_reciprocal_geometry}
\end{figure}

\subsection{Magnetic BM Model}
At finite field we adopt a rectangular magnetic Brillouin with sides lengths $Q_x = (\sqrt{3}/2) k_\theta$ and $Q_y/q = (3/2) k_\theta / q$, shown in Fig. \ref{fig:moire_reciprocal_geometry}. As $2Q_x Q_y = A_{\rm{mBZ}} = (2\pi)^2/A_{\rm{UC}}$ is the area of the moir\'e Brillouin zone in terms of the area of the moir\'{e} unit cell $A_{\rm UC}$, the commensurability condition is
\begin{equation}
	\phi := \frac{B A_{\rm{UC}}}{\Phi_0} = \frac{p}{2q} \text{ or } Q_x Q_y \ell^2 = 2\pi \frac{q}{p}
\end{equation}
where $\Phi_0 = h/e$, $\ell = (\hbar/eB)^{1/2}$. In real space the magnetic unit cell is twice as wide and $q$ times as tall as the moir\'e unit cell.

We adopt Landau gauge $\v{A} = (-By,0)$ so that $\v{\pi} = \v{p}+e\v{A} = (p_x - eBy, p_y)$ where $p_\alpha = -i \hbar \partial_\alpha$ as usual. Solving for the twists $\xi_{\v{R}}(\v{r})$ gives boundary conditions
\begin{align}
&	u(x+a_x,y) = u(x,y) &a_x &= 2\pi/Q_x\\
&	u(x,y+a_y) = e^{i x a_y\ell^2} u(x,y) &a_y &= 2\pi/(Q_y/q).
\label{eq:unit_cell_boundary_conditions}
\end{align}

For numerical convenience, we now adopt a basis of offset superpositions of Landau levels so that (i) the kinetic term is simple (2) the tunnelling terms are easily computable and (3) the boundary condition Eq. \eqref{eq:unit_cell_boundary_conditions} are manifestly satisfied. To this end, we work with eigenstates of the free magnetic Hamiltonian $H_{f}(B) = \pi^2/(2m)$. Define Landau level operators
\begin{equation}
	\hat{a}^\pm := \frac{\ell}{\sqrt{2}\hbar} (\hat{\pi}_x \pm i \hat{\pi}_y)
\end{equation}
and select a guiding center coordinate $\hat{Y} := \hat{y} + \frac{\ell^2}{\hbar} \hat{\pi}_x = \frac{\ell^2}{\hbar} \hat{p}_x$ where $[\hat{a}^\pm, \hat{Y}] = 0$. The eigenstates of $\hat{Y}$ are plane waves $e^{ik_x x}$ and, in the position basis, the raising and lowering operators act along $y$
\begin{equation}
	\frac{\sqrt{2}}{\hbar \ell} \hat{a} = -\partial_y - [y-Y]/\ell^2
	\label{eq:Landau_level_lowering_along_y}
\end{equation}
for each $Y$. The wavefunctions of the free magnetic Hamiltonian are therefore planewaves along $x$ and harmonic oscillators along $y$:
\begin{align}
	\braket{x,y| n, Y} = \ell^{-1/2} e^{i Y x} \phi_n([y-Y]/\ell),\quad  \phi_n(y) = (2^n n! \sqrt{\pi})^{-1/2} H_n(y) e^{-y^2/2},
	\label{eq:Landau_levels}
\end{align}
where $H_n$ are the Hermite polynomials and $n$ is the Landau level index. Whenever possible we will switch to dimensionless quantities $\tilde{y} = [y-Y]/\ell$, $\tilde{p}_y = -i \partial_{\tilde{y}}$, and $\tilde{a} = -2^{-1/2} [\tilde{y} + i \tilde{p}_y]$.

We define our computational basis in terms of these as a discrete Fourier transform in $Y$ in steps of $Q_x \ell^2$. Let $\gamma$ and $\sigma$ label layer and sublattice. Then define
\begin{align}
	Y_{M,j, k_x} &:= k_x \ell^2 + (Mp+j) Q_x \ell^2, \qquad M \in \Z, j \in \Z/p\Z,\\
	\ket{\gamma,\sigma,n,j,k_x,k_y} &:= a_x^{-1/2} \sum_{M \in \Z} e^{i k_y (Mp+j) Q_x \ell^2} \ket{\gamma,\sigma} \otimes \ket{n,Y_{M,j,k_x}},\\
	v_{\v{k}I}(\v{r}) = v_{\v{k}, \gamma,\sigma,n,j} &:= \braket{\v{r}|e^{-i\v{k} \cdot \v{r}}| \gamma, \sigma, n, j, \v{k}} = (a_x \ell)^{-1/2} \sum_{M \in \Z} e^{-i k_y[y-(Mp+j) Q_x \ell^2]} e^{ix (Mp+j) Q_x} \phi_n([y-Y_{M,j,k_x}]/\ell),
\end{align}
where $I = (\gamma,\sigma,n,j)$ is a multi-index. The $v$'s are the Bloch waves of the free electron in a magnetic field. They obey boundary conditions \eqref{eq:unit_cell_boundary_conditions} with standard normalization $\braket{v_{\v{k}I}|v_{\v{k}J}} = \int_{uc} \dr \, \overline{v}_{\v{k}I}(\v{r}) v_{\v{k} J}(\v{r})  = \delta_{IJ}$ over the magnetic unit cell. We consider the eigenproblem in this computational basis with a UV cutoff of a few hundred Landau levels.

As discussed in the main text, three types of matrix elements suffice to describe the Hamiltonian and all other quantities of interest to us. We merely quote their formulas here, and delay their derivation to Sec. \ref{subsec:magnetic_matrix_elements} below. Let $\v{G} = r \v{G}_x + s \v{G}_y = (r Q_x, s Q_y/q)$. Then
\begin{subequations}
\begin{align}
	[\mathcal{T}_{\v{G}}(\v{k})]^{I'}_I := \braket{v_{\v{k} I'}|e^{-i\v{G} \cdot \v{r}}| v_{\v{k}I}}
	&=
	\left[e^{i r \frac{k_y}{G_2} 2\pi/p} e^{-i \frac{k_x}{G_1} s  2\pi q/p} \right] e^{-i (sj - rs/2) 2\pi q/p} \\ 
	&\hspace{2em}\times \delta_{\gamma'\gamma} \delta_{\sigma'\sigma} \delta_{j',j-r} D_{n'n}(\overline{z}_{\v{G}}), \quad (\text{assuming } q|s) \\
	[\lambda_{\v{q}}(\v{k})]^{I'}_I := \braket{v_{\v{k}I'} | v_{\v{k}+\v{q},I}} 
	&= 
	\left[
		e^{-i \frac{k_x}{G_1} \frac{q_y}{G_2} 2\pi/p}
	\right]
	\times
	e^{-i \frac{q_x}{G_1} \frac{q_y}{G_2} 2\pi/p}
	\delta_{\gamma'\gamma} \delta_{\sigma'\sigma} \delta_{j'j} D_{n'n}(\overline{z}_{\v{q}}),
	\\
	[\mathcal{V}_{\v{G}}(\v{k})]^{I'}_{I} := \braket{v_{\v{k}+\v{G},I'}|e^{-i\v{G}\cdot \v{r}}|v_{\v{k},I}}
	&= \left[ 
		e^{i r \frac{k_y}{G_2} 2\pi/p}
	\right]
	\times e^{-i sj' 2\pi/p} \delta_{\gamma'\gamma} \delta_{\sigma'\sigma} \delta_{j',j-r} \delta_{n'n},
\end{align}
\label{eq:magnetic_matrix_elements}
\end{subequations}
where $\overline{z}_{\v{k}} := \frac{\ell}{\sqrt{2}}(k_x - i k_y)$ is a complexified momentum, $G_1 = Q_x$ and $G_2 = Q_y/q$ are the dimensions of the magnetic Brillouin zone, and 
\begin{equation}
	\hat{D}(z) := e^{z a^\dagger - z^* a}
	\label{eq:displcement_operator}
\end{equation}
is the standard displacement operator for the harmonic oscillator, whose matrix elements may be expressed in terms of the associated Laguerre polynomials $L_n^\alpha(x)$ as
\begin{equation}
	D_{mn}(z)= \braket{m|e^{z a^\dagger - z^* a}|n} = 
	\begin{cases}
		\sqrt{\frac{n!}{m!}} (z)^{m-n} L_{n}^{m-n}(\n{z}^2) e^{-\n{z}^2/2} & \text{ if } m \ge n\\
		\sqrt{\frac{m!}{n!}} (-z^*)^{n-m} L_{m}^{n-m}(\n{z}^2) e^{-\n{z}^2/2} & \text{ if } m < n.
	\end{cases}
	\label{eq:Laguerre_factor}
\end{equation}

We now resolve the Hamiltonian in this basis. The kinetic energy for the $K$ valley can be written in terms of the Landau level lowering operator as
\begin{equation}
	(\hat{h}_0)^K_{\pm} = v_F \v{\sigma}_{\pm \theta/2} \cdot \left[\hat{\v{\pi}} \mp \frac{\hbar}{2} \v{q}_0 \right] 
	= \hbar v_F \sigma^+ e^{\pm i \theta/2} \left[ \frac{\sqrt{2}}{\ell} \hat{a} \mp \frac{i k_\theta}{2} \right] + \text{ h.c.}
	\label{eq:magnetic_kinetic_term}
\end{equation}
where the second equality follows from  \eqref{eq:Landau_level_lowering_along_y}. For both valley, this takes the form
\begin{equation}
	\hat{h}_0 = - \frac{v_f \sqrt{2} \hbar}{\ell} \left\{
		[\hat{a} \sigma^+ + \hat{a}^\dagger \sigma^- ] \gamma^z \frac{\tau_0 + \tau_z}{2} +  
	[- \hat{a}^\dagger \sigma^+ - \hat{a} \sigma^-] \gamma^z \frac{\tau_0 - \tau_z}{2}
\right\}
\label{eq:kinetic_term_finite_field}
\end{equation}	
where the matrices are in sublattice space and the combinations of $\tau$'s give projectors to the $K$ and $K'$ valleys respectively. The tunneling matrix elements take the form
\begin{equation}
	\braket{v_{\v{k}I'}|T(\v{r})|v_{\v{k}J}} = \sum_{i=0}^2 (\gamma^+)^{\gamma'}_{\gamma''} (T_i)^{\sigma'}_{\sigma''} [\mathcal{T}_{\v{b}_i}(\v{k})]^{\gamma''\sigma'' j' n'}_{\gamma\sigma j n} + h.c.
\end{equation}
where the reciprocal translations are $\v{b}_{0,1,2} = r\v{G}_x + s\v{G}_y$ with $(r,s) = (0,0), (1,q), (-1,q)$ in terms of the magnetic lattice.

\subsection{Magnetic Matrix Elements}
\label{subsec:magnetic_matrix_elements}

This technical section derives the magnetic matrix elements in Eq. \eqref{eq:magnetic_matrix_elements}. Let $\v{G}  = (G_x,G_y) = (r Q_x, s Q_y/q)$ and $\v{q} = (q_x,q_y)$. For concision, let $\Delta = Q_x\ell^2$. We will often use the fact that $Q_x Q_y \ell^2 = 2\pi q/p$ or $G_1 G_2 \ell^2 = 2\pi /p$. We also recall a fact about the Harmonic oscillator matrix eigenstates $\phi_{n}(y)$
\begin{equation}
	\int_\R dy \, \varphi_{n'}(y) e^{-i q_y y} e^{-i q_x \ell p_y} \varphi_n(y) = e^{-i q_x q_y \ell^2/2} D_{n'n}(\overline{z}_q); \quad \overline{z}_q := \frac{\ell}{\sqrt{2}}(q_x - i q_y),
	\label{eq:harmonc_oscillator_matrix_element_I}
\end{equation}
which follows from Baker-Campbell-Hausdorff and the definition of the displacement operator. 

\textbf{Tunneling matrix elements} This is computed in the supplemental information to \cite{xie2019spectroscopic}. 

\textbf{Form factors of the basis vectors}
\begin{align}
	[\mathcal{T}_{\v{q}}(\v{k})]^{\gamma\sigma n j}_{\gamma'\sigma'n' j'}
	&= \delta_{\gamma\gamma'} \delta_{\sigma\sigma'} \sum_{M,M'} \int_0^{a_x} \frac{dx}{a_x} \int_0^{a_y} 
	\frac{dy}{\ell} e^{i k_y [y-(Mp + j) \Delta]} e^{-ix(Mp+j) Q_x} \varphi_n([y-Y]/\ell)\\
	& \hspace{1in} 
	\times e^{-i (k_y+q_y) [y-(M'p + j') \Delta]} e^{ix(M'p+j') Q_x} \phi_n([y-Y']/\ell)\\
	\intertext{the $x$ integral gives $\delta_{jj'}\delta_{MM'}$, whereupon the $k_y$ phase factors cancel}
	&=  \delta_{\gamma\gamma'} \delta_{\sigma\sigma'}\delta_{jj'} \sum_{M} \int_0^{a_y} \frac{dy}{\ell} \varphi_n([y-Y]/\ell) e^{-i (q_y) [y-(Mp+j)\Delta]} \varphi_{n'}([y-Y-k_x \ell^2]/\ell)
	\intertext{Making the substitution $\overline{y} = [y-Y_{M,j,k_x}]/\ell$ and combining the sum and integral,}
	&= \delta_{\gamma\gamma'} \delta_{\sigma\sigma'}\delta_{jj'} \int_{\R} d\overline{y}\;  \varphi_n(\overline{y}) e^{-i q_y [\ell \overline{y}+\ell^2 k_x]} \varphi_{n'}(\overline{y} - q_x \ell)\\
	&= \delta_{\gamma\gamma'} \delta_{\sigma\sigma'}\delta_{jj'} e^{-i q_y k_x \ell^2} \int_{\R} d\overline{y}\;  \varphi_n(\overline{y}) e^{-i q_y \ell \overline{y}} e^{-i q_x p_y \ell} \varphi_{n'}(\overline{y})\\
	\intertext{The integral is exactly \eqref{eq:harmonc_oscillator_matrix_element_I}, so}
	&= 
	\left[
		e^{-i \frac{k_x}{G_1} \frac{q_y}{G_2} 2\pi/p}
	\right]
	e^{-i \frac{q_x}{G_1} \frac{q_y}{G_2} 2\pi/p}
	\delta_{\gamma\gamma'} \delta_{\sigma\sigma'}\delta_{jj'} D_{nn'}(\overline{z}_{\v{G}})
\end{align}

\textbf{Reciprocal translations}
\begin{align}
	&[\mathcal{V}_{\v{G}}(\v{k})]^{I'}_I = \braket{v_{\v{k} + \v{G},\gamma',\sigma',n',j'}| e^{-i\v{G}\cdot\v{r}}| v_{\v{k},\gamma,\sigma,n,j}}\\
    &=
    \sum_{M,M' \in \Z} \int_0^{a_x} \frac{dx}{a_x} \int_0^{a_y} \frac{dy}{\ell} 
    e^{i(k_y + G_y) [y - (M'p+j') \Delta]} e^{-ix(M'p+j')Q_x} \varphi_{n'}([y- (k_x + G_x)\ell^2 - (M'p+j')\Delta]/\ell)\\
    & \hspace{2em} \times
    e^{-i k_y [y -(Mp+j) \Delta]} e^{ix (Mp+j) Q_x} \varphi_n([y-k_x \ell^2 - (Mp+j)\Delta]/\ell) e^{-i[G_x x + G_y y]}\delta_{\gamma,\gamma'} \delta_{\sigma,\sigma'}.
\end{align}
The $x$ integral gives $\delta_{M'M} \delta_{j',j-r}$. We eliminate the primed indices (and drop $\gamma$, $\sigma$), to get
\begin{align}
    & \delta_{j',j-r} \sum_{M} \int_{0}^{a_y} \frac{dy}{\ell} 
    e^{i(k_y + G_y) [y - (Mp+j-r)\Delta]}
    e^{-i k_y [y -(Mp+j) \Delta]} 
    e^{-i[G_y y]}\\
    & \hspace{4em}\times 
    \varphi_{n'}([y- (k_x + G_x)\ell^2 - (Mp+j-r)\Delta]/\ell)
    \varphi_n([y-k_x \ell^2 - (Mp+j)\Delta]/\ell)
\end{align}
As $- G_x \ell^2 + r \Delta = - r Q_x \ell^2 + r Q_x \ell^2 = 0$, the oscillator wavefunctions have the same argument, $y-Y$ with $Y = k_x \ell^2 + (Mp+j)\Delta$, so
\begin{align}
    & \delta_{j',j-r} \sum_{M} \int_{0}^{a_y} \frac{dy}{\ell} 
    e^{i(k_y + G_y) [y - (Mp+j)\Delta]}
    e^{i(k_y + G_y) r \Delta}
    e^{-i k_y [y -(Mp+j) \Delta]} 
    e^{-i[G_y y]}
    \varphi_{n'}([y- Y]/\ell)
    \varphi_n([y-Y]/\ell).
\end{align}
Cancelling exponential factors leaves us with
\begin{align}
    & \delta_{j',j-r} \sum_{M} \int_{0}^{a_y} \frac{dy}{\ell} 
    e^{iG_y [- (Mp+j)\Delta]}
    e^{i(k_y + G_y) r \Delta}
    \varphi_{n'}([y- Y]/\ell)
    \varphi_n([y-Y]/\ell).
\end{align}
As $G_y Mp \Delta = (s Q_y/q) Mp Q_x \ell^2 = Ms (p/q) 2\pi q/p \in 2\pi \Z$, we are left with
\begin{align}
    & \delta_{j',j-r}
    e^{- iG_y j\Delta}
    e^{i(k_y + G_y) r \Delta}
    \sum_{M} \int_{0}^{a_y} \frac{dy}{\ell}
    \varphi_{n'}([y- Y]/\ell)
    \varphi_n([y-Y]/\ell).
\end{align}
Making the substitutions $\tilde{y} = [y-Y_{M,j,k_x}]/\ell$, we are left with
\begin{align}
    & \delta_{j',j-r}
    e^{- iG_y j\Delta}
    e^{i(k_y + G_y) r \Delta}
    \int_{\R} d\tilde{y}
    \varphi_{n'}(\tilde{y})
    \varphi_n(\tilde{y})
\end{align}
The integral is simply $\delta_{n'n}$. The phase factor simplifies as $-i G_y j\Delta + i G_y r\Delta = -i (j-r) s G_1 G_2 \ell^2 = - i j' s 2\pi/p$. Altogether, we have
\begin{equation}
\braket{u_{\v{k} + \v{G},\gamma',\sigma',n',j'}| e^{-i\v{G}\cdot\v{r}}| u_{\v{k},\gamma,\sigma,n,j}} 
= 
\left[
	e^{i r \frac{k_y}{G_y} 2\pi /p}
\right]
e^{i \frac{k_y}{Q_y} r 2\pi q/p}
\delta_{\gamma',\gamma} \delta_{\sigma'\sigma} \delta_{n',n} \delta_{j',j-r}.
\end{equation}

\section{Hartree-Fock Subtraction}
\label{app:Hamiltonian_construction}

This Appendix defines the Hamiltonian and --- crucially --- the bandwidth of the $\nu=3$ state of TBG we consider as a parent state for fractional Chern insulators. In the presence of interactions, the bandwidth is not uniquely defined, as there are many ways to group terms between the dispersion and interaction. Four common options, suitable for different purposes, are
\begin{subequations}
\begin{align}
	\hat{H} = \hat{H}_A &= \hat{h}_A + V \; \rho \rho\\
	= \hat{H}_B &= \hat{h}_B + V :\rho \rho: & \text{(DMRG)} \\
	= \hat{H}_C &= \hat{h}_C + V\; \delta\rho \delta\rho & \text{(strong-coupling ferromagnets)}\\
	= \hat{H}_D &= \hat{h}_D + V :\delta\rho \delta\rho: & \text{(psuedopotentials)}
\end{align}\label{eq:Hamiltonian_many_forms}\end{subequations}
where $: \hat{\mathcal{O}} :$ denotes normal ordering and $ \widehat{\delta \rho} = \hat{\rho} - \braket{\hat{\rho}}$ is the charge density with respect to a reference state. We note that option D has the form employed by psuedopotential arguments. Physically, $\hat{h}_D$ gives the exact 1-electron excitation spectrum above the normal-ordering reference state--- in our case, the $\nu=3$ spin- and valley-polarized state. This assumes the restriction to the space of active bands alone. Therefore splitting D is the physically relevant choice for studying FCIs and their indicators. The total dispersion used in the main text is defined as $\hat{h}_T := \hat{h}_D$.

In Steps 1-4 below we consider eight active bands, as is standard in \tbg{}. We restrict only in Step 5 to a single active band.

We now proceed to construct $\hat{H}_D$ for the $\nu=3$ state of TBG using the so-called `Hartree-Fock subtraction' procedure to account for the effect of the remote bands at mean-field level. Our main tool is the partial trace with respect to a quadratic density matrix. Suppose we have a Hamiltonian $V : \rho\rho:$ defined for many species of fermions, and suppose the species are partitioned into \textit{active} species $\mathcal{A}$ and \textit{remote} species $\mathcal{R}$. Suppose $P_{\mathcal{R}}(k) = \braket{\d_{\v{k}} \c_{\v{k}}}$ is a quadratic density matrix for the remote bands.

We may then integrate out the remote bands at mean-field level via a partial trace. The result, derived in Eq. \eqref{eq:HF_derivation_final} below, is
\begin{equation}
	\braket{V :\hat{\rho} \hat{\rho}:}_{P_{\mathcal{R}}} = \hat{H}_H[P_{\mathcal{R}}] +\hat{H}_{F}[P_{\mathcal{R}}] + V : \hat{\rho}_{\mathcal{A}} \hat{\rho}_{\mathcal{A}} : +\, \text{constant}
	\label{eq:quadratic_partial_trace}
\end{equation}
where $\hat{H}_H[P]$ and $\hat{H}_{F}[P]$ are the Hartree and Fock contributions respectively, and the last term is the residual interaction for any species that have not been traced out. We use this repeatedly below.

 We provide a detailed construction, following \cite{bultinck2020ground,repellin2020ferromagnetism, hofmann2021fermionic}. We first derive the standard Hamiltonian for the active bands of TBG in Steps 1-4, then specialize to the case of $\nu=3$.
 \begin{enumerate}
	 \item \textbf{Microscopic Definition}
 We take the total Hamiltonian to be 	
 \begin{equation}
	 \hat{H} := \hat{h}_B + V:\hat{\rho} \hat{\rho}: = \hat{h}_{B} + \frac{1}{2 A_s} \sum_{\v{q} \in \R^2} \sum_{\v{k},\v{k}' \in \bz} V(\v{q}) \,
	 [\Lambda_{-\q}(\k)]_{\alpha\delta} [\Lambda_{\q}(\k')]_{\beta\gamma} \d_{\k,\alpha} \d_{\k',\beta} \c_{[\k'+\q],\gamma} \c_{[\k-\q],\delta}
 \end{equation}
 where the first equality defines an abbreviated notation for concision, $A_s$ is the sample area, and $[\k+\q] = \k + \q - \v{G}$ lies in the first Brillouin zone. The quadratic term $\hat{h}_T$ is the total dispersion of bilayer graphene, and the Fermions $\d_{\alpha}$ are vectors running over all bands (not just the flat bands) and all species.

	 \item \textbf{Reference Density}
		 The dispersion $\hat{h}_B$ is unknown. Why? Consider the simpler situation of graphene. The standard tight-binding model $H = \sum t \d \c$ is defined in terms of single hopping parameter $t \approx \SI{2.8}{eV}$. However, this is \textit{not} the bare electron hopping $t_B$, but rather a renormalized hopping that gives the best one parameter tight-binding model in the presence of electron-electron interactions. Schematically, $\sum t_B \d \c~+~ \hat{V}\big|_{MF} =\sum t \d \c$.

 For twisted bilayer graphene, the same logic should apply. Let us \textit{assume} that a Bistritzer-MacDonald model is the best quadratic model of TBG. We then have
 \begin{equation}
	 \hat{H}_{BM} = \hat{h}_0 + \hat{V}\big|_{MF,P_{0}}
	 \label{eq:subtraction_scheme}
 \end{equation}
 where $\hat{V}\big|_{MF,P_{0}}$, defined in Eq. \eqref{eq:mean_field_operator_projection}, is the interaction at mean-field level with respect to a ``reference" density matrix $P_0$ at charge neutrality. The reference density matrix is non-unique, and three common choices in the literature are:
 \begin{enumerate}
	 \item \textbf{Decoupled subtraction}: take $P_0$ to be the density matrix of two layers of twisted graphene without tunneling between them at charge neutrality and \textit{without an hBN mass}.
	 \item \textbf{BM Subtraction}: take $P_0$ to be the density matrix of the BM model at charge neutrality \textit{without an hBN mass}.
	 \item \textbf{Infinite-temperature subtraction}: take $P_\mathcal{A} = \tfrac{1}{2} I_{\mathcal{A}}$ for the active bands, then remote bands filled at negative energy. We will see below that this choice implies $\hat{h}_C = \hat{H}_{BM}$.
 \end{enumerate}
 This choice is physical, and effects the final Hamiltonian at the level of a few \si{meV}. With whatever choice we make for the reference density matrix, we may solve for the dispersion as
 \begin{equation}
	 \hat{h}_B = \hat{H}_{BM} - \left[\hat{V} : \rho \rho:\right]_{MF,P_{0}} = \hat{H}_{BM} - \hat{H}_{HF}[P_0]
	 \label{eq:total_dispersion}
 \end{equation}
 where we have used Eq. \eqref{eq:quadratic_partial_trace} and \eqref{eq:mean_field_operator_projection}.

 One should think of this as a renormalization procedure: we seek a dispersion so that, after adding interactions, integrating out high energy modes, and reducing to mean-field level, we match IR observations from DFT and/or experiments.\footnote{See \cite{vafek2020renormalization} for a careful treatment from a similar perspective.} Under this philosophy, the reference density aims to reduce double-counting related to the high-energy physics and should not be sensitive to the details of the reference state at low energy. In practice, however, the dispersion is extremely sensitive to the IR details of the reference state within the flat bands. For instance, choosing a reference density without $C_2$ symmetry in the flat bands can cancel out the Fock contribution in some situations. To evade these subtleties insofar as possible, we pick the simplest possible density matrix at half-filling: $P_0 = I_\mathcal{A}/2$, option (c). Ideally, one would like to select $P_0$ so as to reproduce a set of experimental observations.

	 \item \textbf{Integrate out Remote Bands}
 Next we integrate out the remote bands at mean-field level. Let 
 \begin{equation}
	 [P_{\mathcal{R}}]_{\alpha\beta} = \delta_{\alpha\beta} \begin{cases}
	 1 & \text{ $n$ remote and below charge neutrality}\\
	 0 & \text{ otherwise}
	 \end{cases}
 \end{equation}
 be a density matrix for the occupied remote bands. Using Eq. \eqref{eq:quadratic_partial_trace}, we have
 \begin{equation}
	 \tr_{\mathcal{R}}\left[\rho[P_\mathcal{R}] \left(\hat{h}_{B} + V :\hat{\rho} \hat{\rho}: \right)\right] = \left. \hat{h}_B\right|_{\mathcal{A}} +\hat{H}_H[P_{\mathcal{R}}] +\hat{H}_{F}[P_{\mathcal{R}}] + V : \hat{\rho}_{\mathcal{A}} \hat{\rho}_{\mathcal{A}} : +\, \text{constant}
 \end{equation}
 where $\big|_{\mathcal{A}}$ is the restriction to the active bands. Using the linearity of Hartree-Fock, we have
 \begin{equation}
	 \hat{H}\big|_{\mathcal{A}} 
	 = \hat{h}_{B}\big|_{\mathcal{A}} + V : \hat{\rho}_{\mathcal{A}} \hat{\rho}_{\mathcal{A}}:, \quad \hat{h}_{B}\big|_{\mathcal{A}} := \hat{H}_{BM}\big|_{\mathcal{A}} + \hat{H}_{HF}[P_\mathcal{R} - P_0].
	 \label{eq:H_B_form}
 \end{equation}
 We now restrict ourselves to the Hilbert space of the active bands. In a slight abuse of notation, we drop all $\big|_{\mathcal{A}}$'s. Note, however, that the quadratic density matrices are still defined over \textit{all} bands. For practical numerics, one usually keeps $M \approx 3-5$ bands below and above the active bands in $P_{\mathcal{R}}$ and $P_0$.\footnote{Depending on the form of $P_0$, $\hat{h}_B$ may contain a logarithmic divergence with $M$, which is non-physical.} We have now arrived at the form $\hat{H}_B$ of the Hamiltonian.

	 \item \textbf{Regroup to `Standard Form'}
		 We now transform between the various forms of the Hamiltonian to arrive at the standard form for analytics, $\hat{H}_C$. First, we remove the normal ordering. By Fermion anticommutation, $\d_\alpha \d_\gamma \c_\delta \c_\beta = \d_\alpha \c_\beta \d_\gamma \c_\delta + \d_\alpha \c_\delta \delta_{\gamma\beta}$, so
 \begin{equation}
	 :\hat{\rho} \hat{\rho}: = [\Lambda]_{\alpha\beta} [\Lambda]_{\gamma\delta} \d_\alpha \d_\gamma \c_\delta \c_\beta = \d_\alpha [\Lambda]_{\alpha\beta} \c_\beta \d_\gamma [\Lambda]_{\gamma\delta} \c_\delta  - \d_\alpha \left( 
	 [\Lambda]_{\alpha\beta} \delta_{\beta\gamma} [\Lambda]_{\gamma\delta}
	 \right) \c_{\delta}
	 = \hat{\rho} \hat{\rho} + \hat{H}_F[\tfrac{1}{2} I_{\mathcal{A}}]
	 \label{eq:normal_ordering_gives_Fock}
 \end{equation}
 where we have used the definition of the Fock Hamiltonian \eqref{eq:Fock_Hamiltonian}. Therefore
 \begin{equation}
	 \hat{H}_B = \hat{h}_{B} + V :\hat{\rho}\hat{\rho}: = \hat{h}_B + \hat{H}_{F}[\tfrac{1}{2} I_\mathcal{A}] + V\hat{\rho}\hat{\rho} = \hat{h}_A + V \hat{\rho}\hat{\rho} = \hat{H}_A \implies \hat{h}_A =  \hat{h}_B + \hat{H}_{F}[\tfrac{1}{2} I_\mathcal{A}].
 \end{equation}

 Until now, we have worked with $\hat{\rho}_{\q}$, which measures the electron density relative to the vacuum (i.e. $\nu=-4$ after integrating out the remote bands). This is unwieldy in the current context, where most of the density comes from completely filled bands which provide only a background charge density. To rectify this, we switch to measuring charge density relative to a reference state, characterized by a third density matrix $P_C$. We then define the relative charge density by
 \begin{equation}
	 \widehat{\delta \rho}_{\q} := \hat{\rho}_{\q} - \braket{\rho_{\q}}_{P_C}.
 \end{equation}
 Rewriting the interaction with respect to $\widehat{\delta\rho}$ amounts to a quadratic shift:
 \begin{equation}
	 V_{\q} \widehat{\delta\rho}_{\q} \widehat{\delta \rho}_{-\q} = V_{\q} \hat{\rho}_{\q} \hat{\rho}_{-\q} - V_{\q} \left( \hat{\rho}_{\q} \braket{\rho_{-\q}} + \braket{\hat{\rho}_{\q}} \rho_{-\q}  \right) + V_{\q} \braket{\rho_{\q}} \braket{\rho_{-\q}}.
 \end{equation}
 The expectation value may be written in terms of the form factor as
 \begin{equation}
	 \braket{\rho_{-\q}}_{P_C} = \sum_{\k \in \bz} \braket{\d_{\k} \Lambda_{-\q} \c_{[\k-\q]}}_{P_c} = \delta(\v{k} - [\k+\q]) \Tr[P_c \Lambda_{-\q}] = \sum_{\v{G}} \delta_{\q,\v{G}} \Tr[P_{C} \Lambda_{G}^\dagger]
 \end{equation}
 where the sum is over the reciprocal lattice and comes from the fact that $P_C$ is defined only on the Brillouin zone and is $\k$-diagonal. Hence
 \begin{equation}
	 \frac{1}{2} \sum_{\q} V_{\q} \left( \hat{\rho}_{\q} \braket{\rho_{-\q}} + \braket{\hat{\rho}_{\q}} \rho_{-\q}  \right)
	 = \frac{1}{2} \sum_{\q} V_{\q} \sum_{\k \in \bz} \d_{\k} \left( \Lambda_{\q} \sum_{\v{G}} \delta_{\q,\v{G}} \Tr[P_C \Lambda_{\v{G}}^\dagger] \right)\c_{[\k+\q]} + (\v{q} \leftrightarrow -\v{q})
	 = \hat{H}_H[P_C]. 
 \end{equation}

 Most commonly, one takes $P_C$ as the density matrix of the BM model at half filling. Writing $P_C = \tfrac{1}{2}[I_\mathcal{A} + Q]$ one can show \cite{bultinck2020ground} that the symmetries $U(2) \times U(2)$, $\mathcal{C}_2 \mathcal{T}$, and particle-hole symmetry together imply that $\Tr[Q \Lambda_{\v{q}}] = 0$, so that $\hat{H}_H[P_C] = \hat{H}_H[\tfrac{1}{2} I_{\mathcal{A}}]$. Therefore $V \widehat{\delta\rho} \widehat{\delta \rho} + \hat{H}_H[\tfrac{1}{2} I_{\mathcal{A}}] = V \rho \rho$. Altogether, we have
 \begin{equation}
	 \hat{H} = \hat{h}_C + V \widehat{\delta\rho} \widehat{\delta \rho}; \quad 	\hat{h}_C = \hat{h}_A + \hat{H}_H[\tfrac{1}{2} I_{\mathcal{A}}] = \hat{H}_{BM} + \hat{H}_{HF}[P_\mathcal{R} - P_0 + \tfrac{1}{2} I_\mathcal{A}].
 \end{equation}
 One can immediately see why the choice of $P_0$ in the ``infinite temperature" subtraction scheme gives $\hat{h}_C = \hat{H}_{BM}$. 

	 \item \textbf{The Top Band}
		 We now consider the case of $\nu=3$ where a single spin- and valley-polarized band is unoccupied. We can partition the active bands as $\mathcal{A} = \mathcal{B} \cup \mathcal{A}'$, where $\mathcal{A}'$ is the ``top" unoccupied band. Our goal is to express the Hamiltonian in this band after integrating out the other bands at mean-field level.

				 The most straightforward method is to consider the $\mathcal{B}$ bands as (occupied) remote bands: $\mathcal{R}' = \mathcal{R} \cup \mathcal{B}$. As we do not use any properties of the active bands themselves, the same logic that led to Eq. \eqref{eq:H_B_form} now gives
				 \begin{equation}
					 \hat{H}^{(\nu=3)} = \hat{h}_B^{(\nu=3)} + V :\hat{\rho}_{\mathcal{A}'} \hat{\rho}_{\mathcal{A}'}:, \quad \hat{h}_{B}^{(\nu=3)} = \hat{H}_{BM}^{(\nu=3)} + \hat{H}_{HF}[P_{\mathcal{R} \cup \mathcal{B}} - P_0].
				 \end{equation}
				 Note that this is normal-ordered with respect to the $\nu=3$ state (which is also the vacuum, after reducing the Hilbert space) where $7$ out of the $8$ active bands are filled. Furthermore, since we have dropped constant terms from the partial trace, $\rho$ is measuring deviations in density from the same state. That is, $\rho_{\mathcal{A}'} = \delta \rho^{(\nu=3)}$. Therefore we have $h_B^{(\nu=3)} = h_D^{(\nu=3)}$, which is the form we sought.

				 Using properties of the active bands allows us to simplify further. In the infinite-temperature subtraction scheme
				 \begin{equation}
					 P^{\nu=3} = P_{\mathcal{R} \cup \mathcal{B}} - P_0 = \operatorname{diag}
				 \begin{pmatrix} 
					1/2 & 1/2 &1/2 &1/2;& 1/2 &1/2& 1/2& -1/2	
				 \end{pmatrix},
				 \end{equation}
				 where the semicolon separates the active bands below and above the CNP. In the case of hBN-alignment, all four bands above the CNP have different spin/valley quantum numbers. Since the form-factors are flavor-diagonal, the Fock contribution to the 8th band come only from the 4th and 8th band, giving $\hat{h}_F$. Furthermore, with particle-hole symmetry, the Hartree contribution is only weakly band- and valley-dependent, so the contributions from all bands are essentially the same. If we let $\hat{h}_H$ be the Hartree potential of a single band, then we have $H_H[P^{(\nu=3)}] \approx \frac{7-1}{2} H_H = \nu H_H$. Altogther,
				 \begin{equation}
					 \hat{h}_D^{(\nu)} = \hat{H}_{BM}^{(\nu)} + \hat{H}_{HF}[P^{(\nu)}] \approx \hat{H}_{BM}^{(\nu)} + \nu \hat{h}_{H} + \hat{h}_F,
					 \label{eq:h_dispersion_nu}
				 \end{equation}
				 where the approximation is good at zero field. In fact, if we fill $\nu \in \st{0,1,2,3}$ spin- and valley-polarized bands above the CNP, then $\hat{h}_D^{(\nu)}$ is the Hamiltonian for the next polarized band. This concludes our detailed construction of the Hamiltonian Eq. \eqref{eq:projected_Hamiltonian} used in the main text.

 \end{enumerate}

\section{Numerical Details of DMRG Calculations}
\label{app:numerical_details}

This Appendix gives additional numerical details on the DMRG simulations employed. We follow the methods developed in \cite{parker2020local, soejima2020efficient, parker2020strain} and perform DMRG calculations using the open-source TeNPy library \cite{tenpy}, developed by one of us.

To apply DMRG to the 2d problem of \tbg{}, we employ an infinite cylinder geometry with real-space along the cylinder and $L_y$ discrete $k_y$ momenta around the cylinder. We take $L_y=8$ cuts through the Brillouin zone at $k_y = - \pi + 2\pi n/L_y$ and Fourier transform along $x$ to produce maximally localized Wannier-Qi states\cite{qi2011generic}. Explicitly, we define creation operators for the Wannier orbital in the $n$th unit cell along  the $x$ direction
\begin{equation}
	\hat{d}^\dagger_{n,k_y} := \int \frac{dk_x}{\sqrt{Q_x}} e^{i\v{k}\cdot\v{R}_n} e^{i\theta(k_x)} \hat{c}_{k_x,k_y}
\end{equation}
where $\v{R}_n = n \v{a}_1$ and $\theta(k_x)$ is chosen to give maximally localized 1d Wannier orbitals along the cylinder. We may then project the Hamiltonian Eq. \eqref{eq:projected_Hamiltonian} into these degrees of freedom, whereupon it becomes a quasi-1D model with long-range interactions. To perform DMRG, this Hamiltonian must be encoded as an matrix product operator (MPO). A naive encoding of the Hamiltonian results in an MPO with bond dimension $\chi_{\text{MPO}} \sim 10^6$; typical DMRG problems have Hamiltonians with $\chi_{\text{MPO}} \sim 10^1$. Ref. \cite{soejima2020efficient} developed a method, which we apply here, to compress the MPO down to $\chi_{\text{MPO}} \approx 500$, while retaining an accuracy of $10^{-3}$ \si{meV}.  

\begin{figure}
    \centering
    \includegraphics[width=\textwidth]{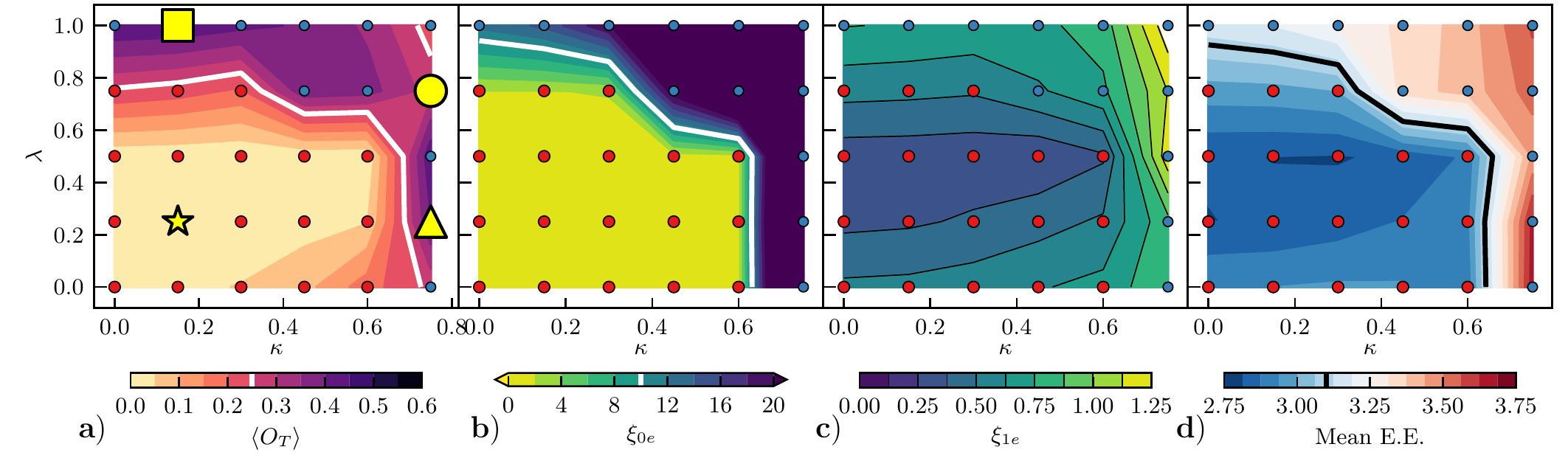}
	\caption{DMRG measurements over the $(\kappa,\lambda)$ plane. Dots show datapoints, and are colored red inside the FCI phase, as determined by the entanglement spectrum. \textbf{a)} Order parameter for translation-breaking along $\v{a}_1$. The yellow shapes are special points where higher bond dimension calculations were performed. \textbf{b)} Largest correlation length of the 0-electron sector of the MPS transfer matrix, i.e. ``the correlation length". \textbf{c)} Largest correlation length from the 1-electron sector of the MPS transfer matrix. \textbf{d)} Mean entanglement entropy within a unit cell. Bold white or black contours are guides fo the eye showing the approximate location of the phase transition. Parameters: $\theta=1.06^\circ$, $\epsilon_r = 10$,  $d=25$\si{\nano\meter}, $\chi=256$.}
	\label{fig:appendix_dmrg_phase_diagrams}
\end{figure}

\begin{figure}
    \centering
    \includegraphics[width=\textwidth]{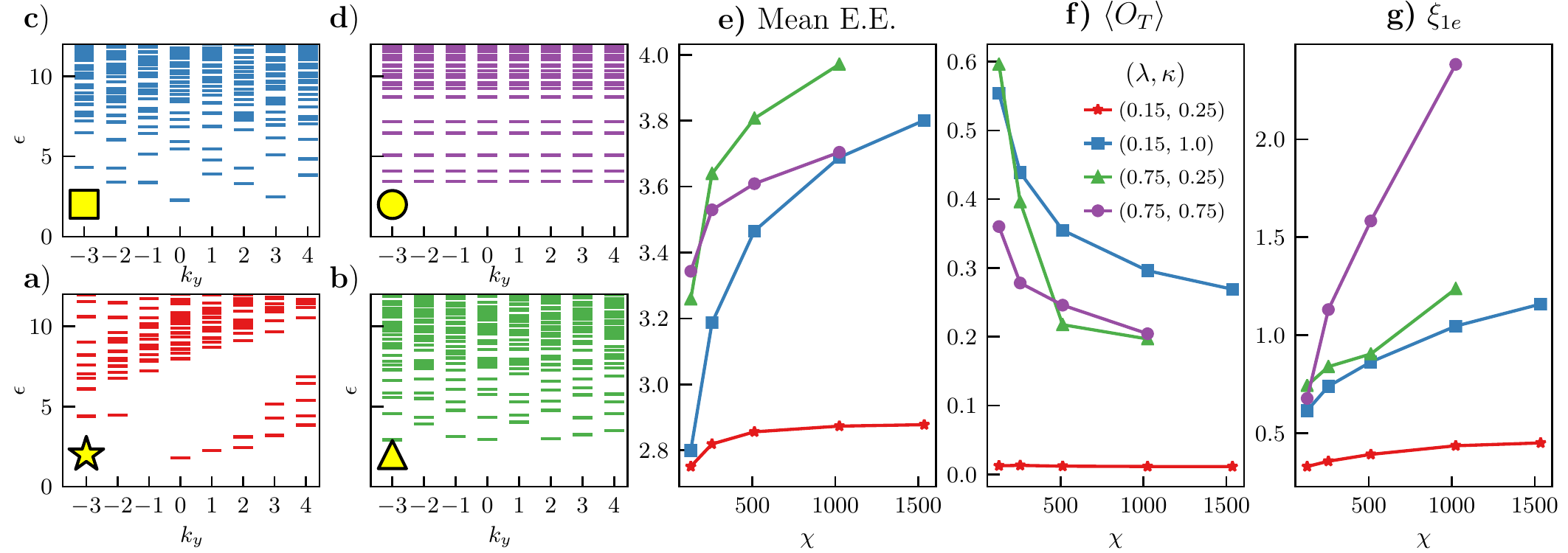}
	\caption{\textbf{abcd)} Entanglement spectra at the four special points [see Fig \ref{fig:appendix_dmrg_phase_diagrams}a or the legend in panel f)] at $\chi_{\text{mps}} =1024$. \textbf{e)} Mean entanglement entropy over a unit cell at the special points as a function of the MPS bond dimension. \textbf{f)} Order parameter for translation-breaking at the special points. \textbf{g)} Largest correlation length of the 1-electron sector of the MPS transfer matrix at the special points. One can see that the extent of the FCI phase is discernable by $\chi=256$, as used in the main text.  Parameters are the same as Fig. \ref{fig:appendix_dmrg_phase_diagrams}.}
    \label{fig:appendix_special_points}
\end{figure}

We then perform DMRG over a range of $\lambda$ and $\kappa$ values at bond dimension $\chi=256$, shown in Fig. \ref{fig:appendix_dmrg_phase_diagrams} and in the main text. (Recall that $\lambda$ directly scales the dispersion.) As there are only $8$ sites per unit cell, and the FCI phase is well-described by a small bond dimension, this is sufficient to capture the essential physics. Each of the observables shown in Fig. \ref{fig:appendix_dmrg_phase_diagrams} separately suggests a phase transition out of the FCI phase found at smaller $\kappa$ and $\lambda$. In particular, the translation order parameter along the $\v{a}_1$ direction is large both at large $\kappa$ and at large $\lambda$, suggesting that the competing phase breaks translation symmetry strongly. Near $(\kappa,\lambda) = (3/4,1)$, i.e. in the upper-right hand corner, DMRG converges to `cat' states, whose transfer matrix has a largest eigenvalue with an 8-fold degeneracy. \Dan{Note $L_y =8$.} This suggests the ground state may break translation symmetry, but in a direction that is not along the cylinder and hence cannot be captured in this geometry. Fig. \ref{fig:appendix_dmrg_phase_diagrams}c shows that the 1-electron correlation length is large and growing with $\chi$ in this region, which suggests the possibility of a subdominant metallic character to the state. (if there is a Fermi surface, then it diverges with $\chi$.) A Hartree-Fock calculation suggests that a gapped state which breaks translation symmetry along $\v{a}_1$ is lower energy than a metallic state at this filling, but this may not be the true ground state.

To ensure these observations are not artifacts of low bond dimension, we have performed higher bond dimension calculations, and show four special points in Fig. \ref{fig:appendix_special_points}. The entanglement spectrum in panels abcd clearly shows a different character inside and outside of the FCI phase. As remarked in the main text, the FCI phase (Laughlin state) is easily distinguished by a chiral entanglement spectrum whose low-lying states have counts of the corresponding chiral CFT partition function. Though we have chosen a point deep in the FCI phase, the entanglement spectrum pattern is also clear even near the boundary. We note that even an ideal Laughlin state on a cylinder has a small amount of translation breaking, and relatively large translation-breaking can coexist with the FCI phase. This makes it difficult to identify the precise phase boundary and the nature of any phase transition out of the FCI phase; we see no clear signs of a continuous transition. Panels efg) of Fig. \ref{fig:appendix_special_points} show how various quantities scale with the MPS bond dimension. It is clear that the FCI is well-converged by $\chi=256$ or $\chi=512$, and has a low translation-breaking, entanglement entropy, and 1-electron correlation length. For the other points, $\langle O_T \rangle$ decreases with $\chi$ but asymptotes to a finite value. As remarked above, $\xi_{1e}$ may diverge with $\chi$ for the points as $\kappa = 3/4$, but it is difficult to determine this precisely as most of the state's ``entanglement budget" is used to make cat states.

Despite these subtleties, we may draw two conclusions. First, the extent of the FCI phase may be clearly determined by the entanglement spectra and, less certainly, from the quantities shown in Fig. \ref{fig:appendix_dmrg_phase_diagrams}. Second, although the precise nature of the competing phase (or phases) is elusive, they exhibit large translation-breaking and thus have CDW order. We note that the same analysis was performed on the $(\theta,\kappa)$ plane, with similar conclusions that were summarized in Fig. \ref{fig:larger_theta} of the main text.

\section{Constant Berry curvature and the GMP algebra}
\label{app:GMP}

In this section we show that if $\sigma[\mathcal{F}] = T[\eta] = 0$, i.e. if the Berry curvature is uniform and the trace condition holds
\begin{equation}
  \mathcal{F} = 2\pi C \, \textrm{Id}_N \quad \& \quad \Tr[\tr g] = \Tr[\mathcal{F}],
  \label{eq:C2_and_C3}
\end{equation}
then the (multiband) projected density operators satisfy the GMP algebra \cite{girvinMagnetorotonTheoryCollective1986}. This was first proven in  Ref.~\onlinecite{royBandGeometryFractional2014}, and elegantly reformulated and applied by Ref.~\onlinecite{varjas2021topological}. Our proof is a direct generalization of the latter work. For concreteness, we take $\zeta = +1$ and accordingly $\F_0 > 0$ (the $\zeta = -1$ case is identical \textit{mutatis mutandis}). Along the way we show that $T[\eta]=\Tr[\tr g] - \Tr[\mathcal{F}] \geq 0$.

\newcommand{\PP}{\mathcal{P}_{\psi}}
\newcommand{\QQ}{\mathcal{Q}_{\psi}}

Consider $N$ bands (separated by a gap), and define the single particle projector
\begin{equation}
    \PP = \sum_{\bk} \sum_{a=1}^N \ket{\psi^a_\bk} \bra{\psi^a_\bk}.
\end{equation}
Note that this is \textit{different} from the band projector $P(\k) = 1 - Q(\k) = \sum_a \ket{u^a_\bk} \bra{u^a_\bk}$ used in the definitions of the band geometry. This distinction is necessary because the projected density operators and, and hence the GMP algebra, are naturally expressed in terms of the single particle wavefunctions $\psi(r)$, rather than the Bloch states. The single particle projected density operator is defined as
\begin{equation}
    \hat{\rho}_\bq= \PP e^{-i \bq \cdot \hat{\br}} \PP.
    \label{densityops}
\end{equation}
Our goal is to show that Eq. \eqref{eq:C2_and_C3} implies that the density operators obey the GMP algebra \cite{girvinCollectiveExcitationGapFractional1985}
\begin{equation}
    [\hat{\rho}_\bq, \hat{\rho}_{\bq'}] = 2i \sin\left( \frac{\F_0 \bq \times \bq'}{2} \right) e^{\F_0 \abs{\bq}^2/2} \hat{\rho}_{\bq + \bq'}.
    \label{GMP}
\end{equation}
We show this by first translating our knowledge of the trace condition on the level of the $\ket{u_{\k}^a}$ to the single-particle $\ket{\psi_{\k}^a}$'s, where holomorphicity places a crucial role. The GMP algebra will then follow by direct computation.

Differentiating $\ket{\psi_\bk^a} = e^{i \bk \cdot \br} \ket{u_\bk^a}$ with respect to $\v{k}$ and rearranging gives
\begin{equation}
    \hat{\br} \ket{\psi_\bk^a} = -i \bnabla_\bk \ket{\psi_\bk^a} - e^{i \bk \br} i \bnabla_\bk \ket{u_\bk^a}.
\end{equation}
(We note that the action of the position operator can be quite subtle in $k$-space  \cite{blount1962formalisms}.) Acting with $\QQ = 1- \PP$ annihilates the first term entirely and allows us to replace $\QQ$ with $Q(\k)$ in the second term, since translation symmetry gives a $\delta$-function in $k$-space:
\begin{equation}
    \QQ \br \ket{\psi_\bk^a} =  -i e^{i \bk \br} Q(\bk)  \bnabla_\bk \ket{u_\bk^a}.
\end{equation}
Taking complex coordinates $z = x+iy$ in real space, we therefore have
\begin{equation}
    \QQ \hat{z} \ket{\psi_\bk} = 0 \iff Q(\bk) \ov{\partial}_k \ket{u_\bk} = 0.
    \label{eq:realspace_fromtrcond}
\end{equation}
The right hand side holds if and only if the (multiband) trace condition is satisfied --- which it is, by hypothesis. To see this, for each $a$, $\k$, we directly compute 
\begin{equation}
    \tr g^{aa}(\k) - \zeta \F^{aa}(\k) = 
    \bra{ \partial_\zeta u_\bk^a } Q(\k) \ket{ \partial_\zeta u_\bk^a} \geq 0 
    \label{trcond_holo}
    \end{equation}
from the definition of the quantum metric. The trace condition holds if and only if the summed inequality $\Tr \tr g \geq \Tr \F$ is saturated which shows that $T[\eta]\geq0$ as we claimed in the main text. Additionally, the sum of inequalities is saturated if and only if each summand is saturated, which means $\tr g^{aa}(\k) - \zeta \F^{aa}(\k) = 0$ for all $a$ and $\k$. Since $Q = Q^2 = Q^\dag$ is positive semidefinite we must have $Q(\bk) \ov{\partial}_k \ket{u_\bk} = 0$ as we claimed above.

Eq. \eqref{eq:realspace_fromtrcond} implies $\QQ \hat{z} \PP = 0$ or $\hat{z}\PP = \PP \hat{z} \PP$. Then using Baker-Campbell-Hausdorff and the fact that $[\PP \hat{x} \PP, \PP \hat{y} \PP] = -i \F = -i \F_0 \in \C$ is a constant by \eqref{eq:C2_and_C3}, we compute
\begin{equation}
    \hat{\rho}_\bq  = P_\psi e^{i \bq \cdot \hat{\br}} P_\psi = P_\psi e^{\frac{i}{2} q \ov{z}}e^{\frac{i}{2} \ov{q} z } P_\psi = e^{\frac{i}{2} q P_{\psi} \ov{z} P_{\psi}}e^{\frac{i}{2} \ov{q} P_{\psi} \hat{z} P_{\psi} } = e^{i \bq \cdot P \hat{\br} P} e^{- \F_0 \abs{\bq}^2/4}.
\end{equation}
The GMP algebra \eqref{GMP} then follows by direct computation of the commutator, again using Baker-Campbell-Hausdorff.

This result highlights how the trace condition, a fact about $\bk$ space geometry, tells us about real space holomorphicity. Namely, multiplying by the holomorphic coordinate $z$, and applying any power series in $z$ (i.e. any holomorphic function), keeps the wavefunction within the bands of interest.

We also note that $[P_\psi \hat{x} P_\psi,P_\psi \hat{y} P_\psi] = -i \F_0$ means that the electrons feel a uniform effective magnetic field, just as in the lowest Landau level. Furthermore, the GMP algebra allows Bloch states to be ``boosted" to any other momentum, which implies that the total bandwidth of the $N$ bands together must vanish.

\end{widetext}

\end{document}